\documentclass[prd,tightenlines,nofootinbib,superscriptaddress]{revtex4}

\makeatletter
\def\l@subsubsection#1#2{}
\makeatother

\usepackage{amsfonts,amssymb,amsthm,bbm,graphicx,mathtools}
\usepackage{subcaption}
\usepackage{amsmath}

\usepackage{hyperref}

\usepackage{subcaption}
\usepackage{color}
\usepackage{tikz}
\usetikzlibrary{calc}
\usetikzlibrary{decorations.pathmorphing}
\usetikzlibrary{decorations.markings}

\newcommand{\C}{{\mathbb C}}
\newcommand{\N}{{\mathbb N}}
\newcommand{\R}{{\mathbb R}}
\newcommand{\Z}{{\mathbb Z}}

\newcommand{\cA}{{\mathcal A}}
\newcommand{\cB}{{\mathcal B}}
\newcommand{\cC}{{\mathcal C}}
\newcommand{\cD}{{\mathcal D}}
\newcommand{\cE}{{\mathcal E}}

\newcommand{\cH}{{\mathcal H}}
\newcommand{\cI}{{\mathcal I}}
\newcommand{\cJ}{{\mathcal J}}

\newcommand{\cL}{{\mathcal L}}
\newcommand{\cM}{{\mathcal M}}

\newcommand{\cO}{{\mathcal O}}
\newcommand{\cP}{{\mathcal P}}

\newcommand{\cR}{{\mathcal R}}

\newcommand{\cU}{{\mathcal U}}
\newcommand{\cV}{{\mathcal V}}

\newcommand{\SU}{\mathrm{SU}}

\newcommand{\su}{{\mathfrak{su}}}

\newcommand{\iso}{{\mathfrak{iso}}}

\newcommand{\lA}{{\mathfrak{a}}}
\newcommand{\lR}{{\mathfrak{r}}}
\newcommand{\deq}{\coloneqq}
\newcommand{\pa}{{\partial}}
\newcommand{\HH}{{\widehat{H}}}

\def\ip{\lrcorner}

\DeclarePairedDelimiter\floor{\lfloor}{\rfloor}

\def\be#1\ee{\begin{equation}#1\end{equation}}
\def\beq#1\eeq{\begin{eqnarray}#1\end{eqnarray}}
\def\bea#1\eea{\begin{align}#1\end{align}}
\def\nn{\nonumber}

\newcommand{\f}{\frac}
\newcommand{\la}{\langle}
\newcommand{\ra}{\rangle}

\newcommand{\tr}{{\mathrm{Tr}}}
\newcommand{\tra}{{\mathrm{t}}}
\newcommand{\ga}{{\mathrm{g}}}

\def\pp{\partial}
\def\rd{\textrm{d}}

\newcommand{\id}{\mathbb{I}}

\def\act{\triangleright}

\def\vphi{\varphi}
\def\eps{\epsilon}

\def\hg{\hat{g}}
\def\hh{\hat{h}}
\def\hy{\hat{y}}
\def\hj{\hat{j}}
\def\hp{\hat{p}}
\def\hJ{\hat{J}}
\def\hP{\hat{P}}
\def\hcD{\hat{\cD}}

\def\hX{\widehat{X}}

\def\hG{\widehat{G}}

\newcommand\hx{\hat{x}}

\def\A{p}
\def\e{ j}
\def\ttheta{\tilde{\theta}}
\def\vJ{\vec{J}}

\def\centerarc[#1](#2)(#3:#4:#5)
{ \draw[#1] ($(#2)+({#5*cos(#3)},{#5*sin(#3)})$) arc (#3:#4:#5); }

\def\centerarcnodes[#1](#2)(#3:#4:#5)(#6,#7)
{\coordinate(#6) at ($(#2)+({#5*cos(#3)},{#5*sin(#3)})$);
	\coordinate(#7) at ($(#2)+({#5*cos(#4)},{#5*sin(#4)})$);
	\draw[#1] ($(#2)+({#5*cos(#3)},{#5*sin(#3)})$) arc (#3:#4:#5); }

\def\angcircle(#1)(#2)(#3:#4) {\coordinate(#1) at ($(#2)+({#4*cos(#3)},{#4*sin(#3)})$); }

\tikzset{->-/.style={decoration={
			markings,
			mark=at position #1 with {\arrow{>}}},postaction={decorate}}}

\tikzset{-<-/.style={decoration={
			markings,
			mark=at position #1 with {\arrow{<}}},postaction={decorate}}}

\begin{document}

\title{The Quantum Gravity Disk: Discrete Current Algebra}

\author{{\bf Laurent Freidel}}\email{lfreidel@perimeterinstitute.ca}
\affiliation{Perimeter Institute for Theoretical Physics, 31 Caroline Street North, Waterloo, Ontario, Canada N2L 2Y5}

\author{{\bf Christophe Goeller}}\email{christophe.goeller@ens-lyon.fr}
\affiliation{Universit\'e de Lyon, ENS de Lyon,  Laboratoire de Physique, CNRS UMR 5672, F-69342 Lyon, France}

\author{{\bf Etera R. Livine}}\email{etera.livine@ens-lyon.fr}
\affiliation{Universit\'e de Lyon, ENS de Lyon,  Laboratoire de Physique, CNRS UMR 5672, F-69342 Lyon, France}

\date{\today}

\begin{abstract}

We study the quantization of the corner symmetry algebra of 3d gravity, that is the algebra of observables associated with 1d spatial boundaries. In the continuum field theory, at the classical level, this symmetry algebra is given by the central extension of the Poincar\'e loop algebra.
At the quantum level, we construct a discrete current algebra based on a  quantum symmetry group given by the Drinfeld double $\cD\SU(2)$.
Those discrete currents depend on an integer $N$,  a discreteness parameter, understood as the number of quanta of geometry on the 1d boundary: low $N$ is the deep quantum regime, while large $N$ should lead back to a continuum picture.
We show that this algebra satisfies two fundamental properties. First, it is compatible with the quantum space-time picture given by the Ponzano-Regge state-sum model, which provides  discrete path integral amplitudes for 3d quantum gravity. The integer $N$ then counts the flux lines attached to the boundary.
Second, we analyse the refinement, coarse-graining and fusion processes as $N$ changes, and we show that the $N\rightarrow\infty$ limit is a classical limit where we recover the Poincar\'e current algebra.
Identifying such a discrete current algebra on quantum boundaries is an important step towards understanding how conformal field theories arise on spatial boundaries in quantized space-times such as in loop quantum gravity.
%



\end{abstract}

\maketitle
\tableofcontents

\section*{Introduction}

The quest for quantum gravity can be understood as the search for the fundamental symmetry algebra of space-time and its relevant (quantum) representation(s). The foremost difficulty is the conceptual clash between the physical picture of local degrees of freedom, with gravitons as described by quantum field theory, and the diffeomorphism invariance of classical general relativity, which erases the physicality of a space-time point and makes localization a truly non-trivial process especially when moving at the quantum level. This apparent paradox is resolved by the deep insight of the holographic behaviour of gravity:  the physical degrees of freedom of the bulk geometry of a region of space-time would be faithfully represented by a boundary field theory living on the region's boundary and quasi-local bulk observables would then be defined as non-local boundary correlations through a stereographic or tomographic localization map (e.g. \cite{Czech:2016xec,Donnelly:2018qya}). Holography mimics the existence of local degrees of freedom while implementing the gauge symmetry under bulk diffeomorphisms.
From this holographic perspective, understanding the boundary symmetry becomes essential. The bulk geometry dynamics can be described by edge modes living and propagating on the space-time boundary (e.g. \cite{Donnelly:2016auv}) and gravitational observables arise as boundary conserved charges, which generate the boundary symmetry algebra. This algebra is supposed to encode the whole physical content of the theory.

Once the fundamental symmetry algebra of classical gravity is properly identified, from a bulk or a boundary perspective, the next crucial question is the quantization of the theory: does quantum gravity result from the straightforward quantization of the symmetry algebra or does it require an extra ingredient?
More precisely, is it enough to find a representation of the diffeomorphism group -and more generally of the classical boundary symmetry algebra- or does quantum gravity require an in-depth modification of the notion of diffeomorphisms which takes into account the intrinsic Planck length scale?
This critical question is about the validity of the quantization process, used to define quantum mechanics and quantum field theory, when applied to the gravitational interaction field and the space-time geometry  in order to derive a fundamental theory of a quantum space-time. Is the quantization procedure, either canonical or by path integral, a paramount principle to define a microscopic quantum theory from its classical macroscopic regime or should the notion of ``quantum'' be affected by our revision of our concept of space-time in quantum gravity?
For instance, classical conformal field theories (CFTs) acquire a central charge and conformal anomalies at the quantum level, which modify their classical symmetry algebra. More generally, condensed matter integrable models are solved by introducing a $q$-deformation of their classical symmetry group, which is the remnant effect of the microscopic origin of their physical degrees of freedom in their continuum field description.
A similar deformation (or extension) of the symmetry algebra is expected in quantum gravity as a result of the ``atomic'' structure of space-time geometry at the Planck scale, as in non-commutative geometry (and related phenomenological models based on quantum deformation of the Poincar\'e group as  in doubly special relativity or relative locality \cite{AmelinoCamelia:2002wr,KowalskiGlikman:2002we,AmelinoCamelia:2011bm}) or in string theory and related approaches.

This leads to consider the continuum theory (general relativity) as a large number of quanta regime $N\rightarrow\infty$ and the fundamental theory (quantum gravity) as a small number of quanta regime $N\sim \cO(1)$, with the actual symmetry algebra explicitly depending on $N$ or more generally encompassing  all possible values of $N$ and shifts in $N$. This number of quanta is then an extra quantum number, not present in the classical theory (e.g. \cite{Ghosh:2011fc}) but crucial to the description of the quantum theory.

Putting those two insights together, the holographic boundary symmetry algebra and the quantum deformation of the symmetry, hints towards a discrete version of holography, similarly to a quasi-local version of the  $1/N$ expansion of AdS/CFT correspondence, which would bridge between string theory's $S$-matrix for asymptotic states and loop quantum gravity's Planck scale quantum geometry.
There has been recently much progress in this direction, for instance on the AdS/CFT correspondence,
on the CFT limit of SYK models and the large $N$ regime of tensor models \cite{Bonzom:2014oua,Gurau:2016lzk}, on the boundary symmetry of general relativity in loop quantum gravity \cite{Freidel:2015gpa,Freidel:2016bxd,Freidel:2019ees,Freidel:2019ofr,Freidel:2020xyx,Freidel:2020svx,Freidel:2020ayo}.
Here, we would like to make precise statements as to the boundary symmetry algebra of three-dimensional gravity, both at the classical level and then at the quantum level.

\medskip

Starting with classical 3d gravity, we revisit the claim that its boundary charges on a space-time cylinder form a BMS algebra.
The solid cylinder consists in a two-dimensional disk evolving in time. The disk boundary is the one-dimensional circle, this is the canonical boundary, or ``corner'' of space-time. The solid cylinder's boundary is the 2d cylinder describing the circle's evolution in time, i.e. the time evolution of the spatial boundary or equivalently the time-like boundary of the evolving spatial slice.
Formulating 3d gravity as a gauge theory in terms of vierbein and connection variables, we show in  section \ref{sec:3dgrav} that the boundary symmetry algebra is actually a current algebra, defined as the Poincar\'e loop algebra. The BMS charges can then be defined by a twisted Sugawara construction and are selected by specific boundary conditions for the classical fields.

We then shift our focus to the quantum theory, which is the heart of the present study.
Considering 3d Euclidean quantum gravity as a discretized quantum gauge field theory, we describe the 1d boundary circle as carrying the holonomy of the connection around the circle with a number $N$ of flux insertions along it.
This definition is consistent with the quantum space-time picture at the Planck scale provided by the Ponzano-Regge model \cite{PR1968,Freidel:2004vi,Freidel:2004nb,Freidel:2005bb, Barrett:2008wh,Goeller:2019apd}. This discrete topological state-sum for the $\SU(2)$ gauge group is the vanishing cosmological constant limit or vanishing $q$-deformation of the Turaev-Viro topological invariant \cite{TuraevViro1992}. As such, it corresponds to the Reshetikhin-Turaev topological invariant for the Drinfeld double $\cD\SU(2)$ \cite{ReshetikhinTuraev1991,Freidel:2004nb}. Pushing in this direction, the Ponzano-Regge partition function was further shown to give the Reidemeister torsion (also equivalent to the Ray-Singer torsion) and thus to provide knot invariants \cite{Barrett:2008wh}. Coming back to its relation to gravity, it has been shown by Ooguri \cite{Ooguri:1991ni} to be equivalent to the quantization of Witten's reformulation of 3d gravity as a Chern-Simon theory \cite{Witten:1988hc} (see \cite{Meusburger:2008bs,Dupuis:2017otn,Dupuis:2019yds} for recent work on this issue).   Finally, it also provides the transition amplitudes, in terms of 6j recoupling symbols, between spin network states for 3d loop quantum gravity \cite{Rovelli:1993kc}, defining the projector onto physical flat connection states, which solve the Wheeler-de Witt equation at the quantum level \cite{Ooguri:1991ni,Bonzom:2011hm,Bonzom:2011nv,Bonzom:2014bua}.

In this context, we show that the algebra of boundary observables of the quantum disk with $N$ insertions is a discrete current algebra. It explicitly depends on the ``number of quanta of geometry'' $N$. Spatial boundary states are then described as representations of this discrete current algebra.
In section \ref{sec:algebra}, we formulate this algebraic structure as quantum doubles. And we show that, for all values of $N$, the 0-modes can be factorized. Indeed, the 0-modes define the symmetry group as the  Drinfeld double $\cD\SU(2)$, while the remaining modes are encoded into a product of Heisenberg doubles which describe the vibration modes of the quantum boundary circle.
In section \ref{sec:exploring}, we introduce coarse-graining and refinement operations, allowing to slide from the fundamental $N=1$ case, where the algebra of observable reduces to the symmetry group $\cD\SU(2)$, to the continuum limit $N\rightarrow \infty$, where we recover the Poincar\'e loop algebra as expected.
We conclude with a discussion of the disk algebra fusion operations.

This work puts on a solid ground the kinematics and excitation modes in 3d quantum gravity of canonical point-like defects, or ``punctures'', and thus of the 1d world-lines they generate by their evolution in time. Describing their algebraic structure by a current algebra, in both the continuous classical and discrete quantum theories, this constitute a solid step forward towards a better understanding of the realization of a quasi-local holography and asymptotic AdS/CFT correspondence and the arising of the BMS symmetry in 3d gravity in the full quantum regime from the picture of a  locally quantized space-time geometry, as described by the Ponzano-Regge topological state-sum.

\section{Classical edge modes as a current algebra}
\label{sec:3dgrav}

In this section, we review how the classical conserved charges of 3d gravity as a classical field theory, in its first order  formulation in terms of vierbein and connection, form a Poincar\'e current algebra.
We particularly insist on the role of boundary conditions, on the resulting boundary theory and edge modes, on how they determine the conserved charges and thus the boundary symmetry algebra.

\subsection{Classical boundary charges as a current algebra}

Gravity in a three-dimensional Euclidean space-time, with vanishing cosmological constant is a topological field theory best described as a BF theory in the first order formalism.
The basic fields are an $\su(2)$-valued coframe field $e$  and an  $\su(2)$-valued connection $A$ with curvature $F(A) =\rd A + \tfrac12 [A,A]$. We can decompose them into components by introducing a basis of the Lie algebra $\su(2)$ given by the matrices $\tau^a =\sigma^a/2i$ in terms of the Pauli matrices  $\sigma^a$. Writing $e= e_{a} \tau^a$ and $A = A_{a} \tau^a$, the action is then an integral over a 3D space-time region $\cM$:
\be\label{bulkL}
S = \int_{\cM} L
\,,\qquad
L= e^a \wedge F_a(A)=-2\, \tr(e \wedge F[A])
\,.
\ee
In the covariant phase space framework, the variation of the Lagrangian $\delta L$ determines both the bulk equations of motion and the symplectic potential $\Theta$:
\be
\delta L= E + \rd \Theta
\qquad\textrm{with}\qquad
\Theta =  \delta A_a\wedge e^a
\quad\textrm{and}\quad
E=\delta e^a \wedge F_a(A) + \delta A_a \wedge \rd_A e^a
\,.
\ee
The equation of motion $E=0$ imposes the flatness of the connection, $F(A)=0$, and the torsionlessness, $\rd_A e=0$.

This theory possesses two types of gauge symmetries: local Lorentz transformations and local translations.
Lorentz transformations correspond to the $\SU(2)$ gauge symmetry and are parametrized by a  $\su(2)$-valued function $\alpha$ with action:
\beq
\delta^{\ga}_\alpha e &=& - [\alpha, e]
\,,\qquad
\delta^{\ga}_\alpha A= \rd_A\alpha
\,.
\eeq
Local translations are also parametrized by a $\su(2)$-valued function $\varphi$, with action
\be
\delta^{\rm{t}}_\varphi e =\rd_A \varphi
\,,\qquad
\delta^{\tra}_\varphi A =0
\,.
\ee
In the Hamiltonian setting, working with a region of space-time foliated as $\R\times\Sigma$ where $\Sigma$ is a codimension-one hypersurface (i.e. a 2d surface) with boundary $S=\partial \Sigma$, the symplectic form  is then defined as $\Omega_\Sigma \deq\int_\Sigma \delta \Theta$.
The gauge  transformations are Hamiltonian, that is they satisfy $I_{\delta^{\ga}_\alpha} \Omega_\Sigma = -\delta J_\alpha$ and
$I_{\delta^{\tra}_{\varphi}}\Omega  =- \delta P_{\varphi}$ with Hamiltonian densities given by
\be
J_{\alpha}
=-\int_\Sigma\alpha_a \rd_A e^a +\oint_{S} \alpha_a e^a, \qquad P_{\varphi}
= -\int_\Sigma\varphi^a F_a(A)+\oint_{S} \varphi^a A_a
\,.
\ee
%
On-shell of the constraint surface, i.e. assuming $F(A) =0$ and $\rd_A e=0$, the bulk component of these generators vanish and we are left with the boundary charges generating the boundary symmetry associated with transformations $(\alpha,\varphi)$  that do not vanish  on $S$:
\be\label{JP}
J_\alpha\hat{=} \oint_{S} \alpha_a e^a, \qquad  P_{\varphi}
\hat{=} \oint_{S} \varphi^a A_a.
\ee
The boundary symmetry algebra is easily computed from the symplectic structure
and  we obtain a closed Lie algebra:
\be\label{eq:LoopA}
\{ J_{\alpha},J_{\beta}\}= J_{[\alpha,\beta]},\qquad
\{J_{\alpha}, P_{\varphi}\}= P_{[\alpha,\varphi]} +  \oint_{S} \varphi^a \rd \alpha_a,\qquad
\{P_{\varphi},P_{\varphi'}\}=0.
\ee
In other words the 3d gravity boundary symmetry algebra is the Poincar\'e loop algebra  $L(\mathfrak{iso}(3))$ with central extension.
While it is often loosely stated that the asymptotic symmetry algebra of 3d gravity is the 3D BMS algebra, this statement is imprecise as explained in \cite{Geiller:2017xad,Grumiller:2017sjh}. The full symmetry algebra of 3D gravity is the Poincar\'e loop algebra. As we review in details in appendix, the BMS algebra can be recovered as a sub-algebra, within the enveloping loop algebra, preserving a restricted set of boundary conditions.

\medskip

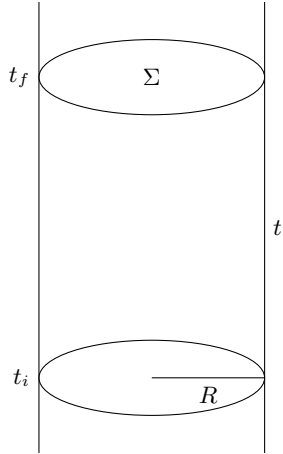
\begin{figure}[!htb]
	\begin{tikzpicture}[scale=1]
	\coordinate (A) at (0,0);
	\coordinate (B) at (0,4);

	\coordinate (A1) at (1.5,0);
	\coordinate (A2) at (-1.5,0);
	\coordinate (B1) at (1.5,4);
	\coordinate (B2) at (-1.5,4);

	\draw (A) ellipse (1.5cm and 0.5cm);
	\draw (B) ellipse (1.5cm and 0.5cm);

	\draw (A) --node[below]{$R$} (A1);

	\draw (A1) --node[right]{$t$} (B1);
	\draw (A2) -- (B2);

	\draw (A1)--+(0,-1);
	\draw (A2)--+(0,-1);
	\draw (B1)--+(0,1);
	\draw (B2)--+(0,1);

	\node[left] at (A2) {$t_i$};
	\node[left] at (B2) {$t_f$};
	\node at (B) {$\Sigma$};

	\end{tikzpicture}
	\caption{Space-time as an infinite solid cylinder, with spatial extension $R$. The canonical surface at $t_i$ and $t_f$ is a two-dimensional slice with the topology of a disk and with boundary at radius $r=R$ a circle.}
	\label{figapp:I:cylinder_space_time}
\end{figure}

In order to study the algebra of boundary charges,  let us consider the space-time as a solid cylinder and parametrize with coordinates $(t,r,\theta)$  where $t\in[t_{i},t_{f}]$ denotes the time development, $r\in [0,R]$ is the radial coordinates, $\theta\in [0,2\pi]$ the angular coordinate, as illustrated on figure \ref{figapp:I:cylinder_space_time}.
The canonical surface is the 2d slice $\Sigma$ at a fixed time $t$: it is a disk with boundary $S$ at radius $r=R$.
The  symmetry currents are given by  $(\A(\theta), \e(\theta))\deq (A_\theta, e_\theta)$ which are functionals on the boundary circle $S$.
It is customary to give the distributional version of the current algebra \eqref{eq:LoopA}:
\be
\{\e^{a}(\theta),\e^{b}(\ttheta)\}
=
\eps^{abc}
\e^{c}(\theta)\delta(\theta-\ttheta)
\,,\qquad
\{\e^{a}(\theta),\A^{b}(\ttheta)\}
=
\eps^{abc}
\A^{c}(\theta)\delta(\theta-\ttheta)
-\delta^{ab}\pp_{\theta}\delta(\theta-\ttheta)
\,,\qquad
\{\A^{a}(\theta),\A^{b}(\ttheta)\}
=
0\,.
\ee
It is also convenient to smear these functionals and define their Fourier components:
\be
J^a_n \hat{=}\oint_S e^{-in \theta} \e^{a} \rd\theta,\qquad
P^a_n \hat{=} \oint_S e^{-in\theta} \A^{a} \rd \theta\,,
\ee
with the normalized measure on the circle, $\oint_S  \rd \theta=1$.
In terms of the Fourier components, the algebra reads
\begin{equation}
\left\{ J_n^a , J_m^b \right\} =  \epsilon^{abc} J^c_{n+m}
\,, \qquad
\left\{ J_n^a , P_m^b \right\}
=
\epsilon^{ a bc} P^c_{n+m} -{i n} \delta^{ab} \delta_{n+m}
\,,  \qquad
\left\{ P_n^a , P_m^b \right\}
=0
\,.
\label{eq:commutators_fourier_component}
\end{equation}
Below, we put this boundary symmetry algebra in perspective with the choice of boundary conditions.

\subsection{Boundary conditions and boundary theory}

The main point of this work is to consider space-time in a bounded region. Therefore, we need to specify the boundary conditions for the field on the cylinder $C_R$ span by $(t,\theta)$ at $r=R$.

To do so, the first step is to ensure the continuity of the bulk at the boundary: the boundary fields are required to satisfy the pull-back of the equations of motion $F(A)=0$ and $\rd_{A}e=0$ on the boundary, which read:
\be
\label{boundaryeqn}
\left|
\begin{array}{lcl}
	\partial_t \A &=& \pp_\theta A_t +[\A,A_{t}]
	\,,
	\vspace*{1mm}\\
	\partial_t \e &=& \pp_\theta e_t + [\A, e_t]+ [\e, A_t]\,,
\end{array}
\right.
\ee
where $\nabla_\theta =\partial_\theta + [A_\theta,\cdot]$ is the covariant derivative. The boundary variables $(p,j)$ and $(A_t, e_t)$ form a Lax pair.
The radial components of the fields $(A_{r},e_{r})$, which are transverse to the boundary, play a different role, which we will comment later in section \ref{sec:tangentdiffeo}.

From a canonical point of view, focusing on the evolution in time of the fields in the bulk, the  fields $A_{t}$ and $e_{t}$ play the role of Lagrange multipliers  respectively for the torsionless constraint $(\rd_{A}e)_{r\theta}=0$ and the flatness constraint $F_{r\theta}=0$.  They control the time evolution of the fields in the bulk. Similarly, the pair $(A_t,e_t)$ plays the role of chemical potentials governing the time dependence of the boundary currents $(\A,\e)$.
Moreover, it is  important to appreciate  that the time evolution of the boundary currents is pure gauge and can be written as :
\be
\partial_t(\A,\e ) =  (\delta^{\ga}_{A_t} + \delta^{\tra}_{e_t})(\A,\e).
\ee
%
%
Not only this clarifies the role of $(A_t, e_t)$ in the evolution, but this simple reformulation of the dynamics could be directly implementable at the quantum level.
It is clear that deriving the time evolution of the boundary currents $(\A,\e)$ requires specifying the value of the boundary fields $(A_t, e_t)$. Such boundary conditions are necessary in order to define closed evolution equations for the bounded system.

\medskip

One could entirely prescribe by hand the time evolution of the boundary fields as:
\be
(A_t, e_t)=(\bar{A}, \bar{e})\,,
\ee
where $\bar{A}$ and $\bar{e}$ can be understood as (possibly time-dependent) external sources. Then inserting those values in the boundary equations of motion \eqref{boundaryeqn} gives straightforward first order linear differential equations for the boundary currents $(\A,\e)$, which we can then integrate.
A more subtle ansatz is to couple the fields $(A_t, e_t)$ and $(\A,\e)$ on the boundary, for instance in a linear ansatz:
\be
\label{linearansatz}
\left|
\begin{array}{lcl}
	e_{t}&=&a\A+c\e+\bar{e}\,,
	\vspace*{1mm}\\
	A_{t}&=&b\e+d\A+\bar{A}\,,
\end{array}
\right.
\ee
where the coefficients $(a,b,c,d)$ are a priori chosen by hand and possibly depend on $\theta$.
Inserting such boundary conditions in the boundary equations of motion \eqref{boundaryeqn} leads to coupled differential equations for the boundary currents $(\A,\e)$. Such non-trivial boundary conditions actually correspond to a non-trivial boundary field theory living on the cylinder $C_R$. Indeed, (admissible) boundary conditions naturally come from boundary Lagrangians. Let us describe this correspondence in more details.

\medskip

More precisely, admissible boundary conditions on the cylinder $C_R$, span by $(t,\theta)$ at $r=R$, are  classified by choices of Lagrangian sub-manifolds, that is, they are classified by a choice of admissible boundary conditions on the time components $(A_t, e_t)$ such that the pull-back of the symplectic form on the boundary is exact,
\be
\Omega|_{C_R}\deq\left(\delta A_\theta \curlywedge \delta e_t+  \delta e_\theta \curlywedge \delta A_t \right)_{r=R} (\rd \theta \wedge \rd t)=\rd \omega_R.
\ee
where $\omega_R$ is a boundary symplectic form and $\curlywedge$ denotes the wedge product in field space.
These boundary conditions ensures that  no symplectic flux is leaking through the boundary cylinder\footnotemark.
\footnotetext{
	More precisely it ensures that the leakage of symplectic flux can be reabsorbed into a boundary redefinition of the symplectic form $\Omega_\Sigma \to\Omega'_\Sigma= \Omega_{\Sigma} +\oint_S \omega_R$.}
A generic boundary condition with $\omega_R=0$ is characterized by a canonical transformation, i.e. a functional $B(e_\theta, A_\theta)$. Indeed the condition  $\Omega|_{C_R} =\delta \Theta_{C_R}=0$ means that locally in field space  one has  $ \Theta_{C_R}= \delta \cB $.
The set of admissible boundary condition is therefore very large. For simplicity one can restricted our study to the case where the functional $\cB$ is  ultra local, i.e. it depends only $(e_\theta, A_\theta)$ and not on their derivatives $(\pa_\theta e_\theta, \pa_\theta A_\theta)$.
Therefore a general ultra-local boundary condition can be written as $\cB = \oint_S B (\A,\e)$,
where one use the identification on the boundary $(A_\theta,e_\theta)=(\A,\e)$.
The most general ultra local boundary condition therefore reads
\be \label{Ultralocalb}
A_{ta}= \frac{\partial B}{\partial \e^a} ,\qquad
e_t^a =  \frac{\partial B}{\partial \A_{a}}.
\ee
These conditions can be obtained by adding a boundary Lagrangian  to the bulk Lagrangian (\ref{bulkL}) and  considering the full action:
\be
S= \int_{\Sigma\times \mathbb{R}} e^i\wedge F_i(A) +\int_{C_R} [ B(A_\theta,e_\theta)- e_\theta A_t] \rd t \wedge \rd \theta.
\ee
One can indeed check  that the boundary equations of motions, that arises from boundary variation of  $e_\theta$ and $A_\theta$, implement as expected the boundary evolution equations \eqref{Ultralocalb}.

\medskip

Let us try the simplest possible boundary Lagrangian, quadratic in the boundary currents:
\be
B(\A,\e)=\f a2 \A^{2}+\f b2 \e^{2}+c \,\A\cdot\e + \bar{e}\cdot\A + \bar{A}\cdot\e\,.
\ee
The corresponding boundary conditions, computed from the resulting boundary equations of motions \eqref{Ultralocalb}, are exactly the linear ansatz \eqref{linearansatz} mentioned above with the coefficients $d=c$:
\be
\label{boundarycondition}
\left|
\begin{array}{lcl}
	e_{t}&=&ap+cj+\bar{e}\,,
	\vspace*{1mm}\\
	A_{t}&=&bj+cp+\bar{A}\,.
\end{array}
\right.
\ee
Now, we are interested in characterizing consistent boundary conditions, that preserves the boundary symmetries, or equivalently such that  the boundary charges are conserved.

\subsection{Consistent boundary conditions and boundary gauge transformations}

One can now establish that the boundary symmetry algebra can be promoted to a symmetry of the general ultra-local boundary condition \eqref{Ultralocalb}.
More precisely, on the one hand, we check which boundary conditions ensure that the boundary charges $(P_\varphi, J_\alpha)$ are conserved, and on the other hand, we check which boundary conditions are compatible with the boundary symmetry transformations $\delta^{t}$ and $\delta^{g}$. These two statements are ultimately equivalent.

Let us start with the conservation of the charges  $P_\varphi$ and $J_\alpha$.
The time evolution of the charges is given by
\beq
\dot{P}_\varphi &=&
\oint_S \Big{[}p\cdot\nabla_t{\varphi} -A_{t}\cdot \partial_\theta\varphi\Big{]}
=\oint_S \Big{[}p\cdot\pp_t{\varphi}\cdot - A_{t}\cdot \nabla_\theta\varphi \Big{]}
\,,
\\
\dot{J}_\alpha &=&
\oint_S \Big{[}j\cdot\nabla_t{\alpha} - e_t \cdot \nabla_\theta\alpha\Big{]}
\,.
\eeq
Plugging the boundary conditions \eqref{boundarycondition} in these evolution equations allows to get rid of the $A_{t}$ and $e_{t}$ fields and get:
\beq
\dot{P}_\varphi &=&
\oint_S
p\cdot\Big{(}\dot\vphi+[\bar{A},\vphi]-c\vphi' \Big{)}-bj\cdot\Big{(}\vphi'+[p,\vphi]\Big{)}+\vphi\cdot\pp_{\theta}\bar{A}
\,,
\\
\dot{J}_\alpha &=&
\oint_S
j\cdot\Big{(}\dot\alpha+[\bar{A},\alpha]-c\alpha' \Big{)}-p\cdot\Big{(}a\alpha'-[\bar{e},\alpha]\Big{)}+\alpha\cdot\pp_{\theta}\bar{e}
\,,
\eeq
where the dots denote the time derivation, $\dot{\vphi}=\pp_{t}\vphi$, and the primes denote the space derivation, $\vphi'=\pp_{\theta}\vphi$.
We would like conserved charges, i.e. $\dot{P}_\varphi=\dot{J}_\alpha=0$, labelled by arbitrary functions $\vphi(\theta)$ and $\alpha(\theta)$ whatever the values of the dynamical fields $(\A,\e)$.

To this purpose, we distinguish three types of terms in the expressions above. The last terms, in $\vphi\cdot\pp_{\theta}\bar{A}$ and $\alpha\cdot\pp_{\theta}\bar{e}$, clearly vanish as soon as we require the external sources $\bar{A}$ and $\bar{e}$ to be constant around the boundary circle. Then we have terms involving a time derivative, which give evolution equations for the smearing functions $\vphi$ and $\alpha$, while terms involving only space derivatives impose constraints on $\vphi(\theta)$ and $\alpha(\theta)$ which reduce the symmetry, for instance, from an arbitrary smearing function $\vphi(\theta)$ on the circle to a specific function determined by its initial condition $\vphi\big|_{\theta=0}$. Thus, those latter terms, involving only space derivatives of the smearing functions, are the obstruction to the conservation of charges.

Exploring the various possibilities, we conclude with three classes of (ultra-local, linear) boundary conditions with conserved charges:
\begin{itemize}
	\item $a=b=0$ and $\bar{e}=0$: {\bf both charges are conserved} $\dot{P}_\varphi=\dot{J}_\alpha=0$ as long as the smearing fields satisfy the following evolution equations
	\be
	\dot\vphi+[\bar{A},\vphi]=c\vphi'
	\,,\qquad
	\dot\alpha+[\bar{A},\alpha]=c\alpha'
	\,.
	\ee

	\item $b=0$ but $a\ne 0$: {\bf the momentum charges are conserved} $\dot{P}_\varphi=0$ as long as the smearing field $\vphi$ satisfy its evolution equation. The angular momentum charge $J_{\alpha}$ would be conserved if and only if the smearing field satisfy both the spatial constraint $a\alpha'=[\bar{e},\alpha]$ and the equation of motion $\dot\alpha+[\bar{A},\alpha]=c\alpha'$. For instance, when $\bar{e}=0$, this means that onlyt the 0-mode angular momentum charge, i.e. $J_{\alpha}$ for $\alpha$ constant on the boundary circle, is conserved.

	\item $a=\bar{e}=0$ but $b\ne 0$: {\bf the angular momentum charges are conserved} $\dot{J}_\alpha=0$ as long as the smearing field $\alpha$ satisfy its evolution equation. The  momentum charge $P_{\vphi}$ would be conserved if and only if the smearing field satisfy both the spatial constraint $\vphi'+[p,\vphi]=0$ and the equation of motion $\dot\vphi+[\bar{A},\vphi]=c\vphi'$. The term $[p,\vphi]$ in the constraint is slightly problematic since it implies that the smearing field would be field-dependent.
	The bigger problem is that nothing ensures that the time evolution equation for $\vphi$ even respects the spatial constraint.

\end{itemize}

\medskip

The second way to proceed is to investigate if the symmetry transformations are compatible with the chosen boundary conditions. Indeed the gauge transformations induce  transformations for the boundary fields:
\be
\left|
\begin{array}{lcl}
	\delta^{g}_{\alpha}e&=&-[\alpha,e]\\
	\delta^{g}_{\alpha}A&=&\rd_{A}\alpha
\end{array}
\right.
\qquad\Rightarrow\qquad
\left|
\begin{array}{lcl}
	\delta^{g}_{\alpha}e_{t}&=&-[\alpha,e_{t}]\,,\\
	\delta^{g}_{\alpha}\e&=&-[\alpha,\e]\,,
\end{array}
\right.
\quad
\left|
\begin{array}{lcl}
	\delta^{g}_{\alpha}A_{t}&=&\dot{\alpha}+[A_{t},\alpha]
	\,,\\
	\delta^{g}_{\alpha}\A&=&\alpha'+[p,\alpha]
	\,,
\end{array}
\right.
\ee
\be
\left|
\begin{array}{lcl}
	\delta^{t}_{\vphi}e&=&\rd_{A}\vphi\\
	\delta^{t}_{\vphi}A&=&0
\end{array}
\right.
\qquad\Rightarrow\qquad
\left|
\begin{array}{lcl}
	\delta^{t}_{\vphi}e_{t}&=&\dot{\vphi}+[A_{t},\vphi]\,,\\
	\delta^{t}_{\vphi}\e&=&\vphi'+[p,\vphi]\,,
\end{array}
\right.
\quad
\left|
\begin{array}{lcl}
	\delta^{t}_{\vphi}A_{t}&=&0
	\,,\\
	\delta^{t}_{\vphi}\A&=&0
	\,.
\end{array}
\right.
\ee
Plugging the boundary conditions \eqref{boundarycondition} in those transformations implies the necessary  compatibility conditions ensuring that these define boundary symmetries:
\be
\begin{array}{ll}
	(\delta^{g}e_{t}) \qquad&
	a\alpha'=[\bar{e},\alpha]
	\vspace*{2mm}\\
	(\delta^{g}A_{t}) \qquad&
	\dot{\alpha}+[\bar{A},\alpha]=c\alpha'
	\vspace*{2mm}\\
	(\delta^{t}e_{t}) \qquad&
	\dot{\vphi}+[\bar{A},\vphi]+b[j,\vphi]=c\vphi'
	\vspace*{2mm}\\
	(\delta^{t}A_{t}) \qquad&
	b\big{(}\vphi'+[p,\vphi]\big{)}=0
\end{array}
\ee
We recover the same equations as for the conservation of the charges, leading to the same conclusion:
\begin{itemize}
	\item If $a=b=0$  and $\bar{e}=0$, then both Lorentz transformations and translations are boundary symmetries.
	\item If $b=0$ but $a\ne 0$, then the constraint $a\alpha'=[\bar{e},\alpha]$ breaks the Lorentz symmetry on the boundary.
	\item If $a=0$ and $\bar{e}=0$ but $b\ne 0$, then the constraint $\vphi'+[p,\vphi]=0$ breaks the translation symmetry on the boundary.
\end{itemize}

To summarize, the only boundary conditions that preserve both Lorentz and translation symmetries on the boundary are:
\be
e_{t}=cA_{\theta}\,,\qquad
A_{t}=ce_{\theta}+\bar{A}\,,
\ee
with $c$ and $\bar{A}$ arbitrary functions on the boundary circle. If we introduce other terms in those linear relations, then some symmetries are broken. For instance, a boundary condition $e_{t}=cA_{\theta}+ae_{\theta}$ would break the translation symmetry, while a boundary condition $A_{t}=ce_{\theta}+b A_{\theta}$ would break the Lorentz symmetry.

It would be interesting to analyze boundary symmetries for general (non-ultra-local) boundary Lagrangian and classify the symmetry algebra according to the corresponding general boundary conditions. Nevertheless, the quadratic boundary Lagrangian and the corresponding linear boundary conditions already offer a pretty representative sample of possible behaviors for the boundary charges.

\subsection{Tangential diffeomorphisms, twist and BMS charges and symmetries}
\label{sec:tangentdiffeo}
\label{app:sugawara}

In this final section of the overview of classical boundary symmetries and boundary conditions, we would like to review how the BMS algebra is embedded as a sub-algebra of the loop algebra $L_k(ISO(3))$.
In order to do so, one focuses on diffeomorphism generators tangent to the canonical slice $\Sigma$. One looks at the vector fields $X\partial_\theta$ tangent to its circle boundary $S$, with $X(\theta)$ an arbitrary function, and considers a twisted extension\footnotemark{} as vector fields  on $\Sigma$ as  $\xi_X\deq  X \partial_\theta+ X' \partial_r$ with $X'=\pa_\theta X$.
These vectors have a simple commutator,
\be
[\xi_X,\xi_Y]= \xi_{[X,Y]}
\,,
\ee
in terms of the Lie bracket of 1D vector fields $[X,Y]=XY'-YX'$.
\footnotetext{
	In order to clearly visualize the deformation of the boundary vector fields, one can introduce an arbitrary parameter $\gamma$ and define the vector fields on $\Sigma$ as $\xi_X\deq  X \partial_\theta+ \gamma X' \partial_r$. This parameter does not affect the action of the diffeomorphisms and, in the end, simply rescale the central charges $\lambda$ and $\mu$.
}

The action of these diffeomorphisms on the boundary phase space variables $(\A, \e)\deq (A_\theta, e_\theta)$ depends on the value of  $(A_r,e_r)$ on $S$.
Through a bulk gauge transformation, one can always gauge-fix the angular dependence of those radial components and assume without loss of generality  that $\partial_\theta A_r= \partial_\theta e_r=0$.  In other words the value of the radial fields is fixed on the boundary circle $S$  to be  constant and we denote these constant values by  $ A_r \deq-\lambda$ and $e_r =-\mu$.
The bulk equations of motions $F_{r\theta}= T_{r\theta}=0$, means that these controls the radial evolution of the boundary fields within the canonical slide $\Sigma$,
\be\label{bdyeom}
\pa_r \A = [\lambda, \A], \qquad \pa_r \e= [\lambda , \e] + [\mu, \A].
\ee
Using these radial evolution equations, we can evaluate the action of the diffeomorphisms $\xi_X$ on the boundary fields.
From the definition of the Lie derivative ${\cal L}_{\xi} =\rd \xi \ip + \xi \ip \rd $, one computes the Lie derivatives $\cL_{\xi}A$ and $\cL_{\xi}e$ on the canonical slice and gets the action of $\xi_{X}$ on the boundary fields from their $\theta$-components
\be
\label{diffeo}
\begin{array}{lcl}
	{\cal L}_{\xi_X} \e&=& (X \e)'  + X' [\lambda, \e] +X' [\mu,\A] - \mu X''\,,
	\vspace*{2mm}\\
	{\cal L}_{\xi_X} \A&=& (X \A)'  + X' [\lambda, \A] - \lambda X''
	\,.
\end{array}
\ee
One recover the usual action of boundary diffeomorphism generators when $\Lambda=\mu=0$. Then non-vanishing values $(\lambda,\mu)$ are understood as a twist of the action of $\mathrm{Diff}(S)$ on the phase space variables $(A,e)$.

In order to ensure that this twisted $\mathrm{Diff}(S)$-action is Hamiltonian, we choose boundary conditions such that  $\delta A_r=\delta e_r=0$ on the boundary circle $S$, i.e. we impose that the boundary values of the radial fields $(A_r,e_r)$ do not belong to the boundary phase space.
Under this assumption, we can evaluate the corresponding diffeomorphism charge $\cD_X$, such that
$ I_{{\cal L}_{\xi_X}}\Omega = -\delta \cD_X$, and obtain
\be
\cD_X =  \oint_S  X (\e \cdot \A + \mu\cdot  \A' +\lambda\cdot  \e') \,\rd\theta
\,.
\ee
It is straightforward to check that the Poisson brackets with this charge leads back to the Lie brackets \eqref{diffeo} computed above and define as wanted the variation under infinitesimal twisted boundary diffeomorphisms:
\be
\delta^{d}_{X}\e\deq\{\cD_{X},\e\}={\cal L}_{\xi_X} \e
\,,\qquad
\delta^{d}_{X}\A\deq\{\cD_{X},\A\}={\cal L}_{\xi_X} \A
\,.
\ee
The expression of the diffeomorphism charges underlines
that twisted diffeomorphisms can be understood as  field dependent gauge transformations. Indeed, the following identity holds for boundary fields $(\A,\e)$:
\be
{\cal L}_{ \xi_X} = \delta^{\ga}_{{\xi_X} \ip A}  + \delta^{\tra}_{ {\xi_X} \ip e}
\,.
\ee
%
%
It appears that, once a vector field $\xi_X$ is selected, it is natural to consider field dependent gauge transformations with  gauge parameters given by ${\xi_X} \ip A$ or ${\xi_X} \ip e$. Accordingly we can also introduce the following transformations:
\be
T_X= \delta^{\tra}_{{\xi_X} \ip A}
\,,\qquad
R_X =  \delta^{\ga}_{{\xi_X} \ip e}
\,.
\ee
The action of these transformations is explicitly given by
\be
\left|
\begin{array}{lcl}
	T_X  \A &=&  0
	\,,
	\vspace*{2mm}\\
	T_X \e &=& (X \A)'   + X' [\lambda, \A] - X'' \lambda
	\,,
\end{array}
\right.
\qquad\qquad
\left|
\begin{array}{lcl}
	R_X \A&=& (X\e)'   + X[\A,\e]+X'[\mu,\A] -X''\mu
	\,,
	\vspace*{2mm}\\
	R_X \e&=& X' [ \mu , \e]
	\,.
\end{array}
\right.
\ee

These variations are also Hamiltonian and are generated by charges $({\cal P}_X,{\cal J}_X)$, which are
also quadratic functionals of the currents:
\be
{{\cal P}}_X \hat{=} \oint_S  X \Big{[}\tfrac12 \A \cdot \A  + \lambda \cdot \A'\Big{]}
\,,\qquad
{\cal J}_X  \hat{=} \oint_S  X \Big{[}\tfrac12 \e \cdot \e + \mu \cdot \e'\Big{]}
\,.
\ee

The construction of the quadratic charges $(\cD_X,\cP_X,\cJ_X)$ corresponds to the {\it twisted Sugawara} construction applied to the Poincar\'e loop algebra. For a application of the twisted Sugawara construction in the case of $\mathfrak{sl}(2,\mathbb{R})$, the interested reader can refer to \cite{Banados:1998ta,Banados:1998gz}.
%

We now evaluate the charge algebra by taking the Fourier transform and defining  the generators  $\cD_n\deq \cD_{X=e^{-in\theta}}$ and similarly for $\cP_{X}$ and $\cJ_{X}$.
One gets
\beq
i\{ \cD_n, \cD_m \} &=&  (n-m) \cD_{n+m} + 2 (\mu\cdot \lambda)\,  n^3 \delta_{n+m,0}
\,,
\eeq
where one recognizes the Virasoro algebra\footnotemark{} with central charge $ c = 24 (\mu\cdot \lambda)$.
\footnotetext{
	If one introduces a shift in the vacuum energy and defines $\hat{\cD}_n =\cD_n - \tfrac{c}{24} \delta_{n,0}$, one recovers the usual expression
	\be
	i\{ \hat{\cD}_n, \hat{\cD}_m \} =  (n-m) \hat{\cD}_{n+m} + \frac{c}{12}  n(n^2-1) \delta_{n+m,0}
	\,.\nn
	\ee
}
Similarly one computes:
\beq
i\{ \cD_n, \cP_m \} &=&  (n-m) \cP_{n+m} + (\lambda\cdot \lambda)\,  n^3 \delta_{n+m,0}
\,,\\
i\{ \cD_n, \cJ_m \} &=&  (n-m) \cJ_{n+m} + (\mu\cdot \mu)\,  n^3 \delta_{n+m,0}
\,.
\eeq
while we have that
\beq
i\{ \cP_n, \cP_m \} =  0
\,,\qquad
i\{ \cJ_n, \cJ_m \} =  0
\,,\qquad
i\{ \cP_n, \cJ_m \} = (n-m) \cD_{n+m} +  (\mu\cdot \lambda)\,  n^3 \delta_{n+m,0}
\,.
\eeq
We see that both $(\cD_n, \cP_n)$ and $(\cD_n, \cJ_n)$ are  BMS sub-algebras
of the algebra generated by $(\cD_n, \cP_n, \cJ_n)$.
The full algebra generated by quadratics in the boundary current is therefore clearly bigger than just the BMS algebra.
We see that there are three types of central charges given by the scalar products $24(\lambda\cdot \mu),12(\lambda\cdot\lambda), 12(\mu\cdot\mu)$. In the gravitational case, when the boundary are pushed to infinity and we obtain an asymptotically flat space-time, it is customary to take $\lambda=0$ (corresponding to the absence of fluxes normal to the boundary) so that only one central charge survives \cite{Oblak:2016eij}.

\medskip


Now that we have established the extended BMS charges as quadratic functional in the boundary current and written the corresponding boundary transformations, the next step is to check in which case those charges are conserved under the evolution of the boundary fields.
Following the same logic as earlier, we compute their time derivative, $(\dot{\cD}_{X},\dot{\cP}_{X},\dot{\cJ}_{X})$, use the boundary equations of motion,
\be
\dot{p}=\pp_{\theta}A_{t}+[p,A_{t}]
\,,\qquad
\dot{j}=\pp_{\theta}e_{t}+[p,e_{t}]+[j,A_{t}]
\,,
\nn
\ee
associated to boundary conditions,
\be
A_{t}=bj+cp\,,\qquad
e_{t}=ap+cj\,,\nn
\ee
where we have set the external off-sets to 0, $\bar{A}=\bar{e}=0$. We further assume that the parameters $a,b,c$ are constant around the boundary circle, $\pp_{\theta}a=\pp_{\theta}b=\pp_{\theta}c=0$. Finally, we also assume that the radial fields are constant in time, $\dot{\lambda}=\dot{\mu}=0$, ensuring that the central charges are constant\footnotemark{}.
\footnotetext{We could imagine a fine-tuned scenario where the radial fields $A_{r}$ and $e_{r}$ still rotate while keeping their scalar products constant, but this does not seem at a first glance to affect the charge conservation in an interesting way.}

Then it is straightforward to check that the three sets of charges are conserved if $a=b=0$ as long as the boundary vector fields $X$ satisfy a simple evolution equation (which we require to be independent from the boundary fields $(p,j)$):
\be
(a=b=0)\qquad\qquad
\dot{X}=cX'
\quad\Rightarrow\quad
\dot{\cD}_{X}=\dot{\cP}_{X}=\dot{\cJ}_{X}=0
\,.
\ee
If $a\ne 0$, then only the quadratic translation charge are conserved $\dot{\cP}_{X}=0$. While, if $b\ne 0$, then only the quadratic rotational charge are conserved $\dot{\cJ}_{X}=0$.

At the end of the day, the boundary conditions $(A_{t},e_{t})=c(e_{\theta},A_{\theta})$ with an arbitrary factor $c$ ensure that the extended BMS charges are all conserved and generate  boundary symmetries with central charges given by the values of the radial fields $(A_{r},e_{r})$.

\medskip

The goal of this paper is to investigate how to define those boundary charges $(\cD_n, \cP_n, \cJ_n)$, both currents and  BMS charges, at the discrete level in 3d quantum gravity.

\section{Quantum Boundary Algebra for 3D Gravity}
 \label{sec:algebra}

\subsection{Discrete Gauge Field: the Disk Algebras}

A customary method to rigorously define and better understand the basic excitations of a field theory,  their dynamics and their quantum transition amplitudes computed from a path integral is to discretize the field degrees of freedom. For a gauge theory, such as Yang-Mills theories or Chern-Simon theories or gravity reformulated in terms of connection variables (see e.g. \cite{Krasnov:2011pp,Freidel:2012np}), we use the framework of lattice gauge theory, where the connection field $A$ is discretized as its holonomies $\cP e^{\int_{\gamma}A}$ along finite curves $\gamma$. In lattice Yang-Mills theories, one usually considers holonomies on a fixed lattice and study its coarse-graining and refinement, while in quantum gravity models, such as matrix models, tensor models, dynamical triangulations or loop quantum gravity, the underlying lattice become dynamical and represents physical degrees of freedom of the fluctuating space-time geometry.

\begin{figure}[!htb]
	\centering
	\begin{tikzpicture}[scale=.8]
	\coordinate (O) at (0,0);
	\coordinate (A1) at ($({1.5*cos(90)},{1.5*sin(90)})$); \coordinate(B1) at ($({1.5*cos(110)},{1.5*sin(110)})$);
	\coordinate (A2) at ($({1.5*cos(130)},{1.5*sin(130)})$); \coordinate(B2) at ($({1.5*cos(150)},{1.5*sin(150)})$);
	\coordinate (A3) at ($({1.5*cos(170)},{1.5*sin(170)})$); \coordinate(B3) at ($({1.5*cos(190)},{1.5*sin(190)})$);
	\coordinate (A4) at ($({1.5*cos(50)},{1.5*sin(50)})$); \coordinate(B4) at ($({1.5*cos(70)},{1.5*sin(70)})$);

	\draw[very thick,->-=0.31,->-=0.43] (O) circle (1.5);
	\node at (A1) {$\bullet$};  \draw[blue,thick] (A1) -- node[pos=0.6,left,scale=1]{$h_1$} +($({0.9*cos(90)},{0.9*sin(90)})$); \node[below,scale=1] at (B1) {$g_1$};
	\node at (A2) {$\bullet$};  \draw[blue,thick] (A2) -- node[pos=0.6,left,scale=1]{$h_2$} +($({0.9*cos(130)},{0.9*sin(130)})$); \node[below right,scale=1] at (B2) {$g_2$};
	\node at (A3) {$\bullet$};  \draw[blue,thick] (A3) -- node[pos=0.6,below left,scale=1]{$h_3$} +($({0.9*cos(170)},{0.9*sin(170)})$); 
	\node at (A4) {$\bullet$};  \draw[blue,thick] (A4) -- node[pos=0.6,right,scale=1]{$h_N$} +($({0.9*cos(50)},{0.9*sin(50)})$); \node[below,scale=1] at (B4) {$g_N$};

	\centerarc[dashed](O)(180:220:1.75);
	\end{tikzpicture}
	\caption{Sunny graph representation of the disk boundary, with $N$ holonomies $g_1,g_2,\dots,g_N$ along the boundary segments and $N$ holonomies $h_1,h_2,\dots,h_N$, in blue, along the edges connecting the boundary to the bulk.}
	\label{fig:sunnygraph}
\end{figure}
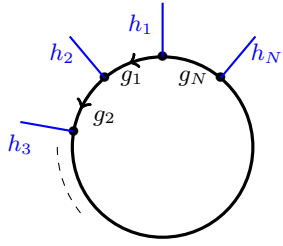

Let us consider as canonical surface a 2D disk. The one-dimensional spatial boundary, or corner, has the topology of a circle.
This can equivalently be thought as the boundary around a point-like defect or ``puncture'' on the 2d canonical spatial slice.
We discretize it in $N$ segments. Each point of junction between those segments is where the corner connects with the exterior region, i.e. where the boundary fields can be excited by external sources or bulk field fluctuations. This setting can be drawn as a ``sunny'' graph, as illustrated on figure \ref{fig:sunnygraph}, where we consider the $N$ holonomies along the boundary segments and $N$ holonomies along edges connecting the boundary to the bulk.

\subsubsection{Wave-functions around a puncture}

For 3D gravity with Euclidean signature, the holonomies live in the $\SU(2)$ Lie group so that the configuration space consists of $N$ group elements $g_{i}\in\SU(2)$ along the boundary and $N$ group elements $h_{i}\in\SU(2)$ extending the boundary into the bulk, up to  gauge transformations acting as $\SU(2)$ transformations at the $N$ insertion points:
\be
(g_{1},..,g_{N},h_{1},..,h_{N})
\sim
(a_{1}g_{1}a_{2}^{-1},..,a_{N}g_{N}a_{1}^{-1},a_{1}h_{1},..,a_{N}h_{N})
\,,\qquad\forall (a_{i})_{i=1..N}\in\SU(2)^{\times N}
\,.
\ee
The configuration space is thus $\SU(2)^{\times  2N}/\SU(2)^{\times N}\sim\SU(2)^{\times N}$. Boundary states are thus gauge-invariant wave-functions $\vphi(\{g_{i},h_{i}\}_{i=1..N})$ satisfying,
\be
\vphi(\{g_{i},h_{i}\})=\vphi(\{a_{i}g_{i}a_{i+1}^{-1},a_{i}h_{i}\})
\,,\qquad\forall a_{i}\in\SU(2)
\,,
\ee
provided with the natural $\SU(2)$-invariant scalar product defined with the $\SU(2)$ Haar measure,
\be
\la \vphi |\tilde{\vphi}\ra
=
\int_{\SU(2)^{2N}}\rd g_{i}\rd h_{i}\,
\overline{\vphi(\{g_{i},h_{i}\})}\,\tilde{\vphi}(\{g_{i},h_{i}\})
\,.
\ee
Such gauge-invariant wave-functions are entirely determined by their section at $h_{i}=\id$:
\be
\vphi(\{g_{i},h_{i}\})=\vphi(\{h_{i}^{-1}g_{i}h_{i+1},\id\})
\,,\qquad\forall g_{i},h_{i}\in\SU(2)
\,,
\ee
thereby identifying the group elements $h_{i}^{-1}g_{i}h_{i+1}$ as the basic gauge-invariant variables on the boundary.

\subsubsection{Spin basis on the sunny graph}

A convenient basis of this Hilbert space of gauge invariant wave-functions around a puncture is given by the spin network basis. Spin networks, originally introduced by Penrose, are  a standard tool for  3d quantum gravity and 3d topological invariants, in particular by the Ponzano-Regge, Reshetikhin-Turaev, Turaev-Viro and related invariants \cite{PR1968,ReshetikhinTuraev1991,TuraevViro1992,Delcamp:2018sef}, and more generally for loop quantum gravity (e.g. see \cite{Rovelli:1995ac} and \cite{Livine:2010zx,Perez:2012wv,Bodendorfer:2016uat} for recent reviews). They even appeared recently in the context of the quantization of Jackiw-Teitelboim gravity (e.g. \cite{Blommaert:2018oro}).
Their definition simply relies on the Peter-Weyl theorem or Plancherel decomposition, which provides a Fourier transform on Lie groups. In our case, it states that functions on the Lie group $\SU(2)$, which are square-integrable with respect to the Haar measure, can be decomposed onto the spin basis defined by the Wigner matrices.

To fix the notations, the $\su(2)$ Lie algebra  has three (Hermitian) generators satisfying the commutation relations:
\be
[J^a,J^b]=i{\eps^{ab}}_c J^c\,,\qquad
(J^a)^\dagger=J^a\,,
\ee
and quadratic Casimir $\mathfrak{C}=\vJ^2\equiv\sum\limits_{a}J^aJ^a$.
The irreducible unitary representations of the $\SU(2)$  Lie group are labelled by a half-integer $j\in\f\N2$ - the spin. These are finite dimensional representation. Let us call $\cV_{j}$ the $(2j+1)$-dimensional Hilbert space carrying the representation of spin $j$.
A standard basis of $\cV_{j}$ is given by the eigenvectors of the operator $J^3$,
diagonalized as  $J^3\,|j,m\ra=m\,|j,m\ra$ with the magnetic moment $m$ running by integer step from $-j$ to $+j$:
\be
\cV_{j}=\bigoplus_{m=-j}^{+j}\C\,|j,m\ra
\,,\qquad
\vJ^2\,|j,m\ra=j(j+1)\,|j,m\ra
\,,\quad
J^3\,|j,m\ra=m\,|j,m\ra
\,.
\ee
The vanishing spin $j=0$ is the trivial representation of dimension 1. The spin $j=\f12$ defines the fundamental two-dimensional representation. The generators $J^a$ are then given by the anti-Hermitian matrices $\tau^a$:
\be
D^{\f 12}(J^a)=i\tau^a=\f{1}{2}\sigma_a
\,,\qquad
\tau^a\tau^b = -\tfrac14 \delta^{ab} + \frac{1}2 {\eps^{ab}}_c \tau^c
\,.
\ee
Group elements are represented as (2$j$+1)$\times$(2$j$+1)-matrices. In the  magnetic basis, these are  the Wigner matrices:
\be
D^j_{mm'}(g)=\la j,m|g|j,m'\ra
\,,\qquad
D^j_{mm'}(g^{-1})=\overline{D^j_{m'm}(g)}
\,.
\ee
Those matrices are well-known and can be explicitly written in terms of the Jacobi polynomials. Their precise expression is not relevant for the present work.
The Peter-Weyl theorem states that the Wigner matrices form an orthonormal basis of this Hilbert space:
\be
L^{2}(\SU(2))=\bigoplus_{j\in\f\N2}\cV_{j}\otimes\cV_{j}^{*}
\,,\qquad
\int \rd g
\,\overline{D^j_{ab}(g)}\,
D^k_{cd}(g)
=
\f{\delta_{jk}}{2j+1}\delta_{ac}\delta_{bd}
\,.
\ee
The Plancherel decomposition follows from the formula for the $\delta$-distribution on the group written in terms of the $\SU(2)$ characters:
\be
\delta(g)=\sum_{j\in\f\N2} (2j+1)\chi_{j}(g)
\,,\qquad
\chi_{j}(g)=\tr D^{j}(g)=\sum_{m}D^{j}_{mm}(g)
\,,
\ee
and reads for an arbitrary square-integrable function $f$:
\be
\forall f\in L^{2}(\SU(2))
\,,
f(g)=\sum_{j,a,b}(2j+1)f^{j}_{ab}D^{j}_{ab}(g)
\,,\qquad
f^{j}_{ab}=\int \rd g\, \overline{D^j_{ab}(g)}\,f(g)
\,.
\ee
\begin{figure}[!htb]
	\centering
	\begin{tikzpicture}[scale=.8]
	\coordinate (O) at (0,0);
	\coordinate (A1) at ($({1.5*cos(90)},{1.5*sin(90)})$); \coordinate(B1) at ($({1.5*cos(110)},{1.5*sin(110)})$);
	\coordinate (A2) at ($({1.5*cos(130)},{1.5*sin(130)})$); \coordinate(B2) at ($({1.5*cos(150)},{1.5*sin(150)})$);
	\coordinate (A3) at ($({1.5*cos(170)},{1.5*sin(170)})$); \coordinate(B3) at ($({1.5*cos(190)},{1.5*sin(190)})$);
	\coordinate (A4) at ($({1.5*cos(50)},{1.5*sin(50)})$); \coordinate(B4) at ($({1.5*cos(70)},{1.5*sin(70)})$);

	\draw[very thick,->-=0.31,->-=0.43] (O) circle (1.5);
	\node at (A1) {$\bullet$};  \draw[thick] (A1) --  +($({0.9*cos(90)},{0.9*sin(90)})$) node[above,scale=1]{$(k_{1},m_{1})$} ; \node[below, scale=1] at (B1) {$j_1$};
	\node at (A2) {$\bullet$};  \draw[thick] (A2) --  +($({0.9*cos(130)},{0.9*sin(130)})$)node[above,scale=1]{$(k_{2},m_{2})$}; \node[below right, scale=1] at (B2) {$j_2$};
	\node at (A3) {$\bullet$};  \draw[thick] (A3) -- +($({0.9*cos(170)},{0.9*sin(170)})$)node[above,scale=1]{$(k_{3},m_{3})$}; 
	\node at (A4) {$\bullet$};  \draw[thick] (A4) --  +($({0.9*cos(50)},{0.9*sin(50)})$)node[above,scale=1]{$(k_{N},m_{N})$}; \node[below, scale=1] at (B4) {$j_N$};

	\centerarc[dashed](O)(180:220:1.75);
	\end{tikzpicture}
	\caption{Spin network basis labels on the sunny graph with $N$ flux insertions, or ``handles'', describing the quantum states of the 1d boundary}
	\label{fig:sunnygraphspin}
\end{figure}
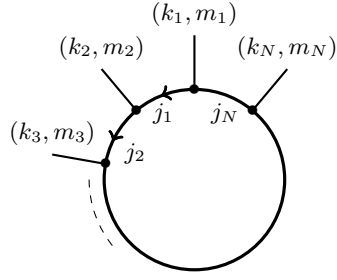

Applying this to wave-functions on the configuration space $\SU(2)^{\times  2N}/\SU(2)^{\times N}$ for a disk boundary with $N$ flux insertions leads to a decomposition in terms of $N$ spins $j_{i}$ around the circle, plus $N$ spins and magnetic indices $(k_{i},m_{i})$ for the flux insertions around the circle, as illustrated on fig.\ref{fig:sunnygraphspin}:
\be
\varphi(\{g_{i},h_{i}\}_{i=1..N})
=
\sum_{\{j_{i},k_{i},m_{i}\}_{i=1..N}}
\varphi^{\{j_{i},k_{i},m_{i}\}}
\la \{g_{i},h_{i}\}|\{j_{i},k_{i},m_{i}\}\ra
\,,
\ee
\be
\la \{g_{i},h_{i}\}|\{j_{i},k_{i},m_{i}\}\ra
=
\prod_{i=1}^{N}
\la (j_{i-1},a'_{i-1})\,(k_{i},m_{i})| (j_{i},a_{i}) \ra
\,
\la j_{i} a_{i}|h_{i}^{-1}g_{i}h_{i+1}|j_{i}a'_{i}\ra
\,,
\ee
where the spin basis wave-functions on the sunny graph $\la \{g_{i},h_{i}\}|\{j_{i},k_{i},m_{i}\}\ra$ are defined in terms of the Clebsh-Gordan coefficients for spin recoupling $C^{j|j',k}_{a|a',m}=\la (j',a')\,(k,m)| (j,a) \ra$.
Now classical boundary observables will act as quantum operators acting on those spin network basis states around the quantum puncture, changing the spins and magnetic indices along the boundary circle. The rest of the paper aims at understanding what is the algebra of those operators representing the deformations and  excitations of the boundary geometry.

\subsubsection{Deformation operators on the sunny graph: the Disk Algebra}


We are interested in the algebra of operators deforming the boundary geometry. We will call the {\it disk algebra} $\cA_{N}$ the algebra of bounded operators acting on the space of smooth gauge-invariant wave-functions on the configuration space  $\SU(2)^{\times  2N}/\SU(2)^{\times N}$. It is given by the semi-direct product
\be
\cA_{N} = \C\left[\SU(2)^{\otimes N} \right] \ltimes_{\mathrm{Ad}} \cC\left(SU(2)^N\right) \,
\ee
where the adjoint action $\mathrm{Ad}$ is defined below. Here $\cC\left(SU(2)^N\right)$ denotes the commutative subalgebra of smooth function, while $\C\left[\SU(2)^{\otimes N} \right]$ denotes the group algebra associated with the group $\SU(2)^{\otimes N}$.
The action of $\C\left[\SU(2)^{\otimes N} \right]$ on $\cC\left(SU(2)^N\right)$ is given by the natural adjoint action on the group on functions over the group. Given $(\otimes_{i=1}^N \HH_i) \in \SU(2)^{\otimes N} $ and $\Phi \in \cC(\SU(2)^N)$, this action reads:
\be
\mathrm{Ad}_{(\otimes_i H_{i})} \Phi (\{g_{i}\})=\Phi (\{H_{i}^{-1}g_{i}H_{i+1}\})
\,.
\ee
The basic operators $(\otimes_{i} \HH_i) \in \SU(2)^{\otimes N} $ and $\Phi \in \cC(\SU(2)^N)$  satisfy the following  composition and commutation rules:
\be
\label{Alg}
(\otimes_i \HH_{i})(\otimes_i \HH'_{i}) =(\otimes_i \widehat{H_{i}H'_{i}})
\,,\qquad
\widehat{\Phi}\widehat{\Phi}' = \widehat{\Phi\Phi}'
\,,\qquad
(\otimes_i \HH_{i}) \widehat{\Phi} (\otimes_i \HH_{i}^{-1})
=
\widehat{\mathrm{Ad}_{(\otimes_i H_{i})} \Phi}\,.
\ee
where ${\Phi\Phi}'$ simply denote the commutative product of functions.
The {\it disk algebra} $\cA_{N}$ admits a star structure and a canonical cyclic state $\omega: \cA_{N} \to \C $ given by
\be
	\left( \widehat{\Phi} (\otimes_i \HH_{i})\right)^* =
	(\otimes_i \HH^{-1}_{i}) \widehat{\overline\Phi},\quad
	\,
	\qquad
	\omega\left( \widehat{\Phi} (\otimes_i \HH_{i})\right)= \Phi(\{H_i\}).
\ee
The pairing induced by $\omega$ shows that we can view the group algebra $\C\left[\SU(2)^{\otimes N} \right]$ as the dual of $\cC(\SU(2)^N)$
\be
	\left[\cC(\SU(2)^N)\right]^* = \C\left[\SU(2)^{\otimes N} \right].
\ee

When representing $\cA_N$ as the algebra of bounded operators acting on the space of smooth gauge-invariant wave-functions, the operators $\HH_i$ and $\widehat{\Phi}$ correspond to two types of basic operation:
\begin{itemize}
	\item Multiplication operators:
	for a function $\Phi$ over $\SU(2)^{ N}$, one defines a multiplication operator acting on gauge-invariant wave-functions,
	\be\label{prod}
	\widehat{\Phi} \,\vphi(\{g_{i},h_{i}\})=
	\Phi(\{h_{i}^{-1}g_{i}h_{i+1}\})\,\vphi(\{g_{i},h_{i}\})
	\,.
	\ee

	\item Translation  operators:
	the translation operators are associated with elements $\otimes_{i=1}^N H_i \in \SU(2)^{\otimes N}$. The operator $(\otimes_{i=1}^N \HH_i) $ acts on gauge-invariant wave-functions as a translation on the bulk-to-boundary holonomies $h_{i}$,
	\be
	(\otimes_{i=1}^N \widehat{H}_i)  \vphi(\{g_{i},h_{i}\})=
	\vphi(\{g_{i},h_{i}H_{i}\}).
	\ee
Such translation operators can naturally be written in term of convolution operators, which allow to extend by linearity their actions to the whole $\C\left[\SU(2)^{\otimes N}\right]$. Indeed, given a distribution $\Psi$ over $\SU(2)^{N}$, one can define the corresponding operator
	\be
	\widehat{\Psi}_c := \int_{\SU(2)^N} \prod_{i=1}^N \rd H_i  \Psi(\{H_i\}) (\otimes_{i=1}^N \widehat{H}_i) \; ,
	\ee
which acts on gauge-invariant wave-functions by convolution on the bulk-to-boundary holonomies $h_{i}$ as
	\be
	\widehat{\Psi}_{c}\,\vphi(\{g_{i},h_{i}\})=
	\int \rd k_{i}\,\Psi(\{k_{i}^{-1}h_{i}\})
	\vphi(\{g_{i},k_{i}\})
	\,.
	\ee
\end{itemize}
While multiplications are abelian, translations do not commute with each other and have non-trivial commutation relations with the multiplications. Together they generate the full disk algebra $\cA_{N}$.

Diving deeper in the algebraic structure of $\cA_N$, we identify a minimal set of generators of this algebra, which are the  multiplication by spin-$\f12$ Wigner matrices and the left-invariant derivations:
\begin{subequations}
	\begin{align}
	&\hg^{AB}_{i} \vphi(g_{1},..,g_{N},h_{1},..,h_{N}) = D^{\f{1}{2}}_{AB}(h_{i}^{-1}g_i h_{i+1})\,\vphi(g_{1},..,g_{N},h_{1},..,h_{N})
	\,, \\
	&\hx^a_i \vphi(g_{1},..,g_{N},h_{1},..,h_{N}) = \left. \frac{\rd }{\rd t} \vphi(g_{1},..,g_{N},h_{1},..,h_{i} e^{tJ^a},..,h_{N})\right|_{t=0}
	\,.
	\end{align}
	\label{eq:action_x_g}
\end{subequations}
The vector operator $\hx_i$ has three components $(\hx_i^a)_{a=1,2,3}$, corresponding to the three generators of the $\su(2)$ Lie algebra, while the matrix operator $\hg_i^{AB}$ with $A,B\in\{1,2\}$ has four components corresponding to the four matrix elements of a $\SU(2)$ group element in its fundamental representation as 2$\times$2 matrices.
The labels $A,B\in\{1,2\}$ corresponds to the two  basis states $m=\pm\f12$  of the representation of spin $j=\f12$.

The derivations operators are the infinitesimal version of the convolution operators. Together, those spin-$\f12$ multiplications and derivation operators are the most fundamental building blocks of $\cA_N$. As we show below, they form a Lie algebra $\lA_N$, whose enveloping algebra is indeed the disk algebra $\cA_N$.

\begin{figure}[h!]
	\centering
	\begin{tikzpicture}[scale=0.8]
	\coordinate(O) at (-4,0);
	\coordinate(p1) at ($(O)+(0,1)$);
	\coordinate(q1) at ($(O)+(0,1.7)$);
	\draw (p1) node {$\bullet$};
	\draw[very thick] (p1)--(q1) ;
	\draw[very thick] (O) circle (1) ;
	\coordinate(A2) at ($(O)+({1*cos(40)},{1*sin(40)})$);
	\coordinate(B2) at ($(O)+({1.7*cos(40)},{1.7*sin(40)})$);
	\draw (A2) node {$\bullet$};
	\draw[very thick](A2)--(B2) ;

	\coordinate(A3) at ($(O)+({1*cos(-10)},{1*sin(-10)})$);
	\coordinate(B3) at ($(O)+({1.7*cos(-10)},{1.7*sin(-10)})$);
	\draw (A3) node {$\bullet$};
	\draw[very thick](A3)--(B3) ;

	\centerarc[dashed](O)(-40:-90:1.4);

	\draw[dotted,thick,blue] ($(O)+(0,1.3)$)--($(O)+(-0.6,1.7)$);
	\coordinate(O) at (0,0);
	\coordinate(A1) at (0,1);
	\coordinate(B1) at (0,1.7);

	\draw (A1) node {$\bullet$};
	\draw[very thick] (A1)--(B1) ;

	\draw[very thick] (O) circle (1) ;
	\coordinate(A2) at ($(O)+({1*cos(40)},{1*sin(40)})$);
	\coordinate(B2) at ($(O)+({1.7*cos(40)},{1.7*sin(40)})$);
	\draw (A2) node {$\bullet$};
	\draw[very thick](A2)--(B2) ;

	\coordinate(A3) at ($(O)+({1*cos(-10)},{1*sin(-10)})$);
	\coordinate(B3) at ($(O)+({1.7*cos(-10)},{1.7*sin(-10)})$);
	\draw (A3) node {$\bullet$};
	\draw[very thick](A3)--(B3) ;
	\centerarc[dashed](O)(-40:-90:1.4);

	\centerarcnodes[blue](O)(50:80:1.2)(C1,D1);
	\centerarcnodes[opacity=0](O)(48:82:1.7)(C2,D2);
	\draw[blue] (C1)--(C2);
	\draw[blue] (D1)--(D2);
	\coordinate(O) at (4,0);
	\coordinate(p1) at ($(O)+(0,1)$);
	\coordinate(q1) at ($(O)+(0,1.7)$);
	\draw (p1) node {$\bullet$};
	\draw[very thick] (p1)--(q1) ;
	\draw[very thick] (O) circle (1) ;
	\coordinate(A2) at ($(O)+({1*cos(40)},{1*sin(40)})$);
	\coordinate(B2) at ($(O)+({1.7*cos(40)},{1.7*sin(40)})$);
	\draw (A2) node {$\bullet$};
	\draw[very thick](A2)--(B2) node[midway,below](C){};
	\centerarc[dashed](O)(-40:-90:1.4);

	\coordinate(A3) at ($(O)+({1*cos(-10)},{1*sin(-10)})$);
	\coordinate(B3) at ($(O)+({1.7*cos(-10)},{1.7*sin(-10)})$);
	\draw (A3) node {$\bullet$};
	\draw[very thick](A3)--(B3) ;

	\draw[dotted, thick,blue] (C)--($(C)+(-0.2,0.8)$);
	\coordinate(O) at (8,0);
	\coordinate(A1) at  ($(O)+(0,1)$);
	\coordinate(B1) at ($(O)+(0,1.7)$);

	\draw (A1) node {$\bullet$};
	\draw[very thick] (A1)--(B1) ;

	\draw[very thick] (O) circle (1) ;
	\coordinate(A2) at ($(O)+({1*cos(40)},{1*sin(40)})$);
	\coordinate(B2) at ($(O)+({1.7*cos(40)},{1.7*sin(40)})$);
	\draw (A2) node {$\bullet$};
	\draw[very thick](A2)--(B2) ;
	\centerarc[dashed](O)(-40:-90:1.4);

	\coordinate(A3) at ($(O)+({1*cos(-10)},{1*sin(-10)})$);
	\coordinate(B3) at ($(O)+({1.7*cos(-10)},{1.7*sin(-10)})$);
	\draw (A3) node {$\bullet$};
	\draw[very thick](A3)--(B3) ;

	\centerarcnodes[blue](O)(0:30:1.2)(C1,D1);
	\centerarcnodes[opacity=0](O)(-2:32:1.7)(C2,D2);
	\draw[blue] (C1)--(C2);
	\draw[blue] (D1)--(D2);
	\coordinate(O) at (12,0);
	\coordinate(p1) at ($(O)+(0,1)$);
	\coordinate(q1) at ($(O)+(0,1.7)$);
	\draw (p1) node {$\bullet$};
	\draw[very thick] (p1)--(q1) ;
	\draw[very thick] (O) circle (1) ;
	\coordinate(A2) at ($(O)+({1*cos(40)},{1*sin(40)})$);
	\coordinate(B2) at ($(O)+({1.7*cos(40)},{1.7*sin(40)})$);
	\draw (A2) node {$\bullet$};
	\draw[very thick](A2)--(B2) ;

	\coordinate(A3) at ($(O)+({1*cos(-10)},{1*sin(-10)})$);
	\coordinate(B3) at ($(O)+({1.7*cos(-10)},{1.7*sin(-10)})$);
	\draw (A3) node {$\bullet$};
	\draw[very thick](A3)--(B3) node[midway,below](C){};
	\centerarc[dashed](O)(-40:-90:1.4);
	\draw[dotted,thick, blue] (C)--($(C)+(0.5,0.7)$);
	\end{tikzpicture}
	\caption{Graphical representation of the deformation operators acting on  boundary states}
	\label{fig:disk_A_N_op_graph_rep}
\end{figure}
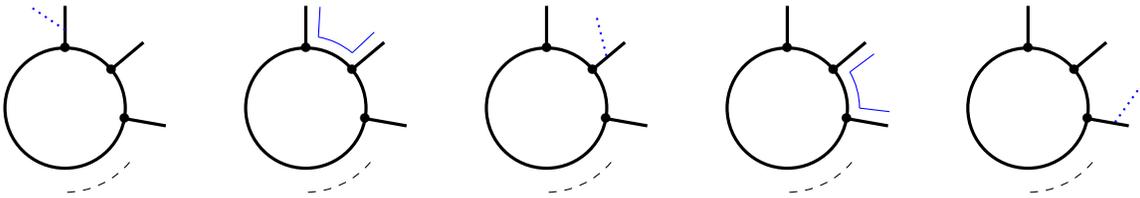

In figure \ref{fig:disk_A_N_op_graph_rep}, we have represented their graphical action on the disk $\cA_N$. The operators $\hg_i$ locally change the value of the holonomy along the boundary disk while the operators $\hx_i$ are grasping operators locally affecting the holonomies linking the boundary to the bulk.

It is then straightforward to check that the operators $\hx_i^a$ and $\hg_i^{AB}$ have closed commutators and that they form a Lie algebra: 
\be
\label{eq:commN1}
\left[ \hg_i \stackrel{\otimes}{,} \hg_j \right]= 0
\,, \qquad
\left[ \hx_i^a , \hx_j^b \right] = i \delta_{ij} {\eps^{ab}}_c \,\hx_i^c
\,,\qquad
\left[ \hx_i^a , \hg_j \right] = i \big( \delta_{i,j}\,  \hg_j \!\cdot \!\tau^a - \delta_{i,j+1}\,  \tau^a \!\cdot \! \hg_j   \big)
\,,
\ee
with the implicit cyclic notation that the indices $i$ and $j$ are defined modulo $N$.
The short hand notation $\tau^a \!\cdot \! \hg_j$ stands for the matrix operator $(\tau^a \hg_j)_A{}^{B}\deq \tau^a_{A}{}^{C}\,\hg_{i C}{}^{B}$. This Lie algebra is also known as the discrete {\it holonomy-flux} algebra $\lA_{N}$, here on the circular graph with $N$ edges. The holonomy-flux algebras are the starting point of the loop quantization of general relativity formulated as a gauge theory (see e.g. \cite{Thiemann:2007zz} for a full review of loop quantum gravity and more generally of the canonical quantization of general relativity in vierbein-connection variables or \cite{Bodendorfer:2016uat} for a more pedagogical introduction). Applied to 3D gravity, they are the basic algebra of observables in the canonical framework of 3D loop quantum gravity for the Ponzano-Regge topological state-sum model \cite{Freidel:2004vi,Freidel:2004nb,Freidel:2005bb,Noui:2004iz,Noui:2004iy,Meusburger:2008bs,Barrett:2008wh,Dowdall:2009eg,Dowdall:2010ej,Barrett:2011qe}.

\medskip

The (completion of the) universal enveloping algebra of this discrete holonomy-flux algebra is the disk algebra  $\cA_N = \cU(\lA_N)$. Indeed the left-invariant derivatives generate all the convolution operators, while spin-$\f12$ Wigner matrix operators generate the multiplication operators.
Let us start with a multiplication operator $\widehat{\Phi}_{m}$. The Peter-Weyl theorem ensures that any square-integrable function on $\SU(2)$ can be decomposed as a  series over the Wigner matrices, which can thus be interpreted as Fourier modes on the Lie group. This extends to $\SU(2)^{\times N}$, as $N$ copies of $\SU(2)$, and the function $\Phi$ can be written as  a series over products of Wigner matrix elements associated to each $\SU(2)$ group element:
\be
\Phi(\{k_{i}\}_{i})=\sum_{\{j_{i},m_{i},m'_{i}\}}\Phi^{\{j_{i}\}}_{\{m_{i}m'_{i}\}}\,
\prod_{i=1}^N
D^{j_{i}}_{m_{i}m'_{i}}(k_{i})
\,,
\ee
where we evaluate this function on the group elements $k_{i}=h_{i}^{-1}g_{i}h_{i+1}$.  By definition of the spin representation, each Wigner matrix $D^{j}_{mm'}(k)$ is a polynomial of degree $2j$ in the Wigner matrix elements $D^{\f12}_{AB}(k)$. This gives a decomposition of the multiplication operator $\widehat{\Phi}_{m}$ as a series of polynomials in the elementary operators $\hg_{i}^{AB}$.

As for the translation operators, the basic operators $\otimes_{i=1}^N \widehat{H}_i$ are simply the exponential of the the derivation operators $\hx^{a}_{i}$. To alleviate the notations, we focus on a single group element $h\in\SU(2)$ and drop the index $i$. Then we have:
\be
H=e^{i\theta_{a}J^a}=e^{-\theta_{a}\tau^a}\in\SU(2)\,,\qquad
\widehat{H}
=
e^{i \theta_a \hat{x}^a}
\,.
\ee
This correspondence $\widehat{H} = e^{i \theta_a \hat{x}^a}$ shows that the completion of $\cU(\su(2))$ is the group algebra $\C(\SU(2))$:
\be
\C(\SU(2)) = \hat\cU(\su(2)),
\ee
where the hat stands for the completion.

An illustrative way to check this is to go to the Fourier space. A convolution operation  is simply a multiplication in Fourier space. Indeed using the Peter-Weyl theorem on both the wave-function $\vphi$ and the convolution function $\Psi$ allows to write the convolution operator as a multiplication operator on the spin components of the wave-function:
\be
\left|
\begin{array}{lcl}
	\vphi(h)&=&\sum_{j,m,m'}\vphi^{(j)}_{m'm}D^j_{mm'}(h)
	\,,\vspace*{1mm}\\
	\Psi(h)&=&\sum_{j,m,m'}\Psi^{(j)}_{m'm}D^j_{mm'}(h)
	\,,
\end{array}\right.
\qquad
\widehat{\Psi}_{c}\triangleright \{\vphi^{(j)}\}_{j}
=
\left\{\f1{2j+1}\,\Psi^{(j)}\vphi^{(j)}\right\}_{j}
\ee
where we look at the Fourier components $\vphi^{(j)}$ as (2$j$+1)$\times$(2$j$+1)-matrices and use the standard matrix multiplication $\Psi^{(j)}\vphi^{(j)}$.
Now comparing the spin decomposition of an operator $\widehat{\delta}_{c}^{H}$,
\be
\HH \triangleright \big\{\vphi^{(j)}\big\}_{j}
=
\big\{D^j(H)\vphi^{(j)}\big\}_{j}
\,,
\ee
with the definition of the elementary derivation operators,
\be
\hx^a\,\vphi(h)=\vphi(h J^a)
\,,\qquad
\hx^a\triangleright \big\{\vphi^{(j)}\big\}_{j}=
\Big\{D^j(J^a)\vphi^{(j)}\Big\}_{j}
\,,
\ee
yields the expected exponential relation.
This allows to conclude that the holonomy-flux Lie algebra $\lA_{N}$ of holonomy and derivation operators around the $N$-segment circle does generate the disk algebra $\cA_{N}$, thus identified as its universal enveloping algebra $\cA_{N}=\cU(\lA_N)$.

\medskip

The rest of the paper is  dedicated to the study of the algebraic structures of the disk algebra $\cA_{N}=\cU(\lA_N)$ focusing on its factorization into elementary algebras and on the role of $N$.
The disk algebra encodes the observables of the 1d boundary of the 2d boundary of the 3d space-time. In the canonical framework, this 1d boundary is the corner of space-time, i.e. the spatial boundary of the canonical surface at the intersection between the space-like and time-like boundaries of the 3d region of space-time.
Exchanging the inside and outside of the 3d region, the disk algebra can also be understood the algebra of observables associated to a puncture around a point-like defect on the 2d boundary. Such  point-like boundary defects  are  the shadow of 1-dimensional  bulk  defects such as particle world-lines.
The algebraic structure of corners and defects reveal the symmetry algebra of the theory and the edge modes induced by the bulk dynamics on the space-time boundary. In the context of 3d quantum gravity, quantized as a topological quantum field theory (TQFT) for instance as the Ponzano-Regge state-sum model \cite{Freidel:2004vi,Freidel:2004nb,Freidel:2005bb,Barrett:2008wh,Dittrich:2018xuk,Dittrich:2017hnl,Dittrich:2017rvb,Goeller:2019zpz}, this is a necessary step in order to go beyond the partition function for pure gravity on a closed space-time manifold and truly understand the local structure of the geometry (through accounting for finite boundaries) and how degrees of freedom (such as matter fields or topological defects) can evolve on/through the 3D quantum gravity.

From this perspective, we propose to interpret the Lie algebra $\lA_{N}$ as a discrete current algebra. The limit case $N=1$,
which can be thought as a ``one-quantum-of-geometry'' regime,  gives the Drinfeld double $\cD\SU(2)$, which is the symmetry algebra of the theory, while the continuum limit $N\rightarrow +\infty$ leads back to the classical current algebra. This begs the question of the physical role of the number of discrete boundary edges  $N$. Whether this number of boundary geometry excitations is a crucial physical degree of freedom or a gauge degree of freedom is to be determined by the dynamics and the analysis of its renormalization flow.
In the meanwhile, the present work focuses on the kinematics: how refining and coarse-graining the boundary can be implemented as super-operators, that is operators sending operators in the disk algebra $\cA_{N}$ onto operators in other disk algebras with smaller or higher $N$,  and how merging defects is mathematically formulated as the fusion of the disk algebras.



\subsection{The Disk Algebra $\cA_1$ as a Drinfeld Double}

We now turn to the in-depth exploration of the algebraic structure of the disk algebras $\cA_{N}$ and discuss their interpretation as a discrete version of the current algebra. It is important to keep in mind the distinction between the whole algebra $\cA_{N}$ of disk boundary observables and the Lie algebra $\lA_{N}$.
The disk Lie algebra $\lA_{N}$ is generated by the tensor operators $\hx_i^a$ and $\hg_{i}^{AB}$,
\be
\left[ \hg_i \stackrel{\otimes}{,} \hg_j \right]= 0
\,, \qquad
\left[ \hx_i^a , \hx_j^b \right] = i \delta_{ij} {\eps^{ab}}_c \,\hx_i^c
\,,\qquad
\left[ \hx_i^a , \hg_j \right] = i \big( \delta_{i,j}\,  \hg_j \!\cdot \! \tau^a - \delta_{i,j+1}\,  \tau^a \!\cdot \! \hg_j   \big)
\,.
\ee
Its enveloping algebra $\cU(\lA_{N})$ is linearly generated by  elements $ P(\{\hat{x}_i \})D^j_{mn}(g)$, where $P$ is  polynomial in $\hat{x}_i$ and $j$ is an arbitrary spin. As we have shown in the previous section the completion of
 $\cU(\lA_{N})$ denoted $\widehat\cU(\lA_{N})$, which contains exponentials of $x^i$ and smooth functions of $g$, is the disk algebra  $\cA_{N}$.

The purpose is to clarify the structure of both  $\lA_{N}$ and  $\cA_{N}$. We show in the next pages that  the disk algebra $\cA_{N}$ can be  decomposed in terms of the Drinfeld double and the Heisenberg double of $\SU(2)$.
In particular, in the case $N=1$, the disk algebra $\cA_{1}$ is exactly the Drinfeld double $\cD \SU(2)$.
In the general case, for a larger number of flux insertions $N\ge 2$, the disk algebra is actually a periodic sequence of coupled Heisenberg doubles:
\be
\cA_{2}=\cH\SU(2)\bowtie\cH\SU(2)\,,\qquad
\cA_{N\ge 2}=\underset{N \textrm{ factors}}{\underbrace{\cH\SU(2)\ltimes\cH\SU(2)\ltimes .. \ltimes\cH\SU(2)\ltimes}}
\quad\,,
\ee
where the opened semi-direct product $\ltimes$ at the end of the expression reflects the periodicity of the disk's boundary and means that the last Heisenberg double acts on the first one. each $\cH\SU(2)$ algebra is generated by the pair $(\hx_i^a,\hg_{i}^{AB})$ for a given $i$.
We will further show that  one can always factor the disk algebra $\cA_{N}$ into its ``centre of mass'' subalgebra $\cA_{1}\sim\cD \SU(2)$ and its vibrational modes or ``relative motion modes'' given by a tower of Heisenberg doubles:
\begin{equation}\label{iso}
	\cA_N =  \cD \SU(2) \times \cR_{N-1}
	\,,\qquad
		\cR_{N-1}\sim\cH\SU(2)^{N-1}
		\,.
\end{equation}
This factorization allows to change the flux insertions in a simple and clear way, thus making possible to cleanly describe the refinement and coarse-graining maps, as we will explain in section \ref{sec:exploring}.

\medskip

Let us start this journey with the study of the simplest case $N=1$ with a single boundary insertion and the corresponding disk algebra $\cA_1$.  This algebra is the enveloping  algebra of the Lie algebra $\lA_{1}$ consisting in the matrix elements $\hG^{AB}$ and the vector elements $\hX^{a}$ with the following commutators
\begin{equation}
\left[\hG_{AB},\hG_{A'B'} \right]
=
0
\; , \quad
\left[ \hX^{a},\hX^{b} \right]
=
i{\eps^{ab}}_c \hX^{c}
\; , \quad
\left[\hX^{a},\hG \right]
=
i \big( \hG\!\cdot \!  \tau^a - \tau^a \!\cdot \!  \hG \big)
\, .
\label{eq:commutator_A1}
\end{equation}
%
Below we explain that this Lie algebra should be recognized as a Drinfeld double, $\lA_{1}\sim\cD\su(2)$.
Identifying this Drinfeld double in the context of 3d quantum gravity is not surprising, it is actually a recurring theme that $\cD\SU(2)$ is the algebra of observables for punctures. But here, it is clearly the first stepping stone for the single-insertion boundary $N=1$ and we will uncover a richer algebra of observables in the following sections as the number of open edges $N$ increases.

 \subsubsection{Lie algebra double}

The structure of the $\lA_{1}$ commutators hints toward a natural interpretation in terms of quantum double \cite{Ashtekar:1989qd,Batista:2002rq,Gutierrez-Sagredo:2018zog}.
More precisely, the operators $\hX^a$ generate an $\su(2)$ Lie sub-algebra. This Lie algebra can be made into a co-commutative Hopf algebra by adding the generator $\hX^{0} = \hat{1}$ (acting as the unit of the Hopf algebra) and with the co-product, the co-unit and the antipode defined respectively by
\begin{equation}
\Delta(\hX^{a}) = \hX^{a} \otimes \hat{1} + \hat{1} \otimes \hX^a \;, \quad \eps(\hX^{a}) = 0 \;, \quad S(\hX^a) = -\hX^a \;.
\end{equation}
The product of the co-commutative Hopf algebra is the usual $\su(2)$ product. To be more precise, this co-commutative Hopf algebra generates the universal enveloping algebra of  $\cU(\su(2))$. The dual of an Hopf algebra is another  Hopf algebra  where the product of one is adjoint to the product of the other. This dual can be used to define the Drinfeld double of the algebra.
We pair the  co-commutative Hopf algebra $\su(2)$ with the commutative algebra $\cC_P[\SU(2)]$ of polynomial functions on $\SU(2)$. It is linearly  generated by polynomials in the coordinate functions $\hG^{AB}$. The coordinate functions act on elements of $\SU(2)$ as follows
\begin{equation}
\hG_{AB} (g) = D^{{\f12}}_{AB}(g) \; ,  \quad \forall \; g \in \; \SU(2).
\end{equation}
The co-product, co-unit and antipode of the commutative Hopf algebra $\C[\SU(2)]$ are respectively
\begin{equation}
\Delta(\hG_{AB}) = \hG_{A C} \otimes \hG^C{}_{ B}\;, \quad \eps(\hG_{AB}) = \delta_{A B} \;, \quad S(\hG_{AB}) = (\hG^{-1})_{AB} \;,
\end{equation}
where the summation over $C$ is implicit and where the inverse is taken at the matrix level.
The two Hopf algebras $\cC_P[\SU(2)]$ and $\cU(\su(2))$ have a non-trivial pairing
given by
\begin{equation}
\la P | \hx^{a_1}\cdots \hx^{a_n} \ra = \left.\f{\rd}{\rd t_1}\cdots \f{\rd}{\rd t_N} P(e^{t_1 J^{a_1} }\cdots e^{t_n J^{a_n} })\right|_{t_1=\cdots = t_N=0}\; , \quad \forall P \in \; \cC_P(\SU(2)).
\end{equation}
At the level of the generators, this reads
\begin{equation}
\la  \hG^{AB} |1 \ra
=
\delta_{AB}
,\qquad
\la  \hG_{AB} |\hX^a  \ra
=
i\tau^a_{AB} \; .
\end{equation}
For the rest of this paper and in particular for the analysis of the Casimir and the representation of $\cA_1$, it is convenient to introduce the basis directly dual to $\hX^\mu=(\hat{1},\hX^a)$ with respect to the previous pairing. It is given by $\hG_a \deq 2 i  \tr(\hG \tau_a )$ where we also introduce $\hG_0\deq \f{1}{2} \tr\,\hG$. The map between the two systems of coordinates is then simply given by
\begin{equation}
\hG = \hG_0 \tau^0 +i \hG_a \tau^a,
\end{equation}
where $\tau^0 = \id$ is the two-dimensional identity matrix. In terms of this basis, the pairing simply becomes
\begin{equation}
\la \hX^\mu | \hG_\nu \ra =  \delta^\mu_\nu,
\end{equation}
where $\mu,\nu=0,1,2,3$. We can easily check that $\hG_0$ is a Casimir while the non-trivial part of the algebra\footnotemark{} is given by
\begin{equation}
[\hG_a,\hG_b]=0,\qquad [\hX^a,\hG_b]  = i \eps^{a}{}_b{}^c \hG_c,\qquad
[\hX^a,\hX^b]= i{\eps^{ab}}_c \hX^c.
\end{equation}
\footnotetext{
As a Lie algebra this is simply the Euclidean algebra $\mathfrak{iso}(3)$, where $\hX$ plays the role of angular momenta while $\hG_a =\lambda P_a $ plays the role of momenta. The parameter $\lambda$ with the physical dimension of a  length  plays the role of the Planck length in $3D$ gravity \cite{Batista:2002rq,Joung:2008mr}.
}
Through this construction, one  recognizes  the Lie algebra $\lA_1$ as the Drinfeld double  $\cD \su(2)$ \cite{Gutierrez-Sagredo:2018zog}.

\subsubsection{ Associative algebra double}

The disk algebra $\cA_{1}$ is the completion of the enveloping algebra  $\cU( \cD\su(2))$ and we have the algebra isomorphism
\be
\cU(\cD\su(2))\simeq \cU[\su(2)] \ltimes   \cC_P(SU(2)),
\ee
where $\cC_P(SU(2))$ is the set  of complex polynomial functions on the group.
We have already seen that the group algebra
$\mathbb{C}[SU(2)]$, which consists of finite sums
$\sum_\alpha c_\alpha G_\alpha$, with $c_\alpha \in \C$ and $G_\alpha\in \SU(2)$,  can be viewed as a completion of the enveloping algebra\footnote{This completion contains only finite sums of exponentials function. It is possible to include a wider class of functions is this completion see \cite{Freidel:2005ec}. For the purpose of this paper we only need to consider the group algebra.}
\be
\mathbb{C}[SU(2)] =\widehat{\cU}(\su(2)).
\ee
We can also complete the polynomial algebra $\cC_P(\SU(2))$ into the algebra $\cC(\SU(2))$ of continuous functions. $\mathbb{C}[SU(2)]$ and $\cC(\SU(2))$ are both sub-algebras of $\cA_1$ and dual to each other with pairing
\be
\omega(\Phi,H) =\Phi(H).
\ee
This means that $\cA_1$ is the braided Hopf algebra $\cU(\cD\su(2))$ and is isomorphic to $\C[SU(2)] \otimes \mathbb{C}[SU(2)]$ as a vector space.
As an algebra it is given by the semi direct product
\be
\cA_1=\hat\cU(\cD\su(2))\simeq \C[SU(2)] \ltimes_{\mathrm{Ad}}  \cC(SU(2))
= \cD\SU(2).
\ee
The action of the group algebra is given by  the adjoint action  defined by
\be
\textrm{Ad}_H\left[ \Phi\right](g) = \Phi(H^{-1}g H) .
\ee
$\cD \SU(2)$ is the Drinfeld double algebra and is an Hopf algebra.
A general element of $\cD \SU(2)$ can be represented as a linear combination of elements $H \otimes \Phi \; \in \; \C[SU(2)] \otimes \cC(SU(2))$.
The Hopf algebra  operations on the double are defined by the following product, co-product, co-unit and antipode
\begin{align}
	( \Phi_1 \otimes H_1 )(\Phi_2 \otimes H_2  ) &= \Phi_1\,\textrm{Ad}_{H_{1}}[\Phi_2] \otimes  H_1 H_2 \; , \label{alg}
	\\
	\Delta(  \Phi \otimes H )(g_1,g_2) &=   \Phi(g_1 g_2) ( H
	 \otimes  H ) \; ,
	\\
	\eps(\Phi \otimes H) &= \Phi(1) \; ,
	\\
	S( \Phi \otimes H)(g) &=  \Phi(g^{-1}) \otimes H^{-1} \; ,
\end{align}
where $1$ is the be understood as the identity function on the group and
where
\be
 \Phi_1\,\textrm{Ad}_{H_{1}}[\Phi_2](g)= \Phi_1(g) \Phi_2(H_1^{-1} g H_1).
\ee
We denote $(  \Phi \otimes H )(g) = \Phi(g) H$.

\begin{figure}[!htb]
	\centering
	\begin{tikzpicture}[scale=0.45,every node/.style={fill=white}]
	\coordinate (A1) at (-6,3); \coordinate (A6) at (6,3);
	\coordinate (A2) at (-4,3); \coordinate (A5) at (4,3);
	\coordinate (A3) at (-2,3); \coordinate (A4) at (2,3);
	\coordinate (B1) at (-6,1); \coordinate (B6) at (6,1);
	\coordinate (B2) at (-4,1); \coordinate (B5) at (4,1);
	\coordinate (B3) at (-2,1); \coordinate (B4) at (2,1);
	\coordinate (C1) at (-6,-1); \coordinate (C6) at (6,-1);
	\coordinate (C2) at (-4,-1); \coordinate (C5) at (4,-1);
	\coordinate (C3) at (-2,-1); \coordinate (C4) at (2,-1);
	\coordinate (D1) at (-6,-3); \coordinate (D6) at (6,-3);
	\coordinate (D2) at (-4,-3); \coordinate (D5) at (4,-3);
	\coordinate (D3) at (-2,-3); \coordinate (D4) at (2,-3);
	\coordinate (O) at (0,0); \node at (O) {$=$};

	\node at (A1) [above] {$1$}; \node at (A2) [above] {$2$}; \node at (A3) [above] {$3$};
	\draw (A1) -- (B1);
	\draw (A2) to[out=-90,in=90] node{~} (B3);
	\draw (A3) to[out=-90,in=90] (B2);

	\draw (B1) to[out=-90,in=90] node{~} (C2);
	\draw (B3) -- (C3);
	\draw (B2) to[out=-90,in=90] (C1);

	\draw (C2) to[out=-90,in=90] node{~} (D3);
	\draw (C3) to[out=-90,in=90] (D2);
	\draw (C1) -- (D1);
	\node at (D1) [below] {$3$}; \node at (D2) [below] {$2$}; \node at (D3) [below] {$1$};

	\node at (A4) [above] {$1$}; \node at (A5) [above] {$2$}; \node at (A6) [above] {$3$};
	\draw (A6) -- (B6);
	\draw (A4) to[out=-90,in=90] node{~} (B5);
	\draw (A5) to[out=-90,in=90] (B4);

	\draw (B5) to[out=-90,in=90] node{~} (C6);
	\draw (B4) -- (C4);
	\draw (B6) to[out=-90,in=90] (C5);

	\draw (C4) to[out=-90,in=90] node{~} (D5);
	\draw (C5) to[out=-90,in=90] (D4);
	\draw (C6) -- (D6);
	\node at (D4) [below] {$3$}; \node at (D5) [below] {$2$}; \node at (D6) [below] {$1$};

	\end{tikzpicture}
	\caption{Graphical representation of the Yang-Baxter equation}
	\label{fig:R_matrix_A1}
\end{figure}
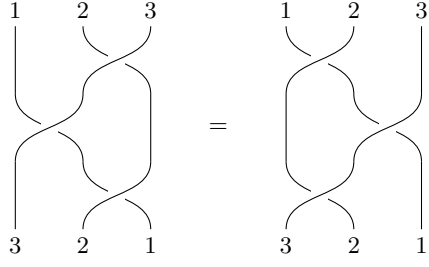

We are therefore able to make use of the extended work already done on the quantum double of $\SU(2)$. In particular, the braiding of the disk algebra, i-e the $R$-matrix, is given by the canonical element  of the Drinfeld double:
\begin{equation}
R = \int_{\SU(2)} \rd H (\delta_H \otimes \mathbb{I}) \otimes (1 \otimes H) \; .
\end{equation}
The $R$-matrix is not a element of $\cD \SU(2) \otimes \cD \SU(2)$, but it possesses well defined action on  $\cD \SU(2) \otimes \cD \SU(2)$. Its action on the coproduct gives
\begin{equation}
R \Delta  = \Delta' R
\end{equation}
where $\Delta'$ is the co-product with transpose target space compared to $\Delta$. This $R$-matrix satisfies the  Yang-Baxter equation, graphically represented figure \ref{fig:R_matrix_A1}, with the standard tensorial notations\footnotemark{},
\begin{equation}
R_{12}R_{13}R_{23} = R_{23} R_{13} R_{12} \;.
\end{equation}
\footnotetext{ With the standard tensorial notations, the three $R$-matrices acting on the triple tensorial product are explicitly
	\be
	R_{12} = \sum_{k}  (\delta_k \otimes \mathbb{I}) \otimes (1 \otimes k) \otimes (1 \otimes \mathbb{I})
	\,,\qquad
	R_{13} = \sum_{k}  (\delta_k \otimes \mathbb{I}) \otimes (1 \otimes \mathbb{I}) \otimes (1 \otimes k)
	\,,\qquad
	R_{23} = \sum_{k}  (1 \otimes \mathbb{I}) \otimes (\delta_k \otimes \mathbb{I}) \otimes (1 \otimes k)
	\,.
	\nn
	\ee
}
and is relevant for a mathematical definition of the fusions of disk algebras that will be explored in future work.

\subsubsection{Reducible and Irreducible representations}
Let us conclude this section with a study of the irreducible representations of the double and their relations to our initial space of spin-networks wave-functions. We have already seen that $\hG_0$ is a Casimir. Moreover, since $\hG$ is an hermitian operator, $\hG^\alpha \hG^\dagger_{\alpha} =1$, we have the relation
\begin{equation}
	(\hG_{0})^2+\sum_{a}\hG_a^2 =\id\,.
\end{equation}
From this equation, one can deduce that the Casimir $\hG_0$ can be labelled by a mass operator $\hat{m}$ such that $\cos\hat{m}= \hG_0$ and such that the quadratic Casimir  $\hG_a\hG^a $ is equal to $\sin^2 \hat{m}$. There is one remaining Casimir on this algebra, given by the spin operator
\begin{equation}
	\hat{s}= \hX^a \hG_a.
\end{equation}
Any other Casimir is a function of $(\hat{m},\hat{s})$ which is $2\pi$ periodic in $\hat{m}$. Note that similarly to the Poincar\'e algebra, the usual $\su(2)$ Casimir $\hX^a \hX_a$ is not a Casimir here.

\begin{figure}[!htb]
	\centering
	\begin{subfigure}[t]{.4\linewidth}
		\begin{tikzpicture}[scale=1.1]

		\coordinate(O) at (0,0);
		\coordinate(A1) at (0,1);
		\coordinate(B1) at (0,1.7);

		\draw (A1) node {$\bullet$};
		\draw[very thick] (A1)--(B1) node[above]{$j,m$};

		\draw[very thick,fill=lightgray] (0,0) circle (1) ;
		\draw (-1,0) node[left]{$k$};
		\end{tikzpicture}
		\caption{One point boundary states on the disk, labeled by the loop spin $k$ and the spin state $j,m$. The open edge corresponds to the flux insertion. }
		\label{fig:disk_alg_N_1_nota_bdry}
	\end{subfigure}
	\hspace*{8mm}
	\begin{subfigure}[t]{.4\linewidth}
		\begin{tikzpicture}[scale=1]
		\coordinate(O) at (0,0);
		\coordinate(A1) at (0,1);
		\coordinate(B1) at (0,1.7);

		\draw (A1) node {$\bullet$};
		\draw[very thick] (A1)--(B1) ;

		\draw[very thick] (O) circle (1) ;
		\centerarcnodes[blue](0,0)(-260:80:1.2)(C1,D1);
		\draw[blue] (D1)--++(0,0.5);
		\draw[blue] (C1)--++(0,0.5);
		\node[above left,blue] at (C1) {$\hG$};

		\coordinate(P) at (4,0);
		\coordinate(p1) at (4,1);
		\coordinate(q1) at (4,1.7);
		\draw (p1) node {$\bullet$};
		\draw[very thick] (p1)--(q1) ;
		\draw[very thick] (P) circle (1) ;
		\draw[dotted,thick] (4,1.3)-- node[pos=0.8,above] {$\hX$} (3.4,1.7);
		\end{tikzpicture}
		\caption{Graphical representation of the two basic operators of $\cA_{1}$ acting on the disk, respectively holonomy $\hG$ and grasping $\hX$ operators.}
		\label{fig:disk_alg_N_1}
	\end{subfigure}
	\caption{Sunny graph for $\cA_1$ with the notations for the boundary wave function (figure \ref{fig:disk_alg_N_1_nota_bdry}) and the operators and their graphical actions (figure \ref{fig:disk_alg_N_1}).}
\end{figure}
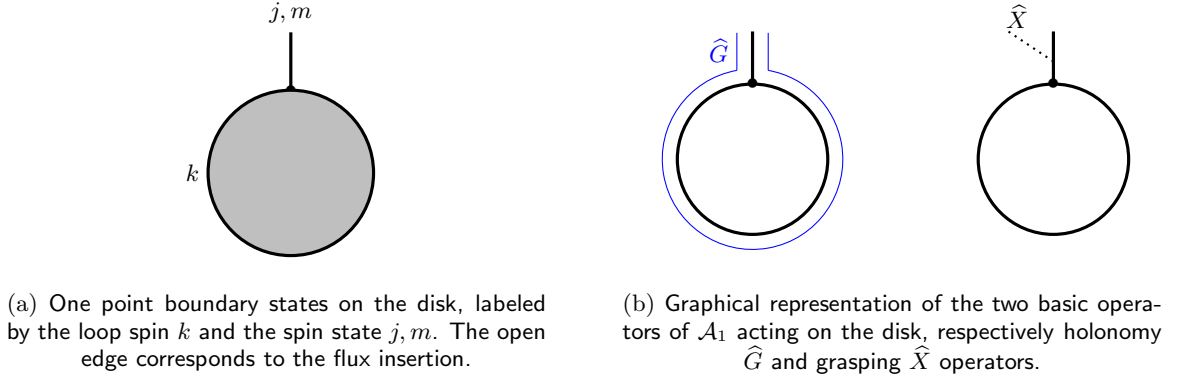

As described in the general case in the previous section, the disk algebra is naturally represented by operators acting on wave-functions $\varphi$ defined on the configuration space $\SU(2) \times \SU(2) / \SU(2)$ of the holonomy around the disk's boundary with a single flux insertion, as drawn in figure \ref{fig:disk_alg_N_1}.
They act on wave-functions $\varphi(g,h)$ satisfying the following gauge invariance:
\begin{equation}
\varphi(g,h) = \varphi(a g a^{-1}, a h) \; , \quad \forall \; a \in \; \SU(2) \;.
\end{equation}
The spin network basis of the Hilbert space of $L^{2}$ wave-functions  is labelled by the spin $k$ around the loop and a spin state $|j,m\ra$ at the insertion. The wave-functions read:
\begin{equation}
\la g,h\,|\,k,j,m\ra
=
\varphi_{k,j,m}(g,h)
=
D^{k}_{\alpha\beta}(g)\,C^{k,j\,|\,k}_{\alpha,\gamma\,|\,\beta}\,D^{j}_{\gamma m}(h)
\,,
\qquad
C^{k,j\,|\,k}_{\alpha,\gamma\,|\,\beta}
=
\la (k,\beta)\,|\, (k,\alpha)(j,\gamma) \ra
\,,
\end{equation}
in terms of Wigner matrices for the two group elements $g$ and $h$ and the Clebsh-Gordan coefficients $C^{k,j\,|\,k}_{\alpha,\gamma\,|\,\beta}$.
%
The basis state for a trivial spin insertion $j=0$ is simply the spin-$k$ Wilson loop observable $\varphi(g,h)=\tr_{k}\,g$, defined as the trace of the group element $g$ around the point in the spin-$k$. Wave-functions with non-trivial $j$ and $m$ are straightforward generalization of Wilson loops with a single boundary insertion as used in the loop quantization of 3D gravity \cite{Ashtekar:1989qd,Ashtekar:1989qc}.

This means that the disk algebra  $\cA_{1}$ hence the Drinfeld double $\cD \SU(2)$  can be represented in terms of mutiplication and translation operators whose actions on the wave-functions $\varphi(g,h)$ read
\begin{align}
\widehat{\Phi}\, \varphi(g,h) &= \Phi(h^{-1}g h) \varphi(g,h) \\
\HH \varphi(g,h) &= \varphi(g,hH) \; .
\end{align}
It is immediate to see that this pair of operators can be associated to elements $\Phi \otimes H$ of $\cD \SU(2)$.
The element $ \Phi \otimes H$ is  is understood to act on wave-functions
$\varphi$ simply as the ordered operator $ \widehat{\Phi}\HH$:
\begin{equation}
(H\otimes \Phi)\varphi(g,h)
=
( \widehat{\Phi} \HH ) \varphi(g,h)
=
\Phi(H^{-1}h^{-1}g h H) \, \varphi(g,hH)  \label{action1}\; .
\end{equation}
From this point, it is straightforward to check that the algebra \eqref{alg} is satisfied.

The infinitesimal version of this  representation means that The operators $\hG^{AB}$ are the components of the holonomy around the disk, while the flux  $\hX^{a}$ acts as a derivation at the insertion.
\medskip

\medskip
We can now show that irreducible unitary representations of $\cD \SU(2)$ are labelled by the quantum number $(\theta,s)$ associated to the previous Casimir, representing respectively the mass and angular momentum.
Since the sub-algebra generated by $\hG$ is commutative we can construct a carrier space at fixed $(\theta,s)$, ${\cal H}_{\theta,s}$, to be a section of the space of square integrable functions $L^2(\mathrm{SU(2)})$. In more detail, they are sections of a line bundle over $\mathrm{SU(2)}$ which carries a representation of spin $s$. An element $\phi$ of ${\cal H}_{\theta,s}$ is then a functional following the equality
\begin{equation}
\phi(  u_t g)= e^{i t s } \phi(g) \; ,
\end{equation}
where $u_{t} = e^{t \tau_3}$. Since $ e^{4\pi  \tau_3}=1$ we need to have $s \in \mathbb{Z}/2$. The action of an arbitrary $\cD \SU(2)$ element $H \otimes \Phi$ on $\phi \in {\cal H}_{\theta,s}$ is then
\begin{equation}
\left((H \otimes \Phi) \act \phi \right)(h) = \Phi(h u_{\theta} h^{-1}) \phi(H^{-1} h) \; .
\end{equation}
One can check that the action of the  mass casimir $\hat{m}$ is fixed to $\theta$
by this action.

A basis of ${\cal H}_{\theta,s}$ is  provided by a decomposition in terms of  Wigner matrices
\begin{equation}
{\cal H}_{\theta,s} = \bigoplus_{j \geq s} \bigoplus_{-j \leq m \leq j} \C \; D^{j}_{m s} \; .
\end{equation}
The question is now, how does our initial space of wave-functions, $\varphi(g,h) = \varphi(a g a^{-1}, a h)$, compare to these irreducible representations ? It is clear that, to compare the two, one must actually fix the class angle of $g$ to $\theta$, or equivalently set $g=u_{\theta}$ and consider the wave-function $\varphi$ as a function on one group element $h$:
\begin{equation}
\phi(h)=\varphi(u_{\theta},h) \,.
\end{equation}

Therefore, decomposing the basis states $\varphi_{k,j,m}$ on $D^{j}_{m s}$ is obtained by a group averaging fixing $g= u_{\theta}$. The particularity of this decomposition is that it does not isolate the spin $k$ going around the disk. More precisely, the projection of the basis state $\varphi_{k,j,m}$ into $\cH_{\theta,s}$ is
\begin{equation}
\la \cH_{\theta,s} |\varphi_{k,j,m}(g,h) \ra = \varphi^{(\theta,s)}_{k,j,m}(h)
=
\int \rd a\, \delta(aga^{-1}u_{\theta})\,D^{j}_{m s}(ah)
=
D^k_{\alpha\beta}(g)D^j_{\gamma s}(h)C^{k,j\,|\,k}_{\alpha,\gamma\,|\,\beta} \,\sum_{-k\le n\le+k}e^{in\theta} C^{k\,|\,k,j}_{n\,|\,n,m} \; .
\end{equation}
One can then recover the full basis state $\varphi_{k,j,m}$ by a simple integration and summation over the mass and angular momentum quantum numbers to explore the full Hilbert space
\begin{equation}\label{eq:decompo_rep_A1}
\varphi_{k,j,m}(g,h) = \sum_{s \in \Z} \int_{0}^{2 \pi} \text{d} \theta  \; \varphi^{(\theta,s)}_{k,j,m}(h) \; .
\end{equation}
Therefore, on can conclude that our initial configuration space of wave-functions $\varphi$ is a reducible representation of $\cD SU(2)$ , whose decomposition into irreducible representation is given by \eqref{eq:decompo_rep_A1}.
\bigskip

In this section, we have shown that the disk algebra for $N=1$ is actually the Drinfeld double of $\SU(2)$, $\cD \SU(2)$. This result is not really surprising, since the quantum double $\cD \SU(2)$ has now long been understood as the quantized version of the algebra $\iso(3)$. What is more interesting is that we still have to understand the role of the parameter $N$. If the $N=1$ case is already sufficient to capture the global quantum symmetries of the theory, what is the use of $N$ ? In the following, we will see that $N$ parametrizes the amount of relative information we can extract from the bulk of the disk. In particular, we will always be able to factorize out a $\cD \SU(2)$ from $\cA_N$, representing the usual quantum symmetries of flat gravity, while the $(N-1)$ other contributions to $\cA_N$ are related to vibrational modes of the 1d boundary.

\subsection{The Disk Algebra  $\cA_2$ and the Heisenberg Double}

In the previous section, we focused our attention on the  disk algebra $\cA_1$ for a single flux insertion  $N=1$. Now, we move one step up and study the case $N=2$.
The  disk algebra $\cA_2$ is the enveloping algebra of the Lie algebra $\lA_2$ built with two pairs of operators $(\hx_1,\hg_1)$ and $(\hx_2,\hg_2)$ with commutators
\begin{equation}
[\hx_1^a,\hx_1^b] = i \eps^{ab}{}_c \hx_1^c \; , \quad [\hg_1,\hg_1 ] = 0 \; , \quad [\hx_1^a,\hg_1] = i \hg_1 \tau^a
\label{eq:com_A2_H1}
\end{equation}
\begin{equation}
[\hx_2^a,\hx_2^b] = i \eps^{ab}{}_c \hx_2^c \; , \quad [\hg_2,\hg_2 ] = 0 \; , \quad [\hx_2^a,\hg_2] = i \hg_2 \tau^a
\label{eq:com_A2_H2}
\end{equation}
\begin{equation}
[\hx_1^a,\hx_2^b] = 0 \; , \quad [\hg_1,\hg_2] = 0 \; , \quad [\hx_1^a,\hg_2] = -i \tau^a \hg_2 \; , \quad [\hx_2^a,\hg_1] = -i \tau^a \hg_1 \;.
\label{eq:com_A2_cross}
\end{equation}
The first two lines corresponds to the closed commutators for each pair of operators  $(\hx_1,\hg_1)$ and $(\hx_2,\hg_2)$ while the last line corresponds to the cross commutators between them. From the commutators structure, it is clear that the algebras $\cE^{(1)}$ and $\cE^{(2)}$, respectively generated  by $(\hx_1,\hg_1)$ and $(\hx_2,\hg_2)$, are isomorphic. Then the cross commutators leads to an action of the two sectors on each other.

\medskip

\begin{figure}[!htb]
	\centering
	\begin{tikzpicture}[scale=1]
	\coordinate (O) at (0,0);
	\coordinate (A1) at (0,1); \coordinate (B1) at (0,1.7);
	\coordinate (A2) at (0,-1); \coordinate (B2) at (0,-1.7);
	\coordinate (C1) at (-1,0); \coordinate (C2) at (1,0);

	\draw[thick,->-=0.5,->-=1] (O) circle (1);
	\node at (A1){$\bullet$};
	\draw (A1)--node[above,pos=1]{$\hx_1$} (B1);
	\node at (A2){$\bullet$};
	\draw (A2)--node[below,pos=1]{$\hx_2$} (B2);
	\node[left] at (C1) {$\hg_1$};
	\node[right] at (C2) {$\hg_2$};
	\coordinate (eq) at (2,0); \node at (eq){$\sim$};
	\coordinate (O1) at (4,0); \coordinate (O2) at (4.5,0);
	\coordinate (A11) at (4,1); \coordinate (B11) at (4,1.7);
	\coordinate (A22) at (4.5,-1); \coordinate (B22) at (4.5,-1.7);
	\coordinate (C11) at (3,0); \coordinate (C22) at (5.5,0);

	\centerarc[thick,->-=0.5](O2)(-90:90:1);
	\centerarc[thick,->-=0.5](O1)(90:270:1);

	\draw[dotted,red,thick] (A11) --+(0.5,0);
	\draw[dotted,red,thick] (A22) --+(-0.5,0);

	\node at (A11){$\bullet$}; \node[below] at (A11){$(1)$};
	\draw (A11)--node[above,pos=1]{$\hx_1$} (B11);
	\node at (A22){$\bullet$}; \node[above] at (A22){$(2)$};
	\draw (A22)--node[below,pos=1]{$\hx_2$} (B22);

	\node[left] at (C11) {$\hg_1$};
	\node[right] at (C22) {$\hg_2$};
	\end{tikzpicture}
	\caption{Graphical representation of $\lA_{2}$ and its splitting into two equivalents part with the pairs of operators $(\hx_1,\hg_1)$ and $(\hx_2,\hg_2)$. The location of the first $(1)$ and second $(2)$ handles are specified, corresponding respectively to the graphical representation of the algebras $\cE^{(1)}$ and $\cE^{(2)}$. The cross-commutators between $\cE^{(1)}$ and $\cE^{(2)}$ are living on the red dashed line linking the two parts.}
	\label{fig:disk_alg_N_2}
\end{figure}
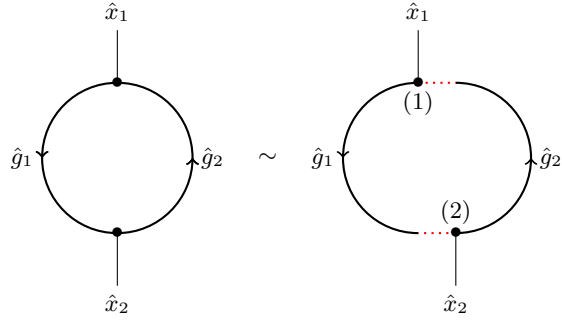

These operators can be naturally represented as multiplication and convolution operators acting on the disk's boundary with two open edges. This provides them with a very handy graphical visualization \ref{fig:disk_alg_N_2}.
They act on  wave-functions $\varphi$ on the configuration space $(\SU(2) \times \SU(2))^{\times 2} / (\SU(2) \times \SU(2))$. The $\SU(2)^{{\times 2}}$ gauge invariance reads
\begin{equation}
\varphi(g_1,g_2,h_1,h_2) = \varphi(a_1 g_1 a_2^{-1},a_{2} g_2 a_1^{-1}, a_1 h_1, a_2 h_2) \; , \quad \forall \; (a_1,a_2) \in \; \SU(2) \otimes \SU(2) \;.
\end{equation}
A basis of the Hilbert space of $L^{2}$ wave-functions is then labelled by spin-networks with two 3-valent nodes as represented in the sunny graph \ref{fig:disk_alg_N_2}. As in the $N=1$ case, one associates  spins $k_1,k_2$ to the edges of the disk and spin states $|j_1,m_1\ra, |j_2,m_2 \ra$ to the two insertions, such that the basis wave-functions reads
\begin{equation}
\la g_1,g_2,h_1,h_2 \,|\,k_1,k_2,j_1,m_1,j_2,m_2\ra
=
D^{k_1}_{\alpha_1\beta_1}(g_1)\, C^{k_1,j_2\,|\,k_2}_{\beta_1,\gamma_2\,|\,\alpha_2} \, D^{j_2}_{\gamma_2 m_2}(h_2) D^{k_2}_{\alpha_2\beta_2}(g_1) \, C^{k_2,j_1\,|\,k_1}_{\beta_2,\gamma_1\,|\,\alpha_1}\, D^{k_1}_{\gamma_2 m_1}(g_1)\,.
\end{equation}
This graphical representation \ref{fig:disk_alg_N_2}  provides the $\lA_{2}$ generators with the physical interpretation of boundary observables.

\medskip

Comparing these $\lA_2$ commutators with the $\lA_1$ commutators \eqref{eq:commutator_A1} allows for a better understanding of the sub-algebra $\cE^{(1)}$ and $\cE^{(2)}$ and how they combine into $\lA_{2}$.
Let us start by focusing on the pair of operators $(\hx_1,\hg_1)$. Similarly to $\lA_1$, the operators $\hx_1$ generate the Lie algebra $\su(2)$ while the operators $\hg_1$ form the abelian algebra $\C[\SU(2)]$. The enveloping algebra $\cE^{(1)}$ is thus isomorphic to $ \C[\SU(2)]\otimes \cC(\SU(2))$. The difference between $\cA_1$ and $\cE^{(1)}$ results from the cross-commutators between the operators $\hx_1$ and $\hg_1$. Instead of the adjoint action of $\SU(2) $ on $\cC(\SU(2))$, group elements now act on group functions by the left action (denoted $\textrm{L}$ thereafter). This algebra is recognized as the  Heisenberg double $\cH\SU(2)$.

In more details, the pair $H_{1}\otimes\Phi_{1}\,\in \C[\SU(2)]\otimes \cC(\SU(2))$ is represented by the composed operator $\widehat{\Phi}_{1}\widehat{H_{1}}$ (where we have dropped the subscript $m$ and $c$ referring to ``multiplication'' and `` convolution'' for alleviate the notations) acting on wave-functions $\vphi$:
\begin{align}
\widehat{\Phi}_{1}\,\vphi(g_{1},g_{2},h_{1},h_{2})
&=
\Phi_{1}(h_{1}^{-1}g_{1}h_{2})\vphi(g_{1},g_{2},h_{1},h_{2})
\,,\\
\widehat{H_{1}}\,\vphi(g_{1},g_{2},h_{1},h_{2})
&=
\vphi(g_{1},g_{2},h_{1}H_{1},h_{2})
\,,
\end{align}
leading to the composed action
\be
(H_{1}\otimes\Phi_{1})\,\vphi(g_{1},g_{2},h_{1},h_{2})
\equiv
\widehat{\Phi}_{1}\widehat{H_{1}}\,\vphi(g_{1},g_{2},h_{1},h_{2})
=
\Phi_{1}(h_{1}^{-1}g_{1}h_{2})\vphi(g_{1},g_{2},h_{1}H_{1},h_{2})
\,,
\ee
and thus yielding the following  multiplication law:
\begin{align}
(H_{1}\otimes\Phi_{1})\,(H'_{1}\otimes\Phi'_{1})\,
\,\vphi(g_{1},g_{2},h_{1},h_{2})
&=
\Phi_{1}(h_{1}^{-1}g_{1}h_{2})\Phi'_{1}(H_{1}^{-1}h_{1}^{-1}g_{1}h_{2})\vphi(g_{1},g_{2},h_{1}H_{1}H'_{1},h_{2})
\nn\\
&=
\big{(}H_{1}H'_{1}\otimes\Phi_{1}\,\textrm{L}_{H_{1}}[\Phi'_{1}]\big{)}\,
\,\vphi(g_{1},g_{2},h_{1},h_{2})\,,
\end{align}
with $\textrm{L}_{H}[\Phi](g)=\Phi(Hg)$.
The  difference between this multiplication law and the one of the algebra $\cA_{1}$ is the left action of group elements instead of the adjoint action. This leads to the Heisenberg double $\cE_{1}= \cH\SU(2)$ instead of the Drinfeld double $\cA_{1}=\cD\SU(2)$. Although this difference might seem innocuous, the consequences are deep, on both mathematical and physical fronts.
Indeed, unlike the Drinfeld double, the Heisenberg double is  not a Hopf algebra since it does not have a co-product.
In fact, one cannot extent the co-product defined on $\cU(\su(2))$ (resp. $\mathbb{C}[\SU(2)]$) consistently to the full algebra without changing the commutators.
On the one hand, the Drinfeld double is not a symplectic space but a Poisson-Lie group, understood as the quantum deformation of a classical symmetry group. It indeed plays the role of the symmetry group of the 0-modes of the discrete disk's boundary. It is endowed with a triangular structure defined by a $R$-matrix satisfying the Yang-Baxter equation.
On the other hand, the Heisenberg double is a symplectic space, understood as a deformation of canonical phase spaces to a momentum space endowed with a non-abelian group structure. It does not admit a $R$-matrix but is instead  provided with a $S$-matrix satisfying the pentagonal equation (which can nevertheless be understood as resulting from the $R$-matrix on the double of the Heisenberg double).
The present work does not make use of the $R$ and $S$ matrices, the interested reader can nevertheless refer to \cite{kashaev1995heisenberg,aghaei2019heisenberg,MaitlandAnson2014}.

%

Now putting the two Heisenberg double sub-algebra together gives us the the disk algebra for $N=2$ flux insertions:
\begin{equation}
\cA_2 = \cH \SU(2) \bowtie \cH \SU(2) \; ,
\label{eq:H2_in_def}
\end{equation}
where both the action of each Heisenberg double on the other is given by right group action of  $\C[\SU(2)]$ on $\cC(\SU(2))$.
Let us check this by using the explicit representation of the second Heisenberg double as operators  acting on the gauge invariant wave-functions:
\begin{align}
\widehat{\Phi}_{2}\,\vphi(g_{1},g_{2},h_{1},h_{2})
&=
\Phi_{2}(h_{2}^{-1}g_{2}h_{1})\vphi(g_{1},g_{2},h_{1},h_{2})
\,,\\
\widehat{\delta^{H_{2}}}\,\vphi(g_{1},g_{2},h_{1},h_{2})
&=
\vphi(g_{1},g_{2},h_{1},h_{2}H_{2})
\,.
\end{align}
We can compare the composed action of the first double then the second and vice-versa:
\begin{align}
(H_{2}\otimes\Phi_{2})\,(H_{1}\otimes\Phi_{1})\,
\,\vphi(g_{1},g_{2},h_{1},h_{2})
&=
(\widehat{\Phi}_{2}\widehat{H_{2}})(\widehat{\Phi}_{1}\widehat{H_{1}})\,\vphi(g_{1},g_{2},h_{1},h_{2})
\nn\\
&=
\Phi_{2}(h_{2}^{-1}g_{2}h_{1})
\Phi_{1}(h_{1}^{-1}g_{1}h_{2}H_{2})
\vphi(g_{1},g_{2},h_{1}H_{1},h_{2}H_{2})
\,,
\end{align}
\begin{align}
(H'_{1}\otimes\Phi'_{1})\,(H'_{2}\otimes\Phi'_{2})\,
\,\vphi(g_{1},g_{2},h_{1},h_{2})
&=
(\widehat{\Phi'}_{1}\widehat{H'_{1}})(\widehat{\Phi'}_{2}\widehat{H'_{2}})\,\vphi(g_{1},g_{2},h_{1},h_{2})
\nn\\
&=
\Phi'_{2}(h_{2}^{-1}g_{2}h_{1}H'_{1})
\Phi'_{1}(h_{1}^{-1}g_{1}h_{2})
\vphi(g_{1},g_{2},h_{1}H'_{1},h_{2}H'_{2})
\,.
\end{align}
This slight difference reflects the non-commutativity of the two Heisenberg doubles within the disk algebra $\cA_2$. This non-trivial braiding is given as anticipated by the right $\SU(2)$ action on $\cC(\SU(2))$:
\be
(H'_{1}\otimes\Phi'_{1})\,(H'_{2}\otimes\Phi'_{2})=(H_{2}\otimes\Phi_{2})\,(H_{1}\otimes\Phi_{1})
\qquad\textrm{with}\quad
\left|
\begin{array}{rcl}
H'_{1}&=&H_{1}\,,\\
H'_{2}&=&H_{2}\,,\\
\Phi'_{1}&=&\textrm{R}_{H_{2}}[\Phi_{1}]\,,\\
\textrm{R}_{H_{1}}[\Phi_{2}']&=&\Phi_{2}\,,
\end{array}
\right.
\ee
where the right group action on functions reads simply $\textrm{R}_{H}[\Phi](g)=\Phi(gH)$.

\medskip

To go one step further, we introduce a graphical representation of the commutators based on the natural involution $\sigma: \cH \SU(2)\to \cH \SU(2)$ given by
\begin{equation}
\sigma(\hy,\hh) = (- \hh \act \hy,\hh^{-1}) \;.
\end{equation}
The action $\act$ of $\hh$ on $\hy$ is, in coordinate
\be
(\hh \act \hy)^{a} :=  2i \, \tr(\tau^a \hh \tau_b \hh^{-1}) \hy^b.
\ee
Since we are dealing with operators, we need to be careful with the ordering. In this case, it happens not to matter since $2i \, \tr(\tau^a \hh \tau^\alpha \hh^{-1})$ and $\hy^\beta$ commute when $\alpha = \beta$. This involution is easily proven by first computing $[\hy,\hh^{-1}]$ via
\begin{equation}
[\hy^a,\hh \hh^{-1}] = 0 \implies [\hy^a,\hh^{-1}] = -i \tau^a \hh^{-1} \;.
\end{equation}
Then, the commutator $[- \hh \act \hy,\hh^{-1}]$ is straightforward, and we find the expected result of the right action of $\cU(\su(2)) $ on $\C[\SU(2)]$
\begin{equation}
[(- \hh \act \hy)^a,\hh^{-1}]= i \hh^{-1} \tau^a
\end{equation}
Finally, one can check that $- \hh \act \hy$ still generates a $\su(2)$ Lie algebra and that $\sigma$ is trivially an involution. Interestingly, one can show that $- \hh \act \hy$ commutes with $\hy$.

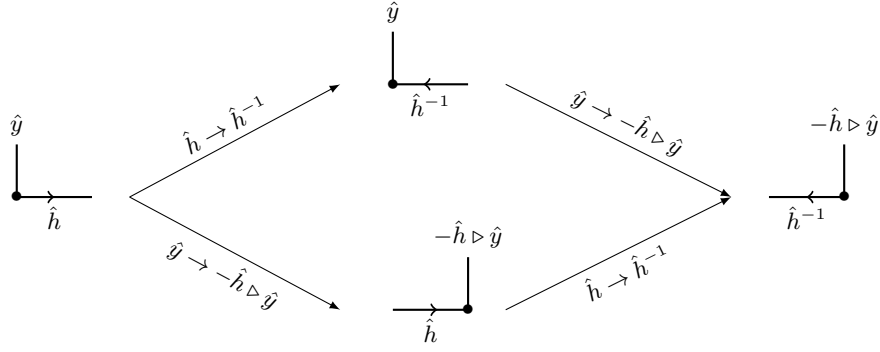
\begin{figure}[!htb]
	\centering
	\begin{tikzpicture}[scale=1]
	\coordinate (O) at (0,0); \coordinate (A) at (1.5,0);
	\coordinate (O11) at (5,1.5); \coordinate (A1) at (4.3,1.5); \coordinate (B1) at (6.5,1.5);
	\coordinate (O12) at (6,-1.5);\coordinate (A2) at (4.3,-1.5); \coordinate (B2) at (6.5,-1.5);
	\coordinate (O2) at (11,0); \coordinate (B) at (9.5,0);

	\draw[->-=0.5,thick] (O) --node[below]{$\hh$} +(1,0);
	\draw[thick] (O) --node[above,pos=1]{$\hy$} +(0,0.7);
	\node at (O){$\bullet$};

	\draw[-<-=0.5,thick] (O11) --node[below]{$\hh^{-1}$} +(1,0);
	\draw[thick] (O11) --node[above,pos=1]{$\hy$} +(0,0.7);
	\node at (O11){$\bullet$};

	\draw[-<-=0.5,thick] (O12) --node[below]{$\hh$} +(-1,0);
	\draw[thick] (O12) --node[above,pos=1]{$-\hh \act \hy$} +(0,0.7);
	\node at (O12){$\bullet$};

	\draw[->-=0.5,thick] (O2) --node[below]{$\hh^{-1}$} +(-1,0);
	\draw[thick] (O2) --node[above,pos=1]{$-\hh \act \hy$} +(0,0.7);
	\node at (O2){$\bullet$};

	\draw[->,>=latex] (A)--node[sloped,above]{$\hh \rightarrow \hh^{-1}$} (A1); \draw[->,>=latex] (A)--node[sloped,below]{$\hy \rightarrow -\hh \act \hy$}(A2);
	\draw[->,>=latex] (B1)--node[sloped,above]{$\hy \rightarrow -\hh \act \hy$}(B); \draw[->,>=latex] (B2)--node[sloped,below]{$\hh \rightarrow \hh^{-1}$}(B);
	\end{tikzpicture}
	\caption{Representation of the step-by-step action of $\sigma$ on the graphical representation of $\cH\SU(2)$. In the above path, we first act on $\hh$ then on $\hy$ and we do the reverse for the below path.}
	\label{fig:HSU_graph_rep}
\end{figure}

The algebra $\cH\SU(2)$ can then simply be seen as a two-valent node, where one edge is associated to $\hy$ and the other to $\hh$ and is outgoing to the node, see figure \ref{fig:HSU_graph_rep}. This orientation is important: it tells us that the commutator between $\hy$ and $\hh$ is the right action of $\cU(\su(2))$ on $\C[\SU(2)]$. We call this particular double a right-hook Heisenberg double. Then, the action of $\sigma$ on $\hh$ is to switch the orientation while its action on $\hy$ is to transport it along the the orientated edge. In both case, we have that the cross-commutator is now the left action of $\cU(\su(2))$ on $\C[\SU(2)]$
\begin{equation}
[\hy^a,\hh^{-1}] = - i \tau^a \hh^{-1} \quad \text{and} \quad [-(\hh \act \hy)^a,\hh]= - i \tau^a \hh
\end{equation}
while the other commutators are left unchanged. This cross-commutator is again in accordance with the orientation. The pairs $(\hy,\hh^{-1})$ and $(-(\hh \act \hy),\hh)$ generate what we call a left-hook Heisenberg double. Graphically, one then see left-hook and right-hook doubles as incoming or outgoing holonomy operator respectively. From this viewpoint, it is immediate to understand why $\hy$ and $-\hh \act \hy$ commute since they do not live at the same node. To understand how this involution helps us construct a graphical representation for the commutators, we first need to split the sunny graph in two parts, see figure \ref{fig:disk_alg_N_2}, where we have split the initial sunny graph for $\lA_2$ into two equivalent parts, denoted by $(1)$ and $(2)$. In this figure, each sub-graph $(1)$ and $(2)$ can be associated to such representation.

It is immediate to see from the commutators \eqref{eq:com_A2_cross} that the gluing of the handles $(1)$ and $(2)$ is then just done following the rule coming from the involution and in accordance with the orientation. As an example, one can see that changing the orientation on the whole graph will simply result in having the pairs $(\hx_1,\hg_2)$ and $(\hx_2,\hg_1)$ as  Heisenberg double instead of the pairs $(\hx_1,\hg_1)$ and $(\hx_2,\hg_2)$. In the rest of this paper, we will always consider the orientation where $\hg_i$ is outgoing with respect to $\hx_i$ such that $(\hx_i,\hg_i)$ is the Heisenberg double $\cH\SU(2)$. This is a right-hook Heisenberg double. The other choice corresponds to a left-hook double.
\bigskip

The goal of the rest of this section is now to show that $\cA_2$ can be expressed using $\cA_1$ as a base. More specifically, we will see that $\cA_2$ is just $\cA_1$ plus a global excitation on the boundary disk, i.e. we will show the following isomorphism
\begin{equation*}
	\begin{tikzpicture}[scale=1]
	\coordinate (Eq1) at (0,1.7);
	\coordinate (Eq2) at (2.2,1.7);
	\coordinate (Eq3) at (4.6,1.7);

	\node at (Eq1){$\cA_{2} \simeq \cD \SU(2) \times \cH \SU(2)$};
	\node at (Eq2){$=$};
	\node at (Eq3){$\cA_1 \times \cR_1 \; \quad \text{or graphically}$};
	\coordinate (O) at (0.,0);
	\coordinate (A1) at (0.,0.6); \coordinate (B1) at (0.,1);
	\coordinate (A2) at (0.,-0.6); \coordinate (B2) at (0,-1);
	\coordinate (C1) at (-0.6,0); \coordinate (C2) at (0.6,0);

	\draw[thick,->-=0.5,->-=1] (O) circle (0.6);
	\node at (A1){$\bullet$};
	\draw (A1)--node[above,pos=1,scale=0.8]{$\hx_1$} (B1);
	\node at (A2){$\bullet$};
	\draw (A2)--node[below,pos=1,scale=0.8]{$\hx_2$} (B2);
	\node[left,scale=0.8] at (C1) {$\hg_1$};
	\node[right,scale=0.8] at (C2) {$\hg_2$};
	\coordinate (Eq4) at (1.8,0);
	\node at (Eq4){$=$};
	\coordinate (O1) at (3.4,0);
	\coordinate (Eq5) at (4.8,0);
	\coordinate (O2) at (5.6,-0.3);
	\coordinate (C1) at (3.4,0.6);
	\coordinate (C2) at (3.4,-0.6);

	\draw[thick,->-=0.7] (O1) circle (0.6);
	\draw (C1) --node[pos=1,above,scale=0.8]{$\hX$} +(0,0.4);
	\node at (C1){$\bullet$};
	\node[below,scale=0.8] at (C2){$\hG$};
	\node at (Eq5){$\times$};
	\node at (O2){$\bullet$};
	\draw[thick,->-=0.5] (O2) --node[pos=0.5,below,scale=0.8]{$\hh$} +(1,0);
	\draw[thick] (O2) --node[pos=1,above,scale=0.8]{$\hy$} +(0,0.6);
	\end{tikzpicture}
\end{equation*}
where $(\hX,\hG)\, \in \, \cA_1 = \cD \SU(2)$ and $(\hy,\hh)\, \in \, \cR_1 = \cH \SU(2)$. Note that this isomorphism is true  not only at the level of vector spaces, but algebras.
\\

There are already two natural candidates for the subalgebra $\cH \SU(2)$ associated with a choice of edge. We can consider either  $\cE^{(1)}$ or  $\cE^{(2)}$. In order for the presentation to be clearer, we consider that $\cR_1 = \cE^{(2)}$ and focus on one choice for the isomorphism. We will see at the end of this section that this isomorphism is not unique, as one could have expected due to obvious multiplicity of choices.

The remaining part of the algebra is encoded in $\cE^{(1)}$ and the non trivial cross-relations
\begin{equation}
[ \hg_2, \hx_1^a] = i\tau^a \hg_2, \qquad [\hx_2,\hg_1]=-i\tau^a \hg_1.
\end{equation}
Let us first construct the commutant $\cD$ of $\cE^{(2)}$ in $\lA_2$. To do so, it is interesting to note the following property
\begin{equation}
[\hx_2,\hg_1 \hg_2] = [\hx_2,\hg_2 \hg_1] = 0
\label{eq:commutation_x_gg}
\end{equation}
and similarly for $\hx_1$. We define the global loop holonomy operator with root at the handle $(1)$ by
\begin{equation}
\hG^{(1)} \deq \hg_1 \hg_2 \; .
\end{equation}
Since the holonomies operators are abelian, it is clear that $\hG^{(1)}$ commutes with $\hg_2$.
Hence $\hG^{(1)}$ belongs to $\cD$. As explained previously, every grasping operators that do not live at the handle $(2)$ will commute with $\hx_2$. In particular, it is the case for both $\hx_1$ and $\hg_2 \act \hx_2$. We now define the global grasping operator with root at the handle $(1)$ by
\begin{equation}
\hX_1 \deq \hx_1 + \hg_2 \act \hx_2 \;.
\end{equation}
It commutes with $\hx_2$ by definition, and one can show that it also commutes with $\hg_2$ using what we learn from the involution $\sigma$ for the commutators
\begin{equation}
\left[\left(X^{(1)}\right)^a,\hg_{2}\right] = [\hx_1^{a}, \hg_2] + [(\hg_2 \act \hx_2)^a, \hg_2] = -i  \tau^a\hg_2 +i  \tau^a\hg_2 = 0 \; .
\end{equation}
Hence $\hX^{(1)}$ also belongs to $\cD$. Since $\cD$ can at most be of dimension $2$, we already have a basis in $(\hX^{(1)},\hG^{(1)})$. Finally, one can easily show that $\cD$ form a $\cD\SU(2)$ algebra
\begin{equation}
[\hG^{(1)}, \hG^{(1)}]=0,\qquad
\left[\left(X^{(1)}\right)^a,\hG^{(1)} \right]=i[\hG^{(1)},\tau^a],\qquad
\left[\left(X^{(1)}\right)^a,\left(X^{(1)}\right)^b\right]=i\epsilon^{ab}{}_{c} \left(\hX^{(1)}\right)^c \; ,
\end{equation}
proving the announced isomorphism.

Physically, these two algebras play a very different roles. The $\SU(2)$ Drinfeld double is easily recognize as the global symmetry group, i.e. its zero mode. The Heisenberg double corresponds to an excitation around the disk. Therefore, it is interesting to understand this decomposition as putting the centre of mass on one side and all the relative movement on the disk on the other side.

In practice, this excitation around the disk should be global and thus the decomposition should not prioritize any choice of root as we have done in the previous demonstration. This is indeed the case. The isomorphism is not unique and depend on how the Heisenberg double is embedded into $\cA_{2}$. Everything starts with a choice of root on the disk on which we attach the global symmetry $\cD\SU(2)$. Previously, we took the handle $(1)$ as root, but we could have taken the handle $(2)$. Once the root is chosen, we have two possible definition for the global loop holonomy and grasping operators, ending up with four possible choices
\begin{align}
(\hX^{(1)}_L, \hG^{(1)}_L)&= (\hx_1 + (\hg_2 \act \hx_2), \hg_1\hg_2)\; ,\cr
(\hX^{(1)}_R, \hG^{(1)}_R)&= (\hx_1 + (\hg_1^{-1} \act \hx_2), \hg_1\hg_2)\; ,\cr
(\hX^{(2)}_L, \hG^{(2)}_L)&= (\hx_2 + (\hg_1 \act \hx_1), \hg_2\hg_1)\; ,\cr
(\hX^{(2)}_R, \hG^{(2)}_R)&= (\hx_2 + (\hg_2^{-1} \act \hx_1 ), \hg_2\hg_1)\; . \nonumber
\end{align}
It is immediate to understand all these choices graphically. Once the root is chosen, we have two possibilities to translate the second grasping operator to the root, either going counter-clockwise or clockwise, corresponding to taking the holonomy or its inverse respectively to transport the remaining grasping operator. Finally, $\cA_2$ is completed by adding the Heisenberg double attached to the remaining node.

In the following, we will see that the decomposition done here is not specific to the case $N=2$ and actually generalizes to any number of flux insertions on the boundary $N\ge 2$.


\subsection{Factorizing $\cA_N$: symmetry and vibration modes of discrete currents}

It is now time to look at the general case with the disk algebra $\cA_N$ for an arbitrary number of insertions $N\ge 2$
The  generating Lie algebra $\lA_N$ is built with $N$ pairs of operators $(\hx_i,\hg_i)$  satisfying the following commutators:
\begin{equation}
[\hx_i^a,\hx_i^b] = i \eps^{ab}{}_c \, \hx_i^c \; , \quad [\hg_i,\hg_i ] = 0 \; , \quad [\hx_i^a,\hg_i] = i \hg_i \tau^a
\label{eq:com_AN_H}
\end{equation}
\begin{equation}
[\hx_i^a,\hx_j^b] \underset{i\neq j}{=} 0 \; , \quad [\hg_i,\hg_j ] = 0  \; , \quad [\hx_i^a,\hg_{i-1}] = -i \tau^a \hg_{i-1} \;.
\label{eq:com_AN}
\end{equation}
with $i \in [1,N]$ abd the implicit periodic conditions $i=k \sim i=N+k$. We immediately  recognize that each pair of operators $(\hx_i,\hg_i)$ generates $\cU(\cH\su(2))$ as presented in the previous section. Hence, $\cA_N$ can been seen as a tower of $N$ Heisenberg doubles,
\begin{equation}\label{AN}
\cA_N = \cH\SU(2) \ltimes \cH\SU(2) \ltimes \; ... \; \ltimes \cH\SU(2)  \ltimes
\end{equation}
where the $i$th double acts on the $(i-1)$th via the right action of $\cU(\su(2))$ on $\C[\SU(2)]$. Note that the notation (with the open $\ltimes$ at the end of the equation) emphasizes the implicit periodic condition on the disk in that the last Heisenberg double in the above notation acts on the first one. When $N=2$, this reduces to equation \eqref{eq:H2_in_def} where the two Heisenberg doubles are in a double cross-product. The goal of this section is to generalize the construction done in $\cA_2$ and show the general algebra isomorphism:
\begin{equation}
\cA_{N} \simeq \cD \SU(2) \times \cH \SU(2)^{\times (N-1)}
=
\cA_1 \times \cR_{N-1} \,,
\qquad
\textrm{with}\quad
\cR_{N-1} = \cH \SU(2)^{\times (N-1)}\,.
\end{equation}
At the level of the Lie algebra, this isomorphism is graphically
\begin{equation*}
\begin{tikzpicture}[scale=1]
\coordinate (Eq0) at (0.4,0);
\node at (Eq0) {$\lA_N$};
\coordinate (Eq4) at (1.8,0);
\node at (Eq4){$=$};
\coordinate (O1) at (3.4,0);
\coordinate (C1) at (3.4,0.6);
\coordinate (C2) at (3.4,-0.6);
\coordinate (Eq5) at (4.8,0);
\coordinate (O2) at (5.3,-0.3);

\draw[thick,->-=0.7] (O1) circle (0.6);
\draw (C1) --node[pos=1,above,scale=0.8]{$\hX$} +(0,0.4);
\node at (C1){$\bullet$};
\node[below,scale=0.8] at (C2){$\hG$};
\node at (Eq5){$\times$};

\node at (O2){$\bullet$};
\draw[thick,->-=0.5] (O2) --node[pos=0.5,below,scale=0.8]{$\hh_{1}$} +(1,0);
\draw[thick] (O2) --node[pos=1,above,scale=0.8]{$\hy_{1}$} +(0,0.6);

\coordinate (Eq6) at (6.8,0);
\node at (Eq6) {$\times$};

\coordinate (O3) at (7.5,0);
\node at (O3) {$\dots \dots$};

\coordinate (Eq7) at (8.2,0);
\node at (Eq7) {$\times$};

\coordinate (O4) at (8.8,-0.3);
\node at (O4){$\bullet$};
\draw[thick,->-=0.5] (O4) --node[pos=0.5,below,scale=0.8]{$\hh_{N-1}$} +(1,0);
\draw[thick] (O4) --node[pos=1,above,scale=0.8]{$\hy_{N-1}$} +(0,0.6);
\end{tikzpicture}
\end{equation*}
and the explicit definition of the isomorphism can be found below in equations \eqref{eq:iso_lN_2} to \eqref{eq:iso_lN_5}. We will denote by $\lR_N$ the infinitesimal version of $\cR_N$.
Note that we will make use of a mathematical shortcut. Indeed we will show the isomorphism at the level of the Lie algebra $\lA_N$ and not directly at the level of the algebra $\cA_N$. The extension of the result at the level of the algebra can be done via the relation between $\lA_N$ and $\cA_N$ exhibited earlier.

\medskip


Let us start with the identification of the global symmetry $\cD\SU(2)$, which amounts to constructing an embedding  of $\cA_{1}$ in $\cA_{N}$. Such an embedding is actually not unique. It depends on the choice of a root handle, or root flux insertion. Once, this root is chosen, we transport all the operators at this handle and define local translational operators and translated grasping operators.

More precisely, given two nodes on the boundary disk denoted by $(i,j)\in [1,N]\times[1,N]$ with $i \leq j$ we define the local translational holonomies $\hG_{ij}$ by the oriented product
\begin{equation}
\hG_{ij} = \overset{\longleftarrow}{\prod_{k=i}^{j-1}}\hg_k \;.
\end{equation}
By definition, this operator transport objects living at the handle $(i)$ to the handle $(j)$, and does not act on objects living at another handle than $(i)$. From this operator, we can then define the translated grasping operator $\hX_{ij}$ from $(i)$ to $(j)$ by
\begin{equation}
\hX_{ij} = \hG_{ij} \act \hx_i \; .
\end{equation}
By definition $\hX_{ij}$ is a grasping operator living at the handle $(j)$. This is immediate to check by computing its commutator with $\hx_{k}$, showing that $\hX_{ij}$ commutes with any grasping operator with $k \neq j$. The definitions of these operators are such that
\begin{equation*}
\hG_{ii}= \hat{1} \; , \quad \hG_{i\,i+1}= \hg_i \quad \text{and} \quad \hX_{ii}= \hx_i \;.
\end{equation*}
These definitions extend to the case $j>i$ via $\hG_{ji} = \hG_{ij}^{-1}$ . Recall also that all these operators are living on the discrete disk, hence the periodic condition\footnote{The comma between the indices is introduced in order to easily distinguish the index. It will appear in the notation when needed.}
\begin{equation}
\hG_{i+N,j+N} = \hG_{ij} \qquad \text{and} \qquad \hX_{i+N,j+N} =\hX_{ij}.
\end{equation}
Using these operators, it is possible to move any grasping operators and handles to any place on the disk.

\medskip

With these new operators in our toolkit, we are ready to define global operators in a straightforward generalization of what we have done for the $N=2$ case. The global translational holonomy at the handle $(i)$ is defined by
\begin{equation}
\hG^{(i)} = \hG_{i,i+N} = \hg_{i+N-1} \dots\hg_{i+1} \hg_{i} \; ,
\end{equation}
while the global grasping operator at the handle $(i)$ is defined by
\begin{equation}
\hX^{(i)} = \sum_{k=i}^{i+N-1} \hX_{ki} = \hx_i + \sum_{k=i+1}^{i+N-1} \hG_{k i} \act \hx_k\;.
\end{equation}
Let us reflect that these observables are part of $\cU(\lA_{N})$ but not part of the Lie algebra $\lA_{N}$ since they are non-linear combinations of the Lie algebra elements.

As a safety check, we can compare these definitions with the ones from the case $N=2$. We actually recover half of the four possibilities for the definition of the global operators. This apparent mismatch is due to the fact that we  only considered, in the definition of $\hX^{(i)}$, a transport following the orientation of the disk, while we could also go in the reverse direction. We will ignore this minor detail in the following, since it does not change at all the logic of the construction of the isomorphism.
%
The relation between two pairs of global operators defined at different handles $(i,j) \in [1,N]\otimes[1,N]$ is
\footnote{
	The action of $\hG_{ij}$ on another elements $\hG_{pq}$ is the action by conjugation
	\begin{equation*}
	\hG_{ij} \act \hG_{pq} = \hG_{ij} \hG_{pq} \hG_{ij}^{-1} \; .
	\end{equation*}
}
\begin{equation}
\hG^{(j)} = \hG_{ij} \act \hG^{(j)} \qquad \text{and} \qquad \hX^{(j)} = \hG_{ij} \act \hX^{(i)} \;.
\end{equation}
This relation simply follows from the definition of the operators $\hG_{ij}$ since they are used to hop from one handle to another.

Similarly to the previous case, we  look at the algebra generated by the global operators at a specific handle $(i_0)$, $(\hX^{(i_0)},\hG^{(i_0)})$. They generate an algebra $\cD^{(i_0)}$ simply identified as the Drinfeld double of $\SU(2)$,
\begin{equation}
\left[\hG^{(i_0)},\hG^{(i_0)} \right] =0,\qquad
\left[\hX^{(i_0)},\hG^{(i_0)} \right] =i [\hG^{(i_0)},\tau^a]\qquad
\left[\left(\hX^{(i_0)}\right)^{a},\left(\hX^{(i_0)}\right)^{b} \right] =i\epsilon^{abc} \left(\hX^{(i_0)}\right)^{c}.
\end{equation}
The key point is then to use the property that $\cD^{(i_0)}$ turns out to be an ideal for $\lA_{N}$,
which follows from the commutation relations:
\begin{equation}
[\hg_j, \hG^{(i_0)}]=0,\qquad [ \hx_j^a, \hG^{(i_0)} ]= i  \delta_{ij} [ \hG^{i_0} , \tau^a ],
\qquad
\left[\hx_j^a , \left(\hX^{(i_0)}\right)^{b} \right] =  i \delta_{ij} \epsilon^{abc} \left(\hX^{(i_0)}\right)^{c}.
\end{equation}
The first commutator is straightforward since holonomy operators are abelian. The last one is also straightforward using the property that grasping operators that do not live at the same handle commutes. The only one that need a bit more work is the middle one. It comes from a generalization of the property \eqref{eq:commutation_x_gg},
\begin{equation}
[\hx_{i},\hg_{i} \hg_{i-1}] = [\hx_{i} , \hg_{i-1} \hg_i] = 0 \; ,
\end{equation}
which is immediate to prove. This property is easily understood graphically. Recall that the grasping operators are ultra-local in their actions. It means that $\hx_i$ only acts on operators living at the handle $(i)$. The operators $\hg_{i} \hg_{i-1}$ and $\hg_{i-1} \hg_i$ only act at the handles $(i+1)$ and $(i-1)$, hence they commute with $\hx_{i}$. This is an important technicality of the holonomy operators since it tells us that translated operators are not correlated to passing handles, but only to the target handle.

Using this property,  we  consider the quotient algebra $\lR_{N}^{(i_0)} = \lA_{N} / \cD^{(i_0)}$. It is generated by the $(N-1)$ pairs of operators $(\hx_j,\hg_j)_{j \neq i_0}$ and is therefore isomorphic to $(N-1)$ Heisenberg doubles at the level of the vector space. Hence, we have proven the isomorphism
\begin{equation}
\begin{tikzpicture}[scale=1]
\coordinate (Eq0) at (-1.5,0);
\node at (Eq0) {$\lA_N \simeq$};

\coordinate (O1) at (0,0);
\coordinate (C1) at (0,0.6);
\coordinate (C2) at (0,-0.6);
\coordinate (Eq1) at (1.4,0);

\draw[thick,->-=0.7] (O1) circle (0.6);
\draw (C1) --node[pos=1,above,scale=0.8]{$\hX^{(i_0)}$} +(0,0.4);
\node at (C1){$\bullet$};
\node[below,scale=0.8] at (C2){$\hG^{(i_0)}$};
\node at (Eq1){$\times$};

\coordinate (O2) at (1.9,-0.3);
\coordinate (O3) at (2.9,-0.3);
\coordinate (Eq2) at (4.4,-0.5); \coordinate (Eq3) at (4.4,0.5);
\coordinate (O4) at (5.9,-0.3);
\coordinate (O5) at (6.9,-0.3);

\node at (O2) {$\bullet$};
\draw[->-=0.5,thick] (O2)--node[pos=0.5,below,scale=0.8]{$\hg_{i_0+1}$}(O3); \draw[thick] (O2) --node[pos=1,above,scale=0.8]{$\hx_{i_0+1}$} +(0,0.6);
\node at (O3) {$\bullet$};
\draw[->-=0.15,->-=0.85,thick] (O3)--node[pos=0.15,below,scale=0.8]{$\hg_{i_0+2}$}(O4); \draw[thick] (O3) --node[pos=1,above,scale=0.8]{$\hx_{i_0+2}$} +(0,0.6);

\node[scale=0.8] at (Eq2) {$\dots$}; \node[scale=0.8] at (Eq3) {$\dots$};

\node at (O4) {$\bullet$};
\draw[->-=0.5,thick] (O4)--node[pos=0.5,below,scale=0.8]{$\hg_{i_0-2}$}(O5); \draw[thick] (O4) --node[pos=1,above,scale=0.8]{$\hx_{i_0-2}$} +(0,0.6);
\node at (O5) {$\bullet$};
\draw[->-=0.5,thick] (O5)--node[pos=0.5,below,scale=0.8]{$\hg_{i_0-1}$} +(1,0); \draw[thick] (O5) --node[pos=1,above,scale=0.8]{$\hx_{i_0-1}$} +(0,0.6);
\end{tikzpicture}
\label{eq:AN_splitting_tower_semi}
\end{equation}
where the relation between the tower of Heisenberg double is still given by the commutators \eqref{eq:com_AN}. Promoting the result to the universal enveloping algebra using what was introduced earlier, we have shown that
\begin{equation}
	\cA_N = \cD\SU(2) \times \left( \cH\SU(2) \ltimes \cH\SU(2) \ltimes \cdot\cdot\cdot \ltimes \cH\SU(2)	\right) \; .
\end{equation}
While this is not the simplest description of the algebra $\cA_N$, it provides an excellent starting point in order to discuss projection and embedding maps $\cA_N$ from $\cA_{N-1}$ or $\cA_{N+1}$ in  coarse-graining or refinement processes. This is the subject of the next section. Before moving on to that topic, let us go one step further, and show that this quotient algebra can in fact be rewritten as a direct product of decoupled Heisenberg doubles,
\begin{equation}
\cR^{(i_0)}_N \simeq \cH\SU(2)^{\times {N-1}}.
\end{equation}
This is proven by a simple reshuffling of the operators $(\hx_j,\hg_j)_{j \neq i_0}$. Indeed, recall that the grasping operators are ultra-local, while it is the abelian holonomy operators that glue the disk as a whole. Consider the pairs $(\hx_{i_0-2},\hg_{i_0-2})$ and $(\hx_{i_0-1},\hg_{i_0-1})$. What we want to do is reshuffling the operators to obtain two commuting pairs of operators that have the same amount of information as our starting two. This is easily done by considering the pairs of operators $(\hx_{i_0-2},\hg_{i_0-2} \hg_{i_0-1})$ and $(\hx_{i_0-1},\hg_{i_0-1})$.  On the one hand, the ultra-locality of the grasping operators tells us that $\hx_{i_0-2}$ commutes with $\hg_{i_0-1}$, hence $(\hx_{i_0-2},\hg_{i_0-1} \hg_{i_0-2} )$ still generates an Heisenberg double. On the other hand, the property of locality of the holonomy operators tells us that they only couple to operators living at their source or target handles, hence $\hg_{i_0-1} \hg_{i_0-2}  = \hG_{i_0-2,i_0}$ commutes with $\hx_{i_0-1}$ since its source handle is $(i_0-2)$ and its target is the fictive handle $(i_0)$.

We  apply this logic on the whole space $\cR_{N}^{(i_0)}$ and define new grasping and holonomy operators $(\hy_j,\hh_j)_{j \neq i_0}$ by
\begin{equation}
\hy_j = \hx_{i_0-j} \qquad \text{and} \qquad \hh_j = \overset{\longleftarrow}{\prod_{k=i_0-j}^{i_0-1}}\hg_k
\,.
\label{eq:iso_lN_1}
\end{equation}
This leads to a new  of the disk algebra as
\begin{equation}
\begin{tabular}{ c c c c c c c c l}
$\cA_N$ & $\simeq$ &$\cD\SU(2)$ & $\times$ & $\cH\SU(2)$  &$\times $& $\cH\SU(2)$ &  $\cdots \quad \times \quad $ &  $\cH\SU(2)$\,. \cr
& & & & & & & &\cr
$
\left|
\begin{array}{l}
\hx_1 \; \; \hx_2 \; \; ... \; \; \hx_N
\vspace{1mm}
\\
\hg_1 \; \; \hg_2 \; \; ... \; \; \hg_N
\end{array}
\right.
$
&	&
$
\left|
\begin{array}{l}
\hX^{(i_0)}
\vspace{1mm}
\\
\hG^{(i_0)} 
\end{array}
\right.
$
&	&
$
\left|
\begin{array}{l}
\hy_1=\hx_{i_0-1}
\vspace{1mm}
\\
\hh_1=\hg_{i_0-1}
\end{array}
\right.
$
& &
$
\left|
\begin{array}{l}
\hy_2=\hx_{i_0-2}
\vspace{1mm}
\\
\hh_2= \hg_{i_0-2}\hg_{i_0-1}
\end{array}
\right.
$
&$\cdots $&
$
\left|
\begin{array}{l}
\hy_{N-1} = \hx_{i_0+1}
\vspace{1mm}
\\
\hh_{N-1}=\hg_{i_0+1}\cdots \hg_{i_0-1}
\end{array}
\right.
$
\end{tabular}
\label{eq:iso_lN_2}
\end{equation}
One can work backwards and get the inverse map expressing the initial operators $(\hat{x},\hat{g})$ in terms of the composite operators $(\hat{y},\hat{h})$. The holonomy operators are obtained by the step by step operations
\begin{equation}
\hg_{i_0-k} = \hh_k \hh_{k-1}^{-1} \qquad \forall \; k \in [2,N-1] \; ,
\label{eq:iso_lN_3}
\end{equation}
while the only missing grasping operator $\hx_{i_0}$ can be re-constructed from $\hX^{(i_0)}$:
\begin{equation}
\hx_{i_0} = \hX^{(i_0)} - \sum_{k=1}^{N-1} \hh_k \act \hy_k \; .
\label{eq:iso_lN_4}
\end{equation}

\bigskip

We summarize all these facts  into one of the main statements of this paper: the existence of $2N$ isomorphisms
\begin{equation}
\cI^{(i)}_{\pm}:  \cA_N \to \cA_1 \times \cR_N^{(i)} \simeq \left( \cD\SU(2) \times \cH\SU(2)^{\times {N-1}} \simeq \cD\SU(2) \times \cH\SU(2)^{\ltimes {N-1}} \right)
\label{eq:iso_lN_5}
\end{equation}
where $\pm$ refers to taking the anti-clockwise or clockwise direction on the disk in order to construct the global grasping operators. In the following, we will keep the handle index silence unless it is explicitly needed. This means that the disk algebra $\cA_N$  decomposes into two distinct parts.

\smallskip

On the one hand, the first sub-algebra consists of the Drinfeld double $\cD\SU(2) \simeq \cA_1$, and can be understood as the global symmetry algebra of the underlying gravitational theory.
It is generated, after choosing a root $i$ along the boundary circle, by  two global observables:  the global holonomy operator $\hG^{(i)}$  around the circle starting and ending at the root and the covariant sum of fluxes $\hX^{(i)}$ transported back to the root $i$.
We think of this 0-mode sub-algebra $\cA_1$ as a one-quantum-of-geometry regime, defining the vacuum of the theory. It is the quantum symmetry of the underlying non-commutative space.
%

Furthermore, this $\cD \SU(2)$ can be seen as a coarse graining $C_i:  \cU(\lA_N) \to \cU(\lA_1)$ of the disk algebra with $N$ insertions to
a single flux insertion algebra.
It turns out that this $\cD \SU(2)$ sub-algebra fully controls the representation theory of the whole disk algebra $\cA_N$: the representation theory of $\cA_N$ is entirely determined by its Drinfeld double factor $\cD \SU(2)$, or in other words, $\cA_N$ and $\cD \SU(2)$ are Morita-equivalent algebras\footnotemark{}.
\footnotetext{
From a broader point of view, this is reminiscent of a celebrated and fundamental result of mathematical physics anticipated by Moore-Seiberg and due to Kazhdan-Lusztig \cite{Kazhdan-Lusztig}, who proved the equivalence as braided tensor category between the category of positive energy representations of the  loop algebra $L_{\ell}(\mathfrak{g})$ at level $\ell$ and the category of representations of the quantum group $U_{q}(\mathfrak{g})$ for $q = e^{i\frac{\pi }{\kappa}}$  with $\kappa=\ell+\check{h} \notin \mathbb{Q}_{\geq 0}$ and  where $\check{h}$ is the dual coxeter number.
We leave for future investigation the precise dictionary and correspondence between the discrete and continuum objects and algebras, which is clearly the missing part of the present study. Our  goal would then be to establish in a mathematically rigorous way the equivalent of the Kazhdan-Lusztig result at the discrete level.
}

\smallskip

On the other hand, the second sub-algebra is as a tower of Heisenberg doubles. It is interesting that we obtain two different kinds of factorization for the Heisenberg doubles. The first factorization couples the Heisenberg doubles, making them all part of a bigger structure where each double is in semi-direct product with the next double and so on. There is still a natural ordering between the double in this particular setup. As we will see in the next section, the main advantage of this description is that it allows for local coarse-graining and refinement of the disk algebras.

The second description of the  tower of Heisenberg doubles is in terms of a direct product, where all the Heisenberg doubles are decoupled. The crucial difference with the previous factorization is that the quantum operators now describe global excitations on the disk's boundary. Indeed, they involve transport back a chosen root handle of the boundary. These excitations comes from bulk fluctuations, coming either from within the disk or from the exterior region, inducing global excitations on the disk's boundary. Interpreting each excitation as a  particle (on the boundary), we understand $\cA_{N}$ as describing $(N-1)$ such  particles.
%

The disk algebra can therefore be understood as the sum of a zeroth mode -the global symmetry- and all possible  vibration modes of the 1d boundary on top of it. This structure is reminiscent of a Kac-Moody current algebra. One further sees that the amount of information on the boundary can easily be refined by adding an Heisenberg double to the mix, hence going from $N$ to $(N+1)$, or coarse-grained by  removing one double. This lead to refinement and coarse-graining processes, essential to quantum gravity, which we define precisely in the next section.

\section{Exploring Discrete Currents}
\label{sec:exploring}

In this section, we focus on exploring in more details discrete currents and understanding them as the analogue in the discrete of the usual Kac-Moody currents. We present the second main statement of this paper: the algebra $\cA_{N}$ provides us with an exact way to interpolate between the global symmetry algebra of the underlying non-commutative space represented by the Drinfeld double of $\SU(2)$ and the classical boundary symmetry algebra of 3D gravity, the Poincar\'e Loop algebra derived in the first section \ref{sec:3dgrav}.
Indeed, we will show that  the continuum limit $\cA_{\infty}$ of the algebra $\cA_{N}$ in the infinite refinement limit  $N\rightarrow+\infty$ leads back to the classical current algebra $L(\mathfrak{iso}(3))$. This validates the interpretation of the disk algebra as a discrete current algebra.

Here, we focus on the role of the number of handles $N$ on the algebraic structure of the disk algebra $\cA_{N}$ and investigate the dual processes of refinement and coarse-graining, respectively embedding the $\cA_{N}$ algebra within the  $\cA_{N+1}$ algebra on the one hand and projecting  the $\cA_{N}$ algebra down to the  $\cA_{N-1}$ algebra on the other. This will effectively allow to slide from the continuum limit $\cA_{\infty}\sim L(\mathfrak{iso}(3))$ at large $N$ back to the deep quantum regime at low $N$ and ultimately to a single insertion $N=1$ defining the global symmetry algebra $\cA_{1}\sim\cD\SU(2)$.

\subsection{From $\cA_1$ to $\cA_\infty$ and vice-versa: Refinement and Coarse-graining}

One of the most important building blocks for a discrete quantum theory is the ability to do refinement and coarse-graining processes. These processes tell us how to modify the underlying discrete setup, moving from the fundamental discrete structure, here a simple disk with one handle, i.e. the $N=1$ case, up to the continuum picture with $N\rightarrow \infty$. For three-dimensional gravity formulated as a solvable theory, such processes can be done exactly. We are therefore interested in doing here the following exact operations on $\cA_N$
\begin{equation*}
\cA_{1} \; \longleftarrow \; \cdots \cdots \longleftarrow \;\cA_{N-1} \; \underset{\text{coarse-grain}}{\longleftarrow} \; \cA_N \; \underset{\text{refine}}{\longrightarrow} \; \cA_{N+1} \; \longrightarrow \; \cdots \cdots \longrightarrow \; \cA_{\infty} .
\end{equation*}
Most of the technical aspects have  already been worked out in the previous section. Indeed, we showed that the fundamental variables in $\cA_N$ are the pairs of operators $(\hx_i,\hg_i)$ and that $\cA_N$ factorizes into its global symmetry, independent of $N$, and its  algebra of excitations which depends on the number $N$ of insertions. Therefore, the coarse-graining and refinement processes can be seen as directly acting on the vibrational part of $\cA_N$, hence corresponding to
\begin{equation*}
\cR_{N-1} \; \underset{\text{coarse-grain}}{\longleftarrow} \; \cR_N \; \underset{\text{refine}}{\longrightarrow} \; \cR_{N+1}
\,.
\end{equation*}

There are two ways to tackle these operations, depending on how one sees $\cR_{N}$. If we consider $\cR_N$ as a tower of independent Heisenberg double then, in a sense, there is nothing much to do. The coarse-graining process corresponds to deleting one Heisenberg double while the refinement process is just adding one. There are, however, two problems with this approach. Firstly, from the viewpoint of the initial algebra $\cA_N$ it is unsure that such an operation leaves the global symmetry algebra untouched. Secondly, by doing so, we miss the exact nature of these two processes. Indeed, just removing or adding an Heisenberg double seems that one is adding or removing information by hand. To study these operations, it is more natural to consider $\cR_{N}$ as a tower of Heisenberg doubles in semi-direct product. Recall that, graphically, the algebra then takes the form \eqref{eq:AN_splitting_tower_semi}
\begin{equation*}
\begin{tikzpicture}[scale=1]
\coordinate (Eq0) at (-1.5,0);
\node at (Eq0) {$\cA_N \simeq$};

\coordinate (O1) at (0,0);
\coordinate (C1) at (0,0.6);
\coordinate (C2) at (0,-0.6);
\coordinate (Eq1) at (1.4,0);

\draw[thick,->-=0.7] (O1) circle (0.6);
\draw (C1) --node[pos=1,above,scale=0.8]{$\hX^{(i_0)}$} +(0,0.4);
\node at (C1){$\bullet$};
\node[below,scale=0.8] at (C2){$\hG^{(i_0)}$};
\node at (Eq1){$\times$};

\coordinate (O2) at (1.9,-0.3);
\coordinate (O3) at (2.9,-0.3);
\coordinate (Eq2) at (4.1,-0.5); \coordinate (Eq3) at (4.1,0.5);
\coordinate (O4) at (8.9,-0.3);
\coordinate (O5) at (9.9,-0.3);

\node at (O2) {$\bullet$};
\draw[->-=0.5,thick] (O2)--node[pos=0.5,below,scale=0.8]{$\hg_{i_0+1}$}(O3); \draw[thick] (O2) --node[pos=1,above,scale=0.8]{$\hx_{i_0+1}$} +(0,0.6);
\node at (O3) {$\bullet$};
\draw[->-=0.10,->-=0.90,thick] (O3)--node[pos=0.09,below,scale=0.8]{$\hg_{i_0+2}$}(O4); \draw[thick] (O3) --node[pos=1,above,scale=0.8]{$\hx_{i_0+2}$} +(0,0.6);

\node[scale=0.8] at (Eq2) {$\dots$};
\node[scale=0.8] at (Eq3) {$\dots$};

\node at (O4) {$\bullet$};
\draw[->-=0.5,thick] (O4)--node[pos=0.5,below,scale=0.8]{$\hg_{i_0-2}$}(O5); \draw[thick] (O4) --node[pos=1,above,scale=0.8]{$\hx_{i_0-2}$} +(0,0.6);
\node at (O5) {$\bullet$};
\draw[->-=0.5,thick] (O5)--node[pos=0.5,below,scale=0.8]{$\hg_{i_0-1}$} +(1,0); \draw[thick] (O5) --node[pos=1,above,scale=0.8]{$\hx_{i_0-1}$} +(0,0.6);

\coordinate (OM1) at (4.8,-0.3);
\coordinate (OM2) at (5.8,-0.3);
\coordinate (OM3) at (6.8,-0.3);
\coordinate (EqM1) at (7.7,-0.5); \coordinate (EqM2) at (7.7,0.5);

\node at (OM1) {$\bullet$}; \node at (OM2) {$\bullet$}; \node at (OM3) {$\bullet$};
\draw[thick,->-=0.5] (OM1) --node[pos=0.5,below,scale=0.8]{$\hg_{j-1}$} (OM2);
\draw[thick,->-=0.5] (OM2) --node[pos=0.5,below,scale=0.8]{$\hg_j$} (OM3);
\draw[thick] (OM1) --node[pos=1,above,scale=0.8]{$\hx_{j-1}$} +(0,0.6);
\draw[thick] (OM2) --node[pos=1,above,scale=0.8]{$\hx_{j}$} +(0,0.6);
\draw[thick] (OM3) --node[pos=1,above,scale=0.8]{$\hx_{j+1}$} +(0,0.6);

\node[scale=0.8] at (EqM1) {$\dots$};
\node[scale=0.8] at (EqM2) {$\dots$};
\end{tikzpicture}
\end{equation*}

Then what is a basic coarse-graining or refining step? It simply corresponds to the merging of two handles or the splitting of one handle into two.
\bigskip

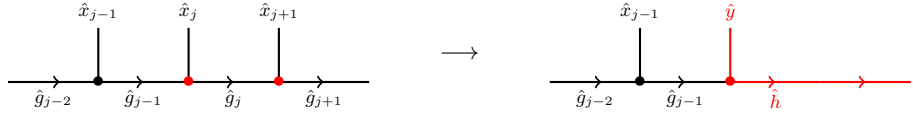
\begin{figure}[!htb]
	\begin{tikzpicture}[scale=1.2]
	\coordinate (OM1) at (0,0);
	\coordinate (OM2) at (1,0);
	\coordinate (OM3) at (2,0);

	\draw[thick,->-=0.5] (OM1) --node[pos=0.5,below,scale=0.8]{$\hg_{j-1}$} (OM2);
	\draw[thick,->-=0.5] (OM2) --node[pos=0.5,below,scale=0.8]{$\hg_j$} (OM3);
	\draw[thick] (OM1) --node[pos=1,above,scale=0.8]{$\hx_{j-1}$} +(0,0.6);
	\draw[thick] (OM2) --node[pos=1,above,scale=0.8]{$\hx_{j}$} +(0,0.6);
	\draw[thick] (OM3) --node[pos=1,above,scale=0.8]{$\hx_{j+1}$} +(0,0.6);
	\draw[thick,-<-=0.5] (OM1) --node[pos=0.5,below,scale=0.8]{$\hg_{j-2}$} +(-1,0);
	\draw[thick,->-=0.5] (OM3) --node[pos=0.5,below,scale=0.8]{$\hg_{j+1}$} +(1,0);
	\node at (OM1) {$\bullet$}; \node[red] at (OM2) {$\bullet$}; \node[red] at (OM3) {$\bullet$};
	\coordinate (Eq0) at (4,0.3);
	\node at (Eq0) {$\longrightarrow$};
	\coordinate (OM1) at (6,0);
	\coordinate (OM2) at (7,0);
	\coordinate (OM3) at (8,0);

	\draw[thick,->-=0.5] (OM1) --node[pos=0.5,below,scale=0.8]{$\hg_{j-1}$} (OM2);
	\draw[thick,->-=0.5,red] (OM2) --node[pos=0.5,below,scale=0.8,red]{$\hh$} (OM3);
	\draw[thick] (OM1) --node[pos=1,above,scale=0.8]{$\hx_{j-1}$} +(0,0.6);
	\draw[thick,red] (OM2) --node[pos=1,above,scale=0.8,red]{$\hy$} +(0,0.6);
	\draw[thick,-<-=0.5] (OM1) --node[pos=0.5,below,scale=0.8]{$\hg_{j-2}$} +(-1,0);
	\draw[thick,->-=0.5,red] (OM3) -- +(1,0);
	\node at (OM1) {$\bullet$}; \node[red] at (OM2) {$\bullet$}; 
	\end{tikzpicture}
	\caption{Coarse-graining step between the handle $(j)$ and $(j+1)$, in bullet red on the left. The process gives rise to two new operators, $\hy$ and $\hh$.}
	\label{fig:coarse_graining_RN_algebra}
\end{figure}

Let us start with the coarse-graining. The graphical representation of the coarse-graining step of the handle $(j)$ with the handle $(j+1)$ is given in figure \ref{fig:coarse_graining_RN_algebra}.
The idea behind the coarse-graining is just to merge the handle $(j+1)$ with the handle $(j)$. To do so, the first step is to remove all the operators living at the handle $(j+1)$. This is realized by transporting the to-be-coarse-grained handle $(j+1)$ toward $(j)$ using the operator $\hG_{j+1,j}$. Hence, we define a new operator $\hy$ by
\begin{equation}
\hy \deq \hx_{j} + \hG_{j+1,j} \act \hx_{j+1} = \hx_{j} + \hg_{j}^{-1} \act \hx_{j+1}
\end{equation}
which consists of the grasping operator already living at the handle $(j)$ plus the one coming from the handle $(j+1)$. For its associated holonomy operator $\hh$, a natural candidate is
\begin{equation}
\hh \deq \hg_{j} \hg_{j+1} \; .
\end{equation}
Indeed, as illustrated in figure \ref{fig:coarse_graining_RN_algebra}, once one has move the handle $(j+1)$ to the handle $(j)$, then there is nothing left between $\hg_{j}$ and $\hg_{j+1}$ and so it is natural to take them together as definition for $\hh$. The pair of operators $(\hy,\hh)$ is naturally a Heisenberg double. Indeed, from the viewpoint of the disk, the grasping operator $\hy$ is living at the handle $(j)$ and hence commutes with $\hg_{j+1}$. The only step that required a bit of computation is to verified that its commutation with $\hg_j$ or with itself returns the right action of $\cU(\su(2))$ on $\C[\SU(2)]$ and the $\su(2)$ algebras commutators respectively, but it can also be obtained graphically.

From the two above relations between the pair of operators $(\hy,\hh)$ and the pairs $(\hx_j,\hg_j),(\hx_{j+1},\hg_j)$, we introduce the coarse-graining map $C^{(N)}_{j,j+1}$ of the handle $(j)$ and $(j+1)$ by
\begin{equation}
	\begin{tabular}{c c c c}
		$C^{(N)}_{j,j+1}$:	&	$\lA_{N-1} \simeq \lA_{1} \times \cR_{N-1}$ 	&	$\quad \longrightarrow$ &  $\quad \lA_{N} \simeq \lA_{1} \times \cR_{N}$ \\
		~		&	$\quad \quad \{(\hy,\hh),(\hx_k,\hg_k)_{i\in[1,N] \neq (j,j+1)}\}$			&	$\quad \longmapsto$			&	 $(\hx_k,\hg_k)_{k\in[1,N]}$
	\end{tabular}
\end{equation}
Note that we did not specified the position of the pair $(\hy,\hh)$ since it can be anywhere on the disk.

In more details, we have introduced two new operators, $\hy$ and $\hh$ such that they form an Heisenberg double. The relation between this Heisenberg double and the rest of the algebra is that of the right action $\cU(\su(2))$ on $\C[\SU(2)]$
\begin{equation}
[\hy^a,\hg_{j-1}] = -i \hg_{j-1} \tau^a \quad \text{and} \quad [\hx_{j+2},\hh] = -i \hh \tau^a
\end{equation}
following the graphical rules introduced earlier such that $C_{ij}$ is indeed a Lie algebra morphism. Hence, the set of operators $\{(\hy,\hh),(\hx_i,\hg_i)_{i \neq (i_0,j,j+1)} \}$ is isomorphic to the algebra $\cR_{N-1}$. In practice, we have merged the two pairs of operators $(\hx_{j},\hg_{j})$ and $(\hx_{j+1},\hg_{j+1})$ to obtain the simple pair $(\hy,\hh)$. The last step in order to finish this coarse-graining process is to check that the global grasping and holonomy operators are effectively left untouched by the operation, i.e. that these operators can be expressed using $(\hy,\hh)$ instead of $(\hx_{j},\hg_{j})$ and $(\hx_{j+1},\hg_{j+1})$. It is obviously the case for $\hG^{(i_0)}$ by its definition since we have
\begin{equation}
\hG^{(i_0)} = \hg_{i_0-1} \cdots \hg_{j} \hg_{j+1} \cdots \hg_{i_0} = \hg_{i_0-1} \cdots \hy \cdots \hg_{i_0} \; .
\end{equation}
For the global grasping operator, one need to use the associativity property
$
\hg_i \act \hg_j \act \hx_k = \hg_i \hg_j \act \hx_k \;
$
to show that
\begin{equation}
\hG_{ji_0} \act \hx_j \;+\; \hG_{j+1,i_0} \act \hx_{j+1} = \hG_{j,i_0} \act \left( \hx_{j} + \hg_{j}^{-1} \act \hx_{j+1} \right) = \hG_{j,i_0} \act \hy
\end{equation}
hence showing that $\hX^{(i_0)}$ is also left untouched.

\medskip

There are a few remarks that need to be done. First, we made the choice to move the handle $(j+1)$ to the handle $(j)$ during the process. We could have also done the reverse, i.e. moving the handle $(j)$ to the handle $(j+1)$. The only change that this possibility brought are technical: instead of having $(\hy,\hh)$ being an Heisenberg double, it would have been $(\hy,\hg_{j+1})$.

Second, with our definition of the operator $C_{j,j+1}$, it seems that  coarse-graining is only possible as an ultra-local operation between two neighbouring handles. This would be a misconception. Recall that we are free to move the handles around the disk using the operators $\hG_{ij}$. Therefore, one can consider any two handles $(j_0)$ and $(j_1)$ to coarse-grain. In order to apply the procedure presented, it is only needed to translate them to the same handle. Note that it is not necessary for this particular handle to be either $(j_0)$ or $(j_1)$. This allows for a straightforward generalization of the coarse-graining operator to an operator $C_{ij}$ by
\begin{equation}
	\begin{tabular}{c c c c}
		$C^{(N)}_{i,j}$:	&	$\lA_{N-1} \simeq \lA_{1} \times \cR_{N-1}$ 	&	$\quad \longrightarrow$ &  $\quad \lA_{N} \simeq \lA_{1} \times \cR_{N}$ \\
		~		&	$\quad \quad \{(\hy,\hh),(\hx_k,\hg_k)_{i\in[1,N] \neq (i,j)}\}$			&	$\quad \longmapsto$			&	 $(\hx_k,\hg_k)_{k\in[1,N]}$
	\end{tabular}
\end{equation}
Finally, one can generalized these coarse-graining map to act at the level of the algebra $\cA_N$ via the relation between $\lA_N$ and $\cA_N$ introduced earlier.
\bigskip

\begin{figure}[!htb]
	\begin{tikzpicture}[scale=1.2]
	\coordinate (OM1) at (0,0);
	\coordinate (OM2) at (1,0);
	\coordinate (OM3) at (2,0);

	\draw[thick,->-=0.5] (OM1) --node[pos=0.5,below,scale=0.8]{$\hg_{j-1}$} (OM2);
	\draw[thick,->-=0.5] (OM2) --node[pos=0.5,below,scale=0.8]{$\hg_j$} (OM3);
	\draw[thick] (OM1) --node[pos=1,above,scale=0.8]{$\hx_{j-1}$} +(0,0.6);
	\draw[thick] (OM2) --node[pos=1,above,scale=0.8]{$\hx_{j}$} +(0,0.6);
	\draw[thick] (OM3) --node[pos=1,above,scale=0.8]{$\hx_{j+1}$} +(0,0.6);
	\draw[thick,-<-=0.5] (OM1) --node[pos=0.5,below,scale=0.8]{$\hg_{j-2}$} +(-1,0);
	\draw[thick,->-=0.5] (OM3) --node[pos=0.5,below,scale=0.8]{$\hg_{j+1}$} +(1,0);
	\node at (OM1) {$\bullet$}; \node[red] at (OM2) {$\bullet$}; \node at (OM3) {$\bullet$};
	\coordinate (Eq0) at (4,0.3);
	\node at (Eq0) {$\longrightarrow$};
	\coordinate (OM1) at (6,0);
	\coordinate (OM2) at (7,0);
	\coordinate (OM3) at (8,0);
	\coordinate (OM4) at (9,0);

	\draw[thick,->-=0.5,red] (OM1) --node[pos=0.5,below,scale=0.8]{$\hh_2$} (OM2);
	\draw[thick,->-=0.5,red] (OM2) --node[pos=0.5,below,scale=0.8]{$\hh_1$} (OM3);
	\draw[thick,->-=0.5] (OM3) --node[pos=0.5,below,scale=0.8]{$\hg_j$} (OM4);

	\draw[thick] (OM1) --node[pos=1,above,scale=0.8]{$\hx_{j-1}$} +(0,0.6);
	\draw[thick,red] (OM2) --node[pos=1,above,scale=0.8]{$\hy_1$} +(0,0.6);
	\draw[thick,red] (OM3) --node[pos=1,above,scale=0.8]{$\hy_2$} +(0,0.6);
	\draw[thick] (OM4) --node[pos=1,above,scale=0.8]{$\hx_{j+1}$} +(0,0.6);

	\draw[thick,-<-=0.5] (OM1) --node[pos=0.5,below,scale=0.8]{$\hg_{j-2}$} +(-1,0);
	\draw[thick,->-=0.5] (OM4) --node[pos=0.5,below,scale=0.8]{$\hg_{j+1}$} +(1,0);

	\node at (OM1) {$\bullet$}; \node[red] at (OM2) {$\bullet$}; \node[red] at (OM3) {$\bullet$}; \node at (OM4) {$\bullet$};

	\end{tikzpicture}
	\caption{Refinement of the handle $(j)$, corresponding to it splitting into two: we go from the pair $(\hx_j,\hg_j)$ to a $(\hy_1,\hh_1)$ two auxiliary operators $\hy_2,\hh_2$. The auxiliary operators are corrected version $\hx_{j}$ and $\hg_{j-1}$ in order to take into account the gluing of $(\hy_1,\hh_1)$.}
	\label{fig:refinement_RN_algebra}
\end{figure}
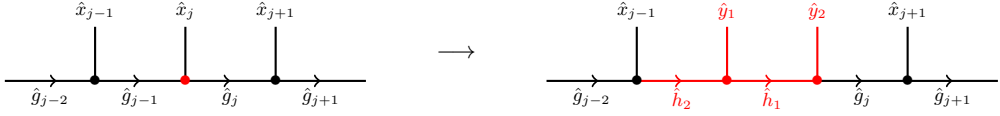

The second operation we will discuss is the refinement. This operation is technically the inverse of the coarse-graining in the sense that we want to go from $\cA_{N}$ to $\cA_{N+1}$. This corresponds to the splitting of a pair of operator in $\cA_N$ to return two pairs of operator. At the end of the day, it amounts to the addition of a pair of operator to the system. Its graphical representation is given figure \ref{fig:refinement_RN_algebra}.

As we said, the idea of the refinement is to add one particles to the system by splitting an handle into two. We consider two operators $(\hy_1,\hh_1)$ coming with an Heisenberg double structure
\begin{equation}
[\hy_1^a,\hy_1^b] = i {\eps^{ab}}_c \hy_1^c \; , \quad [\hh_1,\hh_1 ] = 0 \; , \quad [\hy_1^a,\hh_1] = i \hh_1 \tau^a \;
\end{equation}
and we want to glue it to $\cR_{N}$ in a consistent way, that is, with the right action of $\cU(\su(2))$ on $\C[\SU(2)]$. To do so, let us consider the change of variable
\begin{equation}
\hx_{j} \rightarrow \hy_2 \deq \hx_j - \hh_1 \act \hy_1 \quad \text{and} \quad \hg_{j-1} \rightarrow \hh_2 \deq \hg_{j-1} \hh_{1}^{-1} \; .
\end{equation}
Again, the intuition behind these definitions is to be found in the graphical representation given figure \ref{fig:refinement_RN_algebra}. Since we are splitting the handle $(j)$ into two, there are three possible operators than can be modified to take into account the new particle, $\hg_{j-1}$, $\hx_j$ and $\hg_j$. Similarly to the coarse-graining, there are two possibilities, depending on the relative position between the handle $(j)$ and the new handle. For the sake of the presentation, we will consider the choice depicted in figure \ref{fig:refinement_RN_algebra}.
This suggest the definition of the refining map $R^{(N)}_{j}$ splitting the handle $(j)$ into two handles by
\begin{equation}
	\begin{tabular}{c c c c}
		$R^{(N)}_{j}$:	&	 $\lA_{N+1} \simeq \lA_{1} \times \cR_{N+1}$	&	$\quad \longrightarrow$ &  $\quad \lA_{N} \simeq \lA_{1} \times \cR_{N}$ \\
		~		&	 	$\quad \quad \{(\hx_i,\hg_i)_{i \neq j-1,j}, (\hx_{j-1},\hh_2), (\hy_1,\hh_1), (\hy_2,\hg_j)\}$		&	$\quad \longmapsto$			&	$(\hx_k,\hg_k)_{k\in[1,N]}$
	\end{tabular}
\end{equation}

One can then show that these new operators have the right commutators in order to generated $\cR_{N+1}$:
\begin{equation}
[\hy_2^a,\hg_j] = i \tau^a \hg_j \; , \quad [\hy_2^a,\hh_1] = -i \hh_1 \tau^a\; \quad [\hy_1^a,\hh_2]=-i \hh_2 \tau^a \; , \quad [\hx_{j-1}^a,\hh_2] = i \tau^a \hh_2 \; ,
\end{equation}
and all the others commutators in $\cR_N$ are left untouched. Therefore, the set of operators $\{(\hx_i,\hg_i)_{i \neq i_0,j-1,j },(\hx_{j-1},\hh_2),(\hy_1,\hh_1),(\hy_2,\hg_j)\}$ is the algebra $\cR_{N+1}$ and $R_j$ is a Lie algebra morphism. Similarly to the coarse-graining, we must now check that the operators of the global symmetry algebra can be expressed with this new set of operators. This is again straightforward for the global holonomy while for the global grasping operator we need to use the same property used in the case of the coarse-graining. Again, these maps can be lifted to map on the algebra $\cA_N$.
\medskip

We conclude this section with the important remark that the coarse-graining and the refinement are inverse operations in the sense that
\begin{equation}
R^{(N)}_j \circ C^{(N+1)}_{j,j+1}  = \id\;.
\end{equation}
This mathematical relation between the two operations is at the basis of the exact nature of these processes. That is, the coarse-graining removes exactly the information that the refinement adds. It is not necessary to add a cut-off step to the coarse-graining in order to finish the operator. Also, one can remark that the coarse-graining process bears obvious similarity with the work done in the previous sections to obtain the isomorphisms of $\cA_N$. Indeed, let us look at the isomorphism for $\cA_2$ starting from the construction if the global symmetry algebra. To do so we are moving all the grasping operators to the same handle. In the case of $\cA_2$, this amount to the merging of the only two handles of the disk into one, giving the global holonomy and grasping operators. This corresponds to the coarse-graining detail earlier. Similarly, we could construct the global grasping and holonomy operators for $\cA_N$ by doing $(N-1)$ coarse-graining to obtain $\cA_1$, which is, in some sense, irreducible. That is, there is nothing left to coarse-grain. What is left is therefore only the global symmetry of the vacuum state of the theory.

In the next section, we  focus on the pushing the refinement process to infinity and study the continuum limit $N\rightarrow \infty$ for the algebra $\cA_N$. This will provide us with a complete view of the disk algebra.

\subsection{Sliding from the Center to the Continuum limit}

We already have a good understanding of the behaviour of the disk algebra and on the role of the number $N$ of flux insertions on the boundary. In particular that the $N=1$ case corresponds to a single quantum of geometry, representing the vacuum state of the underlying quantum gravity theory, that is, the global symmetry of the theory given by the Drinfeld double $\cD\SU(2)$. Increasing $N$ increases the amount of information the algebra encodes, and in particular the amount of vibrational modes it describes. What truly misses now is to understand the extreme case when $N$ goes to infinity.
In that infinite refinement limit, we expect to recover continuum theory of 3D gravity with the quantum version of the Poincar\'e loop algebra \eqref{eq:LoopA} derived in the first  section \ref{sec:3dgrav}.

To work out the continuum limit, we start with the commutation relations defining the disk algebra, as given in equation \eqref{eq:commN1}. In the spirit of going to the continuum limit, it is more convenient to work in the basis where the grasping and holonomy operators are dual, similarly to the continuum currents. We recall that it was previously introduced using the simple map
\begin{equation}
\hg_i = \hg^0_i \tau^0 +i \hg^a_i \tau^a,
\end{equation}
where $\hg^a_i = 2 i\tr(\hg_i \tau_a )$ and $\hg^0_i = \f{1}{2} \tr\,\hg_i$. Using this map, our initial commutators \eqref{eq:commN1} becomes
\begin{subequations}
	\begin{align}
	\left[ \hg_i , \hg_j \right] &= 0 \; , \\
	\left[ \hx_i^a , \hx_j^b \right] &= i \delta_{ij} {\eps^{ab}}_c \; \hx_i^c \; , \\
	\left[ \hx_i^a , \hg_j^b \right] &=  \f{i}{2} {\eps^{ab}}_c \; \hg_j^c \left[ \delta_{i,j+1} + \delta_{i,j} \right] + i \delta^{ab} \hg_j^0 \left[\delta_{i,j+1} - \delta_{i,j} \right],\; \\
	\left[ \hx_i^a , \hg_j^0 \right] &= \f{i}{4} \hg_j^a \left[ \delta_{i,j+1} - \delta_{i,j} \right]. \;
	\end{align}
	\label{eq:dual_algebra_comm}
\end{subequations}
The continuum limit is then obtained by defining the discrete angle $\theta_n = \f{2 \pi}{N}$ and viewing $\hx_i$ and $\hg_i$ as the functions $\hx_i \deq \hx(\theta_i)$ and $\hg_i \deq \hg(\theta_i)$ on the boundary disk. But before doing so, we need to understand how the discrete operators scale with $N$. To do so, we recall that the actions of $\hx_i$ and $\hg_i$, given equation \eqref{eq:action_x_g}, are the left derivations and multiplication by Wigner matrices respectively. When $N$ goes to infinity, we expect the continuum grasping operators to have the same interpretation, hence we demand $\hx_i$ to be at least of order $1$ in $\f{1}{N}$. Similarly, the arguments of the Wigner matrices goes to identity when $N$ goes to infinity. We are therefore demanding  $\hg_i$ to be of order $0$ in $\f{1}{N}$. From this, we can easily deduce that the components $\hg_i^0$ is of order $1$ in $N$ while $\hg_i^a$ of order $1$. In formula, the scaling of our discrete operators with $N$ is
\begin{subequations}
	\begin{align}
	\hx_i^a &= \f{2\pi}{N} \hj^a(\theta_i) + O\left( \f{1}{N^2}\right) \; , \\
	\hg_i^a &= \id + \f{2\pi i}{N} \hp^a(\theta_i) \tau^a + O\left( \f{1}{N^2}\right) \quad \implies \quad \hg_i^0 = \hat{1} + O\left( \f{1}{N^2}\right) \;  \quad \text{and} \quad \hg_i^a =\f{2\pi}{N} \hp^a(\theta_i) + O\left( \f{1}{N^2}\right)  ,
	\end{align}
\end{subequations}
where $(\hj^a,\hp^a)$ are operator-valued functions on the circle, acting as first-order scaling functions for the discrete operators.

Let us first focus on deriving the continuum limit directly from the operators $\hx$ and $\hg$. To do so, we introduce the following integrated operators
\begin{equation}
\hJ_\alpha \deq \sum_{i=1}^N \alpha_{i}^a\hx^{a}_i \quad \text{and} \quad  \hP_{\varphi}\deq \sum_{i=1}^{N} \varphi_{i}^a \hg^a_i
\end{equation}
with $\alpha_{i}$ and $\varphi_i$ with $i \in [1,N]$ two families of three-dimensional $\su(2)$-vector acting as smearing parameters. By integrated operators, we mean operator that are living on the whole discrete disk, i.e. at each handles. Note that since the smearing parameters are $\su(2)$ vectors, they are equipped with the non-trivial commutator
\begin{equation}
	[\alpha_i,\beta_i]^c = {\eps_{ab}}^c \alpha_i^a\beta_i^b \; .
\end{equation}
Similarly to the operators, we can freely see them as coming from the continuous functions $\alpha$ and $\varphi$ on the circle at $\theta_i$. Therefore, in the continuum limit, these integrated operators converge to the continuous currents defined in equation \eqref{eq:LoopA}
\begin{align*}
\hJ_{\alpha} &= \f{2\pi}{N}\sum_{i=1}^N \alpha^a(\theta_i) \hj^a(\theta_i) + O\left( \f{1}{N^2}\right) \underset{N\to \infty}{\rightarrow}\oint_{0}^{2\pi} \rd\theta\; \alpha^a(\theta) \hj^a(\theta) \; ,\\
\hP_{\varphi} &= \f{2\pi}{N}\sum_{i=1}^N \varphi^a(\theta_i) \hp^a(\theta_i) + O\left( \f{1}{N^2}\right) \underset{N\to \infty}{\rightarrow}\oint_{0}^{2\pi} \rd\theta\; \varphi^a(\theta) \hp^a(\theta) \; .
\end{align*}
Using these integrated operators, we can easily rewrite the commutators \eqref{eq:dual_algebra_comm} and we get
\footnote{The notation is such that $P_{[\alpha_{.+1} + \alpha_., \varphi]} = \sum\limits_{j=1}^N [\alpha_{j+1}+\alpha_j,\varphi_j]^c \hg_j^c$.}
\begin{subequations}
	\begin{align}
	\left[ \hP_{\varphi}, \hP_{\varphi'}  \right] &= 0 \; , \\
	\left[ \hJ_{\alpha}, \hJ_{\beta} \right] &= i \hJ_{[\alpha,\beta]} \; ,  \\
	\left[ \hJ_{\alpha} , \hP_{\varphi} \right] &=  \f{i}{2} P_{[\alpha_{.+1} + \alpha_., \varphi]} + i \sum_{k=1}^N (\alpha_{j+1}^a-\alpha_j^a) \varphi_j^a \hg_j^0  ,\; \\
	\left[ \hJ_\alpha, \hg_j^0 \right] &= \f{i}{4} \hg_{j}^a (\alpha_{j+1}^a - \alpha_j^a) . \;
	\end{align}
\end{subequations}
where the summation over $a$ is implicit. The continuum limit is then obtained by a straightforward computation\footnote{
	We simply use the fact that
	\begin{equation*}
	\alpha^a(\theta_{j+1}) + \alpha^a(\theta_j) = 2 \alpha^a(\theta_j) + O\left(\f{1}{N}\right) \quad \text{and} \quad \alpha^a(\theta_{j+1}) - \alpha^a(\theta_j) = \f{2 \pi}{N}\partial_{\theta}\alpha^a|_{\theta_j} + O\left(\f{1}{N^2}\right)
	\end{equation*}
	with the fact that $\hg^0 \sim 1$ and $\hg^a \sim \f{1}{N}$ to show that
	\begin{equation*}
	\f{i}{4} \hg_{j}^a (\alpha_{j+1}^a - \alpha_j^a) \sim O\left(\f{1}{N^2}\right) \underset{N\to \infty}{\rightarrow} 0 \; , \quad  P_{[\alpha_{.+1} + \alpha_., \varphi_j]} \sim 2 P_{[\alpha, \varphi]}  \; \quad \sum_{k=1}^N (\alpha_{j+1}^a-\alpha_j^a) \varphi_j^a \hg_j^0 = \f{2\pi}{N} \sum_{k=1}^N \partial_{\theta}\alpha^a|_{\theta_j} \varphi^a(\theta_j) \sim  \oint \varphi \rd \alpha \;.
	\end{equation*}
}
and we get the expect quantum\footnote{with the usual convention where $\hbar = 1$.} version of the continuous classical loop algebra \eqref{eq:LoopA}
\begin{equation*}
\left[ \hP_{\varphi}, \hP_{\varphi'}  \right] = 0 \; , \qquad \left[ \hJ_{\alpha}, \hJ_{\beta} \right] = i \hJ_{[\alpha,\beta]} \; , \quad  \left[\hJ_{\alpha} , \hP_{\varphi}\right] =  i \hP_{[\alpha , \varphi]} + i \oint \varphi^a \rd \alpha^a.
\end{equation*}
This explicitly establish that the disk algebra $\cA_N$ is an algebra that interpolates between the quantum group $\cD\SU(2) = \cA_{1}$ acting as global symmetry of the underlying non-commutative geometry of quantum gravity and the Poincar\'e loop algebra $\cU(L(\mathfrak{isu}(2)))=\cA_{\infty}$, which is the full symmetry algebra of 3D gravity. There is one primordial remark that must be done. The main difference between the discrete algebra at fixed $N$ and the Poincar\'e loop algebra comes from the central nature of the later. Indeed, it is well known that the presence of boundary make the bulk symmetry algebra into a centrally extended algebra at the boundary, as we have seen in the first section of this paper. It is interesting to note that, when introducing a discretization, we break this property: the algebra $\cA_N$ is not central due to the non-commutativity of the would-become central element $\hg^0_j$. Most of the analysis on the boundary symmetries are done thanks to the property of having a centrally extend algebra at the boundary. Therefore, the approach we are using here could be an interesting one to have a better grasp of the impact of discretization on the boundary symmetries. This lack of centrality will be a major problem for the construction of the Virasoro operators, as we will show in the next section.

\subsection{Towards discrete Virasoro operators}

Now that we have understood that the disk algebra interpolates between the quantum symmetry and the continuous current algebra, we tackle another interesting aspect of currents algebra: the construction of Virasoro operators. In the classical setup, we have done so via a twisted Sugawara construction at the beginning of the papers. Here, we will revisit this construction with the Fourier component in the quantum setup. Recall that the quantum continuum algebra of symmetries is given by the quantum version of \eqref{eq:commutators_fourier_component}
\begin{equation}
\left[ J_n^a , J_m^b \right] = i {\eps^{ab}}_c J^c_{n+m} \; , \quad \left[ J_n^a , P_m^b \right] = i {\eps^{ab}}_c P^c_{n+m} + n \delta^{ab} \delta_{n+m,0} \; , \quad \left[ P^a_n, P^b_m \right] = 0 \; .
\label{eq:commutators_fourier_component_continuum}
\end{equation}
For simplicity, we use the same notation for the discrete and continuous currents with the difference that we do not put a hat on the continuous quantum current in order to  make easier to distinguish  the two (even thought they are operators). In Fourier components, the twisted Sugawara construction amount to defining the following operator
\begin{equation}
\cD_{n} = \f{1}{2} \sum_{k \in \Z} :J^{\alpha}_{-k} P^{\alpha}_{n+k} + P^{\alpha}_{-n}J^{\alpha}_{n+k}:
\label{eq:continuum_def_fourier_op}
\end{equation}
where $:.:$ holds for the normal ordering of the operators. Explicitly, we have
\begin{subequations}
	\begin{align}
	\cD_{0} &= \f{1}{2}\left(J_{0}^\alpha P_0^{\alpha}+ P_0^{\alpha} J_{0}^\alpha \right) + \sum_{k > 0} \left( J_{-k}^\alpha P_k^{\alpha}+ P_{-k}^{\alpha} J_{k}^\alpha \right) \; , \\
	\cD_{n} &= \f{1}{2} \sum_{k=0}^{n} \left( J_{k}^{\alpha} P_{n-k}^{\alpha} + P_{k}^{\alpha}J_{n-k}^{\alpha} \right) + \sum_{k>0} \left( J_{-k}^{\alpha} P_{n+k}^{\alpha} + P_{-k}^{\alpha}J_{n+k}^{\alpha} \right) \; .
	\end{align}
	\label{eq:continuum_def_fourier_op_alt}
\end{subequations}
These operators are such that $\cD_{-n} = \cD_{n}^{\dagger}$. Similarly to the classical case, one can show that these operators form a Virasoro algebra with central extension
\begin{equation}
[\cD_{n},\cD_{m}] = (n+m)\cD_{n-m} + \f{c}{12}n(n^2-1) \delta_{n+m,0} \; .
\label{eq:commuation_vir_fourier}
\end{equation}
with central charge $c=3$, representing the three dimensions of the manifold. A defining property of $\cD_n$ is its commutation relations with $J_{m}^a$ and $P_m^a$. These relations read
\begin{equation}
[\cD_n, J_{m}^{a}] = \sum_{k \in \Z} [J_{n-k}^{\alpha}, J_{m}^{a}] P_{k}^{\alpha} + J_{k}^{\alpha}  [P_{n-k}^{\alpha}, J_{m}^{a}]  \quad \text{and} \quad [\cD_n, P_{m}^{a}] =  \sum_{k \in \Z} [J_{n-k}^{\alpha}, P_{m}^{a}] P_{k}^{\alpha} \; .
\label{eq:def_prop_vir_cont}
\end{equation}
One can then introduce back the explicit commutators and obtains the weights associated to each Fourier mode of the Virasoro operators. The goal of this section is to work toward a discrete definition of these operators. As we will see, it is possible to introduce a discrete zeroth mode Virasoro operator while it seems compromised for the other modes.
\bigskip

In order to use the construction of the Virasoro operators previously introduced, we first need to express the algebra \eqref{eq:dual_algebra_comm} with Fourier modes. The reason we are are using a Fourier decomposition to construct the Virasoro operators in the discrete is because the discreteness of the disk make the construction with the direct currents more cumbersome. The Fourier modes are defined by\footnote{The floor function $\floor{.}$ is defined by $\floor{\f{2p+1}{2}}= p $ and $\floor{\f{2p}{2}} = p $ for  an integer $p$.}
\begin{equation}
\hJ^\mu_n = \sum_{k=-\floor{N/2}}^{\floor{N/2}} e^{-i \f{2\pi}{N}n k} \hx^\mu_k \quad \text{and} \quad \hP_n^\mu = \sum_{k=-\floor{N/2}}^{\floor{N/2}} e^{-i \f{2\pi}{N}n k} \hg^a_k \quad \text{for} \; \; n \in  \left[-\floor*{\f{N}{2}},\floor*{\f{N}{2}}\right]
\end{equation}
where $\mu = 0,1,2,3$.
There are clear ambiguities in this definition attempt: the chosen periodicity - or winding number- around the the boundary circle, for which we make the simplest and natural choice of winding once,  and the split between left and right movers, for which we take the arbitrary cut at half the flux insertions. More refined attempts to identify correct Virasoro generators should probably involve a properly pondered sum over different winding numbers and slides of the  cut between left and right movers.

Here the Fourier modes are symmetrized around the zeroth mode. Therefore, when $n$ is even, we do not sum over $k=0$ and the mode $n=0$ is not defined. In the following, we will consider the case where $N$ is odd for simplicity. However, the whole discussion will still be valid for even $N$. Note that this choice of symmetrisation of the Fourier mode is purely for the technical part of the computation, and to allow or simplest comparison with the continuum case. One has the freedom to shift this choice as one wants. However, we such a choice, we can safely define components with negative $n$ to be the Hermitian of the associated positive mode.  Recall that $\hx^0_k  = \hat{1}$, therefore $ \hJ^0_n = \delta_{0,n} \hat{1}$. These Fourier operators are also $N$ periodic due to the cyclicity condition of the disk, and their algebra reads\footnote{All operators trivially commutes with $\hJ^0_n$.}
\begin{subequations}
	\begin{align}
	[\hP^\mu_n,\hP^\nu_m] &= 0 \\
	[\hJ^a_n,\hJ^b_m] &= i {\eps^{ab}}_c \hJ^c_{n+m} \\
	[\hJ^a_n,\hP^b_m] &= \f{i}{2} {\eps^{ab}}_c (1+e^{-i \f{2\pi}{N}n}) \hP^c_{n+m} + i \delta^{ab} (e^{-i \f{2\pi}{N}n}-1) \hP^0_{n+m} \\
	[\hJ^a_n,\hP^0_m] &= \f{i}{4} (1-e^{-i \f{2\pi}{N}n}) \hP^a_{n+m} \; .
	\end{align}
\end{subequations}
It is interesting to note that the continuum limit can also be taken from this set of commutators. We immediately recover the quantum version of the Fourier algebra \eqref{eq:commutators_fourier_component}. Already from the commutators, one could argue that the definition of Virasoro operators is not possible in the discrete. Indeed, Virasoro operators are by definition quadratic in the basic operators. The reason why they form a non-trivial close algebra in the continuum is entirely due to the fact that the algebra \eqref{eq:commutators_fourier_component_continuum} is central. Therefore, a commutator between two Virasoro operators produce terms of order three and two in $J$ and $P$, and the definition of $\cD_n$ is such that only terms of order two survive. The central extension is then coming from re-writing the operators respecting the normal ordering. In the discrete case however, we lost the property of centrality of the current algebra. Therefore, in that case, quadratic operators only produces terms of order three. As such, it seems rather impossible to construct quadratic operators that are stable under commutation. In the following, we will therefore focus not on constructing the well-known commutation relation \eqref{eq:commuation_vir_fourier} by on looking for quadratic operators such that the defining property \eqref{eq:def_prop_vir_cont} holds. For simplicity, we will still call the discrete operators we are going to define Virasoro operators.

There are two straightforward, simplistic even, ways to introduce these discrete Virasoro operators such that they have the right continuum limit. These two ways depend on if one is taking \eqref{eq:continuum_def_fourier_op} or \eqref{eq:continuum_def_fourier_op_alt} as defining definition. While these two are obviously equivalent in the continuum, this is not the case in the discrete. Once a choice is done, then the discrete definition follow by a simple discretization of the formula with the natural cut-off $N$. Starting from \eqref{eq:continuum_def_fourier_op}, we can define the operators $\hcD_{n}$ to be
\begin{equation}
\hcD_{n} = \f{1}{2} \sum_{k=-\floor{N/2}}^{\floor{N/2}} :\hJ^{\alpha}_{-k} \hP^{\alpha}_{n+k} + \hP^{\alpha}_{-n} \hJ^{\alpha}_{n+k}: \; .
\end{equation}
Note that the implicit sum over $\alpha$ includes the zeroth component of the Fourier operators. Explicitly, this definition amount to
\begin{equation}
\hcD_{n} = \f{1}{2} \sum_{k=0}^{n} \left( \hJ_{k}^{\alpha} \hP_{n-k}^{\alpha} + \hP_{k}^{\alpha} \hJ_{n-k}^{\alpha} \right) + \sum_{k=1}^{\floor{N/2}} \left( \hJ_{-k}^{\alpha} \hP_{n+k}^{\alpha} + \hP_{-k}^{\alpha} \hJ_{n+k}^{\alpha} \right) - \sum_{k=\floor{N/2}-n+1}^{\floor{N/2}} \left( \hJ_{-k}^{\alpha} \hP_{n+k}^{\alpha} + \hP_{-k}^{\alpha} \hJ_{n+k}^{\alpha} \right)  \; .
\end{equation}
with $\hcD_{-n} = \hcD_{n}^{\dagger}$. From this formula, it is immediate to see that the first two terms arises from a direct discretization of the second possible defining definition \eqref{eq:continuum_def_fourier_op_alt}, while the last one is specific to the first definition. This term vanishes in two case: when $n=0$ and obviously when $N \rightarrow \infty$. One can then compute the commutators between $\hcD_{n}$ and the basics operators, and the details of the computations are given in the appendix \ref{app:discrete_vir_computation}. At the end of the day, we obtain the following commutators
\begin{subequations}
	\begin{align}
	[\hcD_n, \hJ_{m}^{a}] &= \sum_{k = -\floor{N/2}}^{\floor{N/2}} [J_{n-k}^{\alpha}, J_{m}^{a}] P_{k}^{\alpha} + J_{k}^{\alpha}  [P_{n-k}^{\alpha}, J_{m}^{a}] \; -  \sum_{k = \floor{N/2}-n+1}^{\floor{N/2}} [J_{n-k}^{\alpha}, J_{m}^{a}] P_{-k}^{\alpha} + J_{-k}^{\alpha}  [P_{n-k}^{\alpha}, J_{m}^{a}]
	\\
	[\hcD_n, \hP_{m}^{\mu}] &=  \sum_{k = -\floor{N/2}}^{\floor{N/2}} [J_{n-k}^{\alpha}, P_{m}^{\mu}] P_{k}^{\alpha} \; - \sum_{k = \floor{N/2}-n+1}^{\floor{N/2}} [J_{n+k}^{\alpha}, P_{m}^{\mu}] P_{-k}^{\alpha}  \; .
	\end{align}
\end{subequations}
These commutators have two parts. The first one corresponds to the straightforward discretization of the continuum commutators, and is the part that we wanted. However, the second part corresponds to a deformation of the commutators due to the discreteness of the disk. At first glance, it might seems that to get rid of the second part, it is enough to modified the definition of the discrete Virasoro operators, and takes \eqref{eq:continuum_def_fourier_op_alt} as defining definition. This is not the case, as we explicitly show in the appendix \eqref{app:discrete_vir_computation}. We need half of the contribution of the deformation in the definition of the discrete Virasoro operators to obtain the full sum from $-[N/2]$ to $[N/2]$. Getting rid of it will return an over-complete sum (see equation \eqref{eq_app:alt_com}). As expected however, this deformation also vanishes in the continuum limit. The key point is that it also vanishes for $\hcD_0$. Hence, we define the zeroth mode discrete Virasoro operator by
\begin{equation}
	\hcD_{0} = \f{1}{2} \left( \hJ_{k}^{\alpha} \hP_{-k}^{\alpha} + \hP_{k}^{\alpha} \hJ_{-k}^{\alpha} \right) + \sum_{k=1}^{\floor{N/2}} \left( \hJ_{-k}^{\alpha} \hP_{k}^{\alpha} + \hP_{-k}^{\alpha} \hJ_{k}^{\alpha} \right) \; .
\end{equation}
Therefore, even thought it seems complicated to find quadratic operators that span a closed algebra, we have exhibited operators that behave almost like the continuum Virasoro operators. The behaviour is exact for the zeroth mode, and contains a deformation for all others modes. Note that this deformation is not really surprising: in insight, the deformation deletes some mode contribution in the primary sum, telling us that the quadratic operators $\hcD_n$ do not form a complete set of information around the disk. This is in accordance with the fact that quadratic operators cannot span a closed algebra in the discrete setup.

\subsection{Fusion of defects}

In the previous sections, we have studied in details the structure of the disk algebra $\cA_{N}$  and shown that it iinterpolates between the global symmetry algebra $\cD\SU(2)$ and the continuum current algebra. What remains now is to understand how two such discrete current algebra can be fused. More specifically, recall that from the viewpoint of quantum gravity, the disk algebra encodes the observables of point-like defects on the 2D boundary. This defects are the shadow of one-dimensional bulk defects. A natural question is then, if one has two set of non-connected point-like defects, how can they be described in the same setup. In other word, how can one fused the two respective disk algebra into a bigger one that describes the two defects together.

To construct the fusion of disk algebras, we will follow a simple guiding rule. Recall that the share structure between all disk algebra is the global symmetry algebra of the underlying non-commutative space, its "centre of mass" $\cD\SU(2) \sim \cA_{1}$. In a sense, $\cA_{1}$ describes a boundary without any defects nor excitation in the bulk, i.e. a vacuum state. Considering two sub-regions without any defects, therefore described by $\cA_{1}$, we expect that the fused region will also be without any defect, and hence also described by $\cA_{1}$. In formula, the fusion of two disk algebra $\cA_{1}$ is still the disk algebra $\cA_{1}$ for $N=1$ handle:
\begin{equation}
	\cA_{1} \boxdot \cA_{1} = \cA_{1} \; .
\end{equation}
In some sense, this a fusion without any constraints. It is simple to generalise this rule to arbitrary valency $N$ for the disk algebra. Recall that the disk algebra $\cA_{N}$ can be split between its $\cA_{1}$ part and a tower of Heisenberg double in direct product. Since the fusion of disk algebras is also a disk algebra and that we do not apply any constraints on the fusion, we expect the tower of Heisenberg double to be invisible to the fusion product. Therefore, for any valencies $N$ and $\tilde{N}$, the unconstrained fusion of $\cA_{N}$ and $\cA_{\tilde{N}}$ reads
\begin{equation}
	\cA_{N} \boxdot \cA_{\tilde{N}} \simeq (\cA_{1} \boxdot \cA_{1}) \times \cH\SU(2)^{\times (N + \tilde{N} -2)} = \cA_{1}  \times \cH\SU(2)^{\times (N + \tilde{N} -2)} \simeq \cA_{N+\tilde{N}-1} \;.
\end{equation}

The explicit fusion process is quiet straightforward. Consider two algebras $\cA_{1}$ with operators $(\hx,\hg)$ and $(\hy,\hh)$. The fusion process follows the pattern depicted in figure \ref{fig:fusion_A1_A1}. Explicitly, we define the fused loop holonomy and grasping operators by
\begin{equation}
	\hX = \hx + \hy = \hy + \hx \quad \text{and} \quad \hG = \hg \hh = \hh \hg \; .
\end{equation}
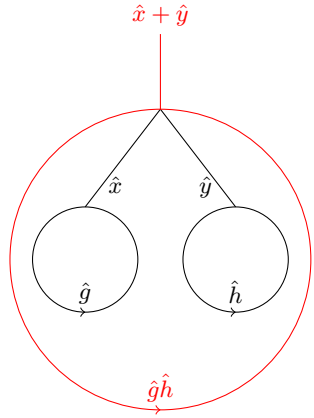
\begin{figure}
	\begin{tikzpicture}[scale=1]
		\coordinate (O1) at (-1,0);
		\coordinate (O) at (0,0); \coordinate (A) at ($(O)+(0,2)$);
		\coordinate (O2) at (1,0);

		\draw[->-=0.75] (O1) circle (0.7); \draw ($(O1)+(0,0.7)$) --node[pos=0.4,below]{$\hx$} (A); \node[above] at ($(O1)+(0,-0.7)$) {$\hg$};
		\draw[->-=0.75] (O2) circle (0.7); \draw ($(O2)+(0,0.7)$) --node[pos=0.4,below]{$\hy$} (A); \node[above] at ($(O2)+(0,-0.7)$) {$\hh$};

		\draw[red,->-=0.75] (O) circle (2); \draw[red] (A) --node[pos=1,above,red]{$\hx + \hy$} +(0,1); \node[above,red] at ($(O)+(0,-2)$){$\hg \hh$};
	\end{tikzpicture}
	\caption{Schematic representation of the fusion between two algebras $\cA_1$. In red are represented the fused operators, and in black the starting operators of the two algebras $\cA_{1}$.}
	\label{fig:fusion_A1_A1}
\end{figure}
It is immediate to check that the algebra spans by $(\hX,\hG)$ is again $\cA_1$
\begin{equation}
	[\hX^a,\hX^b] = [\hx^a,\hx^b]+[\hy^a,\hy^b] = i \eps^{abc} \hX^c \; , \quad [\hG,\hG] = 0 \; , \quad [\hX^a,\hG] = [\hx,\hg]\hh + \hg [\hy,\hh] = i [\hG,\tau^a] \; .
\end{equation}
In broad words, the fusion amount to detaching the boundary links of the starting disk algebras of its target node for one algebra and source node for the other, and then to glue these two boundary links together. In insight, this fusion process follows the product of $\cD\SU(2)$. Indeed, at the infinitesimal level, the product of $\cD\SU(2)$ reads
\begin{equation}
	(\hx,\hg).(\hy,\hh) = (\hx + \hy, \hg \hg^{-1} \hh \hg) = (\hx + \hy, \hh \hg) = (\hx + \hy,\hg \hh) \; .
\end{equation}

\begin{figure}
	\begin{tikzpicture}[scale=1]
	\coordinate (O1) at (-0.8,0);
	\coordinate (O) at (0,0); \coordinate (A) at ($(O)+(0,2)$);
	\coordinate (O2) at (0.8,0);

	\draw[->-=0,->-=0.38,->-=0.64] (O1) circle (0.7);
	\draw ($(O1)+(0,0.7)$) --node[pos=0.2,above left,scale=0.9]{$\hx_1$} (A);
	\draw ($(O1) + (-0.7,0)$) --node[above,pos=0.5,scale=0.9]{$\hx_2$} +(-0.5,0); \draw[red] ($(O1)+(-0.7,0)+(-0.5,0)$) --node[pos=1,left,red]{$\hx_2$} +(-1,0);
	\draw ($(O1) + (0,-0.7)$) --node[left,pos=0.3,scale=0.9]{$\hx_3$} +(0,-1.15); \draw[red] ($(O1)+(0,-0.7)+(0,-1.15)$) --node[pos=1,below,red]{$\hx_3$} +(0,-0.5);
	\node[above,scale=0.9] at ($(O1)+(-0.25,0.15)$) {$\hg_1$};
	\node[above,scale=0.9] at ($(O1)+(-0.25,-0.65)$) {$\hg_2$};
	\node[above,scale=0.9] at ($(O1)+(0.50,-0.15)$) {$\hg_3$};

	\draw[->-=0.14,->-=0.52,->-=0.88] (O2) circle (0.7);
	\draw ($(O2)+(0,0.7)$) --node[pos=0.2,above right,scale=0.9]{$\hy_1$} (A);
	\draw ($(O2) + (0.7,0)$) --node[above,pos=0.5,scale=0.9]{$\hy_3$} +(0.5,0); \draw[red] ($(O2)+(0.7,0)+(0.5,0)$) --node[pos=1,right,red]{$\hy_3$} +(1,0);
	\draw ($(O2) + (0,-0.7)$) --node[left,pos=0.3,scale=0.9]{$\hy_2$} +(0,-1.15); \draw[red] ($(O2)+(0,-0.7)+(0,-1.15)$) --node[pos=1,below,red]{$\hy_2$} +(0,-0.5);
	\node[above,scale=0.9] at ($(O2)+(-0.4,-0.15)$) {$\hh_1$};
	\node[above,scale=0.9] at ($(O2)+(0.25,-0.65)$) {$\hh_2$};
	\node[above,scale=0.9] at ($(O2)+(0.30,+0.15)$) {$\hh_3$};

	\draw[red,->-=0.13,->-=0.38,->-=0.59,->-=0.90,->-=0.75] (O) circle (2);
	\draw[red] (A) --node[pos=1,above,red]{$\hx_1 + \hy_1$} +(0,1);
	\node[below,red] at ($(O)+(0,-2)$){$\hg_3 \hh_1$};
	\node[red] at ($(O)+(-1.7,1.5)$){$\hg_1$};
	\node[red] at ($(O)+(-1.75,-1.4)$){$\hg_2$};
	\node[red] at ($(O)+(1.77,-1.4)$){$\hh_2$};
	\node[red] at ($(O)+(1.77,1.4)$){$\hh_3$};
	\end{tikzpicture}
	\caption{Schematic representation of the fusion between two algebras $\cA_3$ resulting in $\cA_{5}$. In red are represented the fused operators, and in black the starting operators of the two algebras $\cA_{3}$.}
	\label{fig:fusion_N_Ntilde_to_NNtildeinf1}
\end{figure}
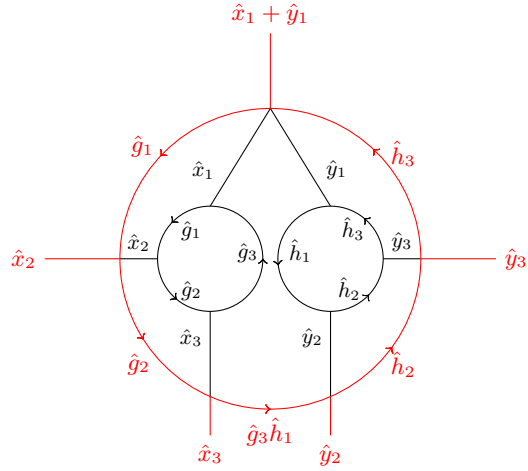

The fusion pattern for two algebra of arbitrary valency can as easily be represented. The only difference is that one need to make a choice of root for the fusion. This is in accordance with the fact that one need also a choice of root for the construction of the isomorphisms of $\cA_N$. An example of fusion for $N = \tilde{N} = 3$ is given figure \ref{fig:fusion_N_Ntilde_to_NNtildeinf1}.

The new operators defined by the fusion process are $\hx_1 + \hy_1$ and $\hg_3 \hh_1$. Via this fusion process, we have implicitly chosen that the root for the isomorphism in both starting algebra was the handle $(1)$. It is immediate to check that the operators $(\hx_1+\hy_1,\hg_1),(\hx_2,\hg_2),(\hx_3,\hg_3 \hh_1),(\hy_2,\hh_2),(\hy_3,\hh_3)$ span the algebra $\cA_{5}$. The only complicated part is at the fused points, and the structure of the sunny graph given figure \ref{fig:fusion_N_Ntilde_to_NNtildeinf1} directly gives us the correct commutators. Note that nothing stop us to now redefine the isomorphism for the fused $\cA_{5}$ with a root that is not the handle $(1)$.

\begin{figure}[!htb]
	\begin{tikzpicture}[scale=1]
		\coordinate (O1) at (-1.4,0);
		\coordinate (O2) at (1.4,0);

		\coordinate (A1) at (-0.74,0.45); \coordinate (A2) at (0.74,0.45);
		\coordinate (B1) at (-0.74,-0.45); \coordinate (B2) at (0.74,-0.45);
		\coordinate (C1) at (-0.6,0); \coordinate (C2) at (0.6,0);

		\draw[->-=0,->-=0.5] (O1) circle (0.8);
		\draw[->-=0,->-=0.5] (O2) circle (0.8);

		\draw (A1) -- node[pos=0.25,above,scale=0.8]{$\hx_{i+1}$} node[pos=0.75,above,scale=0.8]{$\hy_{j}$} (A2);
		\draw (B1) -- node[pos=0.25,below,scale=0.8]{$\hx_{i}$} node[pos=0.75,below,scale=0.8]{$\hy_{j+1}$} (B2);

		\node[left,scale=0.8] at (C1) {$\hg_{i}$}; \node[right,scale=0.8] at (C2) {$\hh_{j}$};

		\centerarc[dashed](O1)(-45:-95:1); \centerarc[dashed](O1)(50:100:1);
		\centerarc[dashed](O2)(225:275:1); \centerarc[dashed](O2)(90:140:1);
	\end{tikzpicture}
	\caption{Example of other possible fusion. In that case, a flatness constraint fusion, with the constraint $\hg_i \hg_j= \hat{\id}$. The value of the fused grasping operators, in red, must be studied in order for the operators of the fused disk algebra to be weak Dirac observables.}
	\label{fig:alt_fusion}
\end{figure}
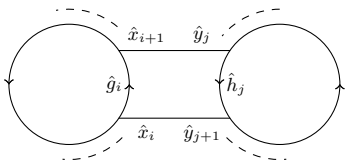

The fusion we have defined here is, in some sense, the simplest one in that it conserves the global symmetry algebra. More refined fusions can be introduced by adding constraint to the processes. For example, one could try to fuse along a face. That is, consider two pairs of two consecutive grasping operators of the two starting algebra, $\hx_i,\hx_{i+1}$ and $\hy_j,\hy_{j+1}$ and let us make the fusion where $\hx_i$ (resp. $\hx_{i+1}$) is fused $\hx_{j+1}$ (resp. $\hx_j$), see figure \ref{fig:alt_fusion}. In such a fusion process, a natural flatness constraint arises: $\hg_i \hg_j= \hat{\id}$. For such a fusion to be well define, one then must check our operators are (weak) Dirac observables with respect to the constraint. This might imply constraints on the fused grasping operators.
We let the study of all the possible class of fusion for a later work. This study is necessary in order to completely capture the behaviour of neighbouring closed region for quantum gravity.

\section*{Conclusion}

The line of research pursued in the present work is the treatment of quantum space-time boundaries for local regions in quantum gravity, their definition, the associated observables and symmetries, the dynamics of the edge modes propagating on the boundary and how they holographically reflect the bulk geometry dynamics.
We tackled this large problem in the context of three-dimensional quantum gravity. It is a topological and solvable theory, which can be formulated exactly
at the discrete level, for instance by the Ponzano-Regge topological state-sum (understood to be equivalent to both  Witten's reformulation of 3d gravity as a Chern-Simon theory and to 3d loop quantum gravity).
Concretely, the present work studied the algebraic structure of points defects in canonical 3d quantum gravity, or equivalently of 1d spatial boundaries (which surround the puncture created by a point defect).  We identified the algebra of observables at the discrete level as the discrete analogue  of a current algebra. We call it ``disk algebra'' and explained how it depends on the number  $N$ of flux insertions to the 1d boundary. This number can be interpreted as the number of geometrical  building blocks making the boundary. We defined the coarse-graining operation of the defect down to the 0-mode symmetry group $\cD\SU(2)$ given by the $N=1$ case, and reversely defined the refinement operation up to the $N\rightarrow\infty$ continuum limit where one recovers the classical boundary current algebra of 3d gravity.

More precisely, we studied the boundary symmetry
of the Ponzano-Regge model
for a 3d space-time region consisting of a solid cylinder with a disk base: the canonical spatial slice is a 2d disk, whose boundary (or space-time corner) is a 1d circle. We started by reviewing the classical theory. The boundary symmetry algebra, generated by the conserved charges, is given by the Poincar\'e loop algebra with central extension. This current algebra admits the Virasoro algebra and the BMS algebra as sub-algebras of conserved charges for specific boundary conditions.
The core of this paper  then consists in the analysis of the quantum theory and identification of the basic excitations of the fields and of the quantum equivalent of this boundary current algebra.
At the quantum level, a discrete space-time picture
naturally induces a discretization of the spatial disk and of its boundary circle: the quantum boundary is described by a holonomy looping around the circle along which we have a certain number $N$ of flux insertions. This number $N$
reflects the number of basic quanta of geometry living on the boundary in the deep quantum regime. The limit $N\rightarrow\infty$ is the infinite refinement limit of the boundary, in which we should recover the continuum regime and the classical Poincar\'e loop algebra.
We identify the algebra of operators acting on the quantum boundary. They are generated by the basic excitations and deformations of the quantum geometry. We call this algebra of quantum observables the disk algebra $\cA_{N}$, emphasizing its clear dependence on the discreteness index $N$.
The  results of this paper are two-fold.

First of all, we show that the disk algebra for arbitrary $N$  always factorizes as the global symmetry algebra (the ``centre of mass'') times the algebra of local deformations of the boundary:  the disk algebra $\cA_{N}$ is isomorphic to the Drinfeld double of $\SU(2)$, representing the global symmetry of the underlying non-commutative space, in direct product with a tower of $(N-1)$ Heisenberg doubles. Each Heisenberg double parametrizes a  excitation propagating around the disk.
In other word,
This discreteness index $N$ encodes the amount of bulk information seen from the boundary. In the vacuum state, without any defect in the bulk, it is enough to consider $N=1$.
Second of all, we  show that, in the continuum limit as $N$ is sent to $\infty$, we do recover the Poincar\'e loop algebra from the disk algebra. That is, we obtain a tower of discrete current algebras $\cA_{N}$ labeled by $N$, which  interpolate between the global quantum symmetry of the theory, given by the Drinfeld double $\cD\SU(2)$ and the  boundary symmetry algebra, given by the Poincar\'e loop algebra.
This allows to slide from the deep quantum regime at low $N$ to the semi-classical regime at high $N$ using coarse-graining and refinement maps between the discrete current algebras $\cA_{N}$ for different $N$'s.

\medskip

Then we have initiated the investigation of two crucial points.
First, we  looked into the possibility of a discrete Sugawara construction, i.e. how to build a discrete equivalent of the Virasoro and BMS algebras as quadratic operators in the current algebra formed by the boundary disk operators. We saw that, due to the intrinsic discretization of the disk, the discrete current algebra lost its central property: there is no natural candidate at finite $N$ for the  operator meant to become central in the continuum limit and it is not possible to construct quadratic operators that  form a closed Lie algebra.
More work is therefore required to identify a discrete Virasoro algebra or discrete BMS algebra at finite $N$ and study how they arise from the discrete current algebra. Understanding the conformal symmetry in a discrete setting is indeed a complicated task, as one realizes from the work in condensed matter and integrable systems (e.g. \cite{Faddeev:1993pe,Faddeev:1996xc,Faddeev:1997hp,Faddeev:2000if,Zou:2019dnc}, see in particular the recent works on Koo-Saleur generators \cite{Grans-Samuelsson:2020kvl,Grans-Samuelsson:2020jym}).


%
Second, we  looked into possible fusion operations for the disk algebras: how to glue two boundaries together, or equivalently, how to merge two defects into a single defect.  We have introduced a simple fusion operation that conserve the global symmetry $\cD\SU(2)$ and allow us to properly defined discrete sub-region. The downside of our definition is that it does not introduce any correlation between the un-fused disk algebra. More work is required to catalogue possible fusion operations and clearly identify the required physical properties that we should demand.

\medskip

More generally, we envision three lines of research that stem from the present work.
First of all, it seems essential to investigate possible fusion operations for the disk algebras, and actually study the algebra of fusion operators between disk algebras. Since it does not seem feasible to define a discrete Virasoro algebra within a given disk algebra at fixed $N$, it seems natural to look for Virasoro operators in the larger framework of a fusion algebra, which would involve ``super-operators'' shifting the number of discrete quanta $N$. From this perspective, the work on building CFTs and a AdS/CFT duality from multi-bodied entanglement on tensor networks (e.g. \cite{Milsted:2018vop}) and quantum error-correcting codes (e.g. \cite{Pastawski:2015qua}) might bring further inspiration towards this goal.
Second, we would like to take the present work as a first step towards revisiting the link between 3d quantum gravity and 2d CFT  \cite{Freidel:2002hy,Krasnov:2000zq,Krasnov:2003qt,Krasnov:2004gm,Freidel:2009nu}. For instance, starting from the discrete boundary current symmetry algebra should most certainly help  eludicate the arising of (regularized) BMS character from Ponzano-Regge amplitudes as shown in \cite{Dittrich:2018xuk,Dittrich:2017hnl,Dittrich:2017rvb,Goeller:2019zpz} and help understand the conformal invariance and criticality at the finite discrete level for quasi-local boundaries in 3d quantum gravity. It would probably allow a new perspective on the duality uncovered between the Ponzano-Regge model for 3d quantum gravity and the 2d Ising model on space-time boundaries \cite{Dittrich:2013jxa,Bonzom:2015ova}. Finally, one should move on the dynamics of the discrete disk, either through discrete step dynamics (as in the Ponzano-Regge model) or continuous Hamiltonian evolution (through a version of the Wheeler-de Witt equation with space-time corners). This implies studying discrete quantum boundary conditions and the corresponding theories induced by the bulk dynamics onto the boundary lattice. This might shed a light on the correspondence between 3d quantum gravity in the deep quantum regime and the renormalization flow of 2d CFTs, as uncovered for the continuum path integrals for $T\bar{T}$ deformation \cite{Mazenc:2019cfg}, and extend it to the discrete space-time picture of the Ponzano-Regge model.

\appendix

\section{Commutators of the tentative discrete Virasoro operators $\hcD_n$}
\label{app:discrete_vir_computation}

In this appendix, we compute the commutators between the discrete Virasoro operators $\hcD_{n}$ defined by
\begin{equation}
	\hcD_{n} = \f{1}{2} \sum_{k=0}^{n} \left( \hJ_{k}^{\alpha} \hP_{n-k}^{\alpha} + \hP_{k}^{\alpha} \hJ_{n-k}^{\alpha} \right) + \sum_{k=1}^{\floor{N/2}} \left( \hJ_{-k}^{\alpha} \hP_{n+k}^{\alpha} + \hP_{-k}^{\alpha} \hJ_{n+k}^{\alpha} \right) - \sum_{k=\floor{N/2}-n+1}^{\floor{N/2}} \left( \hJ_{-k}^{\alpha} \hP_{n+k}^{\alpha} + \hP_{-k}^{\alpha} \hJ_{n+k}^{\alpha} \right)  \; .
\end{equation}
and the basic operators $\hJ_m$ and $\hP_{m}$. We separate the contribution of the three terms. We call them $\hcD_{n}^{i}$ with $i=1,2,3$ respectively. We are also interested in looking at the alternative definition of the discrete Virasoro operators
\begin{equation}
	\hcD_n^{\text{alt}} = \f{1}{2} \sum_{k=0}^{n} \left( \hJ_{k}^{\alpha} \hP_{n-k}^{\alpha} + \hP_{k}^{\alpha} \hJ_{n-k}^{\alpha} \right) + \sum_{k=1}^{\floor{N/2}} \left( \hJ_{-k}^{\alpha} \hP_{n+k}^{\alpha} + \hP_{-k}^{\alpha} \hJ_{n+k}^{\alpha} \right)
	\label{eq_app:alt_com}
\end{equation}
\bigskip

Let us start with computing the commutators with $\hP_{m}^{\mu}$. The computation is straightforward in the sense that we do not need the actual value of the commutators to do it. Explicitly, we have
\begin{align*}
	[\hcD_{n}^1,\hP_m^\mu]
	&= \left[\f{1}{2} \sum_{k=0}^{n} \left( \hJ_{k}^{\alpha} \hP_{n-k}^{\alpha} + \hP_{k}^{\alpha} \hJ_{n-k}^{\alpha} \right),\hP_m^\mu \right] \cr
	&= \f{1}{2} \sum_{k=0}^{n} [\hJ_{k}^{\alpha},\hP_m^\mu] \hP_{n-k}^\alpha + 	\hP_{k}^{\alpha} [\hJ_{n-k}^{\alpha},\hP_m^\mu] \cr
	&= \sum_{k=0}^{n} \hP_{k}^{\alpha} [\hJ_{n-k}^{\alpha},\hP_m^\mu] \; ,
\end{align*}
where we have used that $\hP_m^{\mu}$ is abelian from the first line to the second line and performed the change of variable $k \rightarrow n-k$ on the first term from the second line to the third line. The computation of the remaining two contributions are almost the same as this one, since the only difference is the range of the sum and the switch $k \rightarrow -k$. While the switch is not a problem, the range of the sum is more troublesome due to the discreteness of the disk. We have
\begin{align*}
	[\hcD_{n}^2,\hP_m^\mu]
	&=\left[ \sum_{k=1}^{\floor{N/2}} \left( \hJ_{-k}^{\alpha} \hP_{n+k}^{\alpha} + \hP_{-k}^{\alpha} \hJ_{n+k}^{\alpha} \right),\hP_m^\mu \right] \cr
	&= \sum_{k=1}^{\floor{N/2}}  [\hJ_{-k}^{\alpha},\hP_m^\mu] \hP_{n+k}^\alpha + \hP_{-k}^{\alpha} [\hJ_{n+k}^{\alpha},\hP_m^\mu] \cr
	&= \sum_{k=n+1}^{\floor{N/2}+n}  \hP_{k}^\alpha [\hJ_{n-k}^{\alpha},\hP_m^\mu]  +  \sum_{k=-\floor{N/2}}^{-1} \hP_{k}^{\alpha} [\hJ_{n+k}^{\alpha},\hP_m^\mu]
\end{align*}
and
\begin{align*}
	[\hcD_{n}^2,\hP_m^\mu]
	&=\left[ \sum_{k=\floor{N/2}-n+1}^{\floor{N/2}} \left( \hJ_{-k}^{\alpha} \hP_{n+k}^{\alpha} + \hP_{-k}^{\alpha} \hJ_{n+k}^{\alpha} \right),\hP_m^\mu \right] \cr
	&= \sum_{k=\floor{N/2}-n+1}^{\floor{N/2}}  [\hJ_{-k}^{\alpha},\hP_m^\mu] \hP_{n+k}^\alpha + \hP_{-k}^{\alpha} [\hJ_{n+k}^{\alpha},\hP_m^\mu] \cr
	&= \sum_{k=\floor{N/2}+1}^{\floor{N/2}+n}  \hP_{k}^\alpha [\hJ_{n-k}^{\alpha},\hP_m^\mu]  +  \sum_{k=-\floor{N/2}}^{-\floor{N/2}+n-1} \hP_{k}^{\alpha} [\hJ_{n+k}^{\alpha},\hP_m^\mu]
\end{align*}
where, in both case, we have performed the change of variable $k \rightarrow k-n$ on the first term and $k \rightarrow -k$ on the second term. From this, we obtain that
\begin{equation*}
	[[\hcD_{n}^{\text{alt}},\hP_m^\mu]] = \sum_{k=-\floor{N/2}}^{\floor{N/2}}  \hP_{k}^\alpha [\hJ_{n-k}^{\alpha},\hP_m^\mu] + \sum_{k=\floor{N/2}+1}^{\floor{N/2}+n} \hP_{k}^\alpha [\hJ_{n-k}^{\alpha},\hP_m^\mu] \;.
\end{equation*}
Hence, we have to much terms compared to what we wanted. Adding the last term to the commutators will cancel this term, but also some others
\begin{equation*}
	[\hcD_{n},\hP_m^\mu] = \sum_{k=-\floor{N/2}}^{\floor{N/2}}  \hP_{k}^\alpha [\hJ_{n-k}^{\alpha},\hP_m^\mu] - \sum_{k=-\floor{N/2}}^{-\floor{N/2}+n-1} \hP_{k}^\alpha [\hJ_{n-k}^{\alpha},\hP_m^\mu] \;.
\end{equation*}

In both cases however, we recover the expected results from $n=0$ and in the continuum limit.
\bigskip

The computation of the commutator with $\hJ_m^a$ is more complicated, since it involves the actual value of the commutators. We will focus on computing $[\hcD_{n}^{1},\hJ_m^a]$, since the other two can be directly deduced from it and what we learn previously. We define $e_m = e^{i \f{2\pi}{N}k m}$. We have
\begin{align*}
	[\hcD_{n}^1,\hJ_m^a]
	&= \left[\f{1}{2} \sum_{k=0}^{n} \left( \hJ_{k}^{\alpha} \hP_{n-k}^{\alpha} + \hP_{k}^{\alpha} \hJ_{n-k}^{\alpha} \right),\hJ_m^a \right] \cr
	&= \f{1}{2} \sum_{k=0}^{n} [\hJ_{k}^{\alpha},\hJ_m^a] \hP_{n-k}^\alpha + \hP_{k}^\alpha  [\hJ_{n-k}^{\alpha},\hJ_m^a] + \hP_{k}^{\alpha} [\hJ_{n-k}^{\alpha},\hJ_m^a] + [\hP_{k}^{\alpha} ,\hJ_m^a]\hJ_{n-k}^{\alpha} \cr
	&= \f{1}{2} \sum_{k=0}^{n} \big( i \eps^{\alpha a \beta} \left(\f{1}{2}(1+e_m)J_{k}^{\alpha}P^{\beta}_{n+m-k} + J^{\beta}_{m+k} P^{\alpha}_{n-k} + P^{\alpha}_k J^\beta_{n+m-k} + \f{1}{2}(1+e_m) P^{\beta}_{k+m} J^{\alpha}_{n-k} \right) \cr
	&- i \delta^{\alpha a} \left( (e_m-1)J_k^\alpha P^{0}_{n+m-k} + (e_m-1)P^{0}_{m+k} J^{\alpha}_{n-k} \right) \cr
	&- \f{i}{4} \left( (1-e_m) P^{a}_{n+m-k} \delta_{k,0} + (1-e_m)P^a_{m+k} \delta_{n-k,0}\right) \big)
\end{align*}
We will now make use of the basic commutators to rewrite all the terms such that $\hJ$ is in front of $\hP$, and we regroup all the terms in $\hP^{a}_{n+m}$. Doing so, we obtain
\begin{align*}
	[\hcD_{n}^1,\hJ_m^a] &=
	\f{1}{2} \sum_{k=0}^{n} \big( i \eps^{\alpha a \beta} \left(\f{1}{2}(1+e_m)J_{k}^{\alpha}P^{\beta}_{n+m-k} + J^{\beta}_{m+k} P^{\alpha}_{n-k} + J^\beta_{n+m-k} P^{\alpha}_k + \f{1}{2}(1+e_m)  J^{\alpha}_{n-k} P^{\beta}_{k+m} \right) \cr
	&- i \delta^{\alpha a} \left( (e_m-1)J_k^\alpha P^{0}_{n+m-k} + (e_m-1)J^{\alpha}_{n-k} P^{0}_{m+k} \right) \cr
	&- \f{i}{4} \left( (1-e_m) P^{a}_{n+m-k} \delta_{k,0} + (1-e_m)P^a_{m+k} \delta_{n-k,0}\right) \big) \cr
	&+\f{1}{4}\left( -(1+e_{n-k})(1+e_m) + 2 (1+e_{n+m-k}) + (e_m-1)(1-e_{n-k})  \right) \hP^{a}_{n+m}
\end{align*}
The term in front of $\hP^a_{n+m}$ vanishes. Doing our now usual transformation $k \rightarrow n-k$ on the $n+m-k$ terms return finally
\begin{align*}
	[\hcD_{n}^1,\hJ_m^a]
	&= \sum_{k=0}^{n} J^\alpha_{n-k} \left( \f{i}{2} \eps^{\alpha a \beta} (1+e_m) P^{\beta}_{k+m} - i \delta^{\alpha a} (e_m-1) P^{0}_{k+m} \right) + \f{i}{4}P^{a}_{k+m}\delta{n,k} + i \eps^{\alpha a \beta} J^{\beta}_{k+m} J^\alpha_{n-k} \cr
	&= \sum_{k=0}^{n} J^\alpha_{n-k} [P^\alpha_k,J^a_m] + [J^\alpha_k,J^a_m]J^{\alpha}_{n-k}
\end{align*}
and we obtain the desired result.

\bibliographystyle{bib-style}
\bibliography{3dQG}

\end{document}